\documentclass[11pt]{article}

\usepackage[left=3cm,right=2.5cm,top=2.5cm,bottom=3cm]{geometry}

\usepackage{amsmath}
\usepackage{amsfonts}
\usepackage{amssymb}
\usepackage{amsthm}
\usepackage{array}
\usepackage{bbold}
\usepackage{dsfont}
\usepackage{calc}
\usepackage{xcolor}
\usepackage{framed}
\usepackage{mdframed}
\usepackage{braket}
\usepackage{ulem}
\usepackage{comment}
\usepackage{mathrsfs}  
\usepackage{bm}  
\usepackage{stmaryrd}

\newtheorem*{conjecture}{Conjecture}

\definecolor{mypink}{RGB}{237, 152, 152}
\definecolor{myblue}{RGB}{161, 182, 242}

\usepackage{hyperref}
\hypersetup{colorlinks=true, citecolor=blue, linkcolor=blue,urlcolor= blue}

\usepackage{caption}
\usepackage{float}

\usepackage[utf8]{inputenc}
\usepackage[T1]{fontenc}
\usepackage[french, english]{babel}

\usepackage{csquotes}
\MakeOuterQuote{"}

\usepackage{graphicx}

\numberwithin{figure}{section}
\numberwithin{equation}{section}
\numberwithin{table}{section}

\newcommand{\ud}{\mathrm{d}}
\newcommand{\e}{\mathrm{e}}
\newcommand{\g}{\boldsymbol{\gamma}}
\newcommand{\R}{\mathds{R}}

\newcommand{\sm}{\mathrm{s}}
\newcommand{\um}{\mathrm{u}}
\newcommand{\Sr}{S_{\mathrm{r}}}

\newcommand{\apeupn}{\underset{N \rightarrow \infty}{\asymp}}
\newcommand{\apeupT}{\underset{\tt \rightarrow \infty}{\asymp}}
\newcommand{\df}{\mathfrak{D}}
\newcommand{\un}{N}

\newcommand{\bpi}{\boldsymbol{\pi}}
\newcommand{\A}{\boldsymbol{A}}
\newcommand{\bA}{\bar{\A}}
\newcommand{\D}{\mathcal{D}}
\newcommand{\K}{\boldsymbol{K}}
\newcommand{\ba}{\boldsymbol{a}}
\newcommand{\kk}{\boldsymbol{k}}
\renewcommand{\u}{\boldsymbol{u}}
\renewcommand{\S}{\boldsymbol{F}}
\newcommand{\bkappa}{\boldsymbol{\kappa}}
\renewcommand{\u}{\boldsymbol{u}}
\newcommand{\bl}{\boldsymbol{\lambda}}

\newcommand{\z}{{\boldsymbol{z}}}
\newcommand{\Z}{{\boldsymbol{Z}}}
\newcommand{\f}{{\boldsymbol{f}}}
\newcommand{\p}{{\boldsymbol{p}}}
\renewcommand{\P}{{\boldsymbol{P}}}

\newcommand{\n}{\boldsymbol{n}}
\newcommand{\m}{\boldsymbol{m}}
\newcommand{\N}{\boldsymbol{N}}

\renewcommand{\d}{\bm{\mathfrak{D}}}

\newcommand{\U}{\bm{U}}

\newcommand{\bmu}{\bm{\mu}}
\newcommand{\om}{\bm{\omega}}

\newcommand{\cano}{\mathrm{cano}}

\renewcommand{\L}{\mathscr{L}}
\renewcommand{\H}{\mathscr{H}}
\renewcommand{\tt}{T}
\renewcommand{\t}{\tau}
\newcommand{\ebd}{\mathcal{E}}

\renewcommand{\r}{\textsc{r}}
\newcommand{\ur}{{\mathrm{r}}}
\newcommand{\er}{\epsilon \r}
\newcommand{\pa}{\mathfrak{K}_{+1}}
\newcommand{\ma}{\mathfrak{K}_{-1}}
\newcommand{\pb}{\mathfrak{K}_{+2}}
\newcommand{\mb}{\mathfrak{K}_{-2}}
\newcommand{\Hg}{H_{\g}}

\newcommand{\scgf}{\Gamma}
\newcommand{\scgfn}{\bar \Gamma}

\newcommand{\fp}{\mathfrak{L}}
\newcommand{\fpr}{\mathcal{L}}

\newcommand{\uP}{\mathds{P}}
\newcommand{\uPk}{\mathds{P}_{\tilde{\kk},\bpi(0)}[n]}
\newcommand{\uPkcond}{\mathds{P}_{\tilde{\kk},\bpi(0)}\left[n ~ \middle| ~ \tilde\A_t\left[ n \right] = \textbf{a} \right]}
\newcommand{\uPkconddiff}{\mathds{P}_{\Lambda,\varrho(0)}\left[x_t ~\middle|~ \tilde\A_t = \textbf{a} \right]}

\newcommand{\uPmicro}{\mathds{P}^{\text{micro}}_{\textbf{a},\bpi(0)}[n]}
\newcommand{\uPmicrodiff}{\mathds{P}^{\text{micro}}_{\textbf{a},\varrho(0)}[x_t]}

\newcommand{\formicro}{linear operator formalism}

\title{{Effective Hamiltonians and Lagrangians for conditioned Markov processes at large volume}}

\author{Lydia Chabane$^1$ \and Alexandre Lazarescu$^2$ \and Gatien Verley$^{1}$\footnote{E-mail: gatien.verley@universite-paris-saclay.fr (corresponding author)}}
\date{%
    $^1$Université Paris-Saclay, CNRS/IN2P3, IJCLab, 91405 Orsay, France \\%
    $^2$Institut de Recherche en Math\'ematiques et Physique, UCLouvain, Louvain-la-Neuve, Belgium\\[2ex]% 
    \today
}

\begin{document}
\maketitle

\selectlanguage{english}
\newpage
\begingroup
\hypersetup{linkcolor=black}
\tableofcontents
\endgroup
\newpage

\begin{abstract}

When analysing statistical systems or stochastic processes, it is often interesting to ask how they behave given that some observable takes some prescribed value. 
This conditioning problem is well understood within the linear operator formalism based on rate matrices or Fokker-Planck operators, which describes the dynamics of many independent random walkers. 
Relying on certain spectral properties of the biased linear operators, guaranteed by the Perron-Frobenius theorem, an effective process can be found such that its path probability is equivalent to the conditional path probability.  
In this paper, we extend those results for nonlinear Markov processes that appear when the many random walkers are no longer independent, and which can be described naturally through a Lagrangian--Hamiltonian formalism within the theory of large deviations {at large volume.} 
We identify the appropriate spectral problem as being a Hamilton-Jacobi equation for a biased Hamiltonian, for which we conjecture that two special global solutions exist, replacing the Perron-Frobenius theorem concerning the positivity of the dominant eigenvector.
We then devise a \textit{rectification} procedure based on a canonical gauge transformation of the biased Hamiltonian, yielding an effective dynamics in agreement with the original conditioning. 
Along the way, we present simple examples in support of our conjecture, we examine its consequences on important physical objects such as the fluctuation symmetries of the biased and rectified processes as well as the dual dynamics obtained through time-reversal. 
We apply all those results to simple independent and interacting models, including a stochastic chemical reaction network and a population process called the Brownian Donkey.

\end{abstract}

\section{Introduction}

Conditioning is ubiquitous in thermodynamics and equilibrium statistical physics. It appears in the choice of an equilibrium ensemble when defining the external conditions or constraints on the studied macroscopic system. For instance in the microcanonical ensemble, the energy of an isolated systems is fixed, whereas in the canonical ensemble, this constraint on energy is replaced by a constraint on the temperature of the reservoir coupled to the system, allowing the energy to fluctuate~\cite{Huang1987, Book_Kubo1998, Balian2007, Matsoukas2018}. 
As argued by Gibbs in his book \textit{Elementary Principles in Statistical Mechanics}~\cite{Gibbs1902}, the canonical and microcanonical ensembles are equivalent in the thermodynamic limit when the microcanonical entropy is strictly concave. In this case, the mean energy in the canonical ensemble with the appropriate temperature equals the fixed value of energy in the microcanonical ensemble. This equivalence exists due to the Legendre structure of equilibrium statistical physics: the same equilibrium state is reached both from the maximum entropy state in the constant energy shell or from the minimum free energy state in the constant temperature shell~\cite{Callen1985_vol}. 

Beyond pure static theory yielding the system equilibrium state and the statistics of thermodynamic observables in that state, it may be physically relevant (or prospectively pertinent) to determine the succession of states that leads to a given fluctuation. Searching such a dynamical fluctuation requires information on the system dynamics that itself may be more or less detailed according to the modeling choice. Physicists rely on many types of stochastic processes in this view. Their definition in agreement with fundamental physical principles has become a discipline within statistical physics~\cite{Kampen2007}. Once using the probabilistic framework, the problem of conditioning a stochastic process is well-defined thanks to conditional probabilities. In the end, one looks for a new process with no conditioning that will have a law reproducing the original process under conditioning. In this formulation, the analogy between thermodynamic ensemble equivalence and stochastic processes equivalence is rather clear and turns out to be technically useful. With this conditioning-free process at hand, the statistics of any other observable can be determined under the chosen constraint.

The choice of the simplest modeling for a physical system is at the core of physical sciences. As aforementioned, reducing the description of a system with many degrees of freedom into dynamical equations for a small set of mesoscopic variables is of great practical interest, as for instance in hydrodynamics~\cite{Sasa2014vol112}, elasticity theory~\cite{Barrat2018vol504} and more generally in field theories~\cite{Hartmann2001vol32}. Regarding thermodynamics, from the microscopic dynamical description at the level of elementary degrees of freedom to the macroscopic thermodynamic description based on few thermodynamic variables~\cite{Book_Peliti2011}, a considerable simplification is achieved with a great theoretical consistency. This simplification holds even at the fluctuations level~\cite{Callen1985_vol}, since thermodynamic potentials are formally large deviation function (LDF) or scaled cumulant generating functions (SCGF) in the formalism of large deviation theory~\cite{Touchette2009}. 
Thermodynamic theory remains fully consistent even when considering few degrees of freedom modeled by stochastic processes, as demonstrated by stochastic thermodynamics~\cite{sekimoto2010stochastic,Seifert2012_vol75,Esposito2012_vol85}. However, the notion of thermodynamic potential fall apart when breaking equilibrium beyond the linear regime even in stationary states, except in very special cases~\cite{Verley2016_vol93}. In this situation, the system modeling relies on dynamical equations with appropriate implementation of thermodynamic properties through local detailed balance~\cite{Seifert2012_vol75} and the fluctuations of the associated stochastic processes are considered using dynamical large deviation theory~\cite{Barato2015_vol,Freidlin2012vol,Touchette2013vol,Bertini2015_vol87,Weber2017_vol80,assaf2017wkb,Monthus2019}. The literature on equilibrium~\cite{Touchette2003_vol,Tailleur2007_vol99,Lecomte2010vol184,Mielke2014_vol41, vroylandt2019}, close to equilibrium~\cite{Colangeli2011_vol44,Maes2007_vol48,Maes2008_vol82} and far from equilibrium dynamical fluctuations~\cite{Derrida1998_vol301,Lecomte2005_vol2005,Wynants2010_vol,Baule2010_vol2010,Gorissen2012_vol109,Lazarescu2011_vol44,Lazarescu2015_vol48,Garrahan2018_vol504,Maes2020vol850} has flourished over the last decades, and many historical references can be found in~\cite{Varadhan2008_vol36}. 

For processes with few degrees of freedom, large deviations in time are used while for extensive systems large deviations in size or both in size and time are possible. For the former, the conditioning of Markov processes was initiated for diffusive or jump processes for simplicity reasons. Historically, at least in the physics literature, investigations on the fluctuation relations for physical currents motivated the biasing of such processes to determine their currents statistics~\cite{Kurchan1998_vol31,Lebowitz1999_vol95}.
Another motivation was to understand the peculiarities of activity fluctuations across a dynamical phase transition in glasses~\cite{Lecomte2007_vol,Garrahan2007vol98,Lecomte2007_vol127,Garrahan2009_vol42} and numerical algorithm were designed in this view~\cite{Giardina2006_vol96,Giardina2011_vol145,Ferre2018vol172}. The exponential biasing of the trajectory probability by a product of conjugated variables (the chosen observable and its counting field) is called \textit{exponential tilting}, \textit{Gibbs conditioning} or \textit{canonical path ensemble} in analogy with equilibrium ensemble terminology~~\cite{Evans2004, Evans2004b, evans2010statistical}. However, the linear operator associated with the process, either a biased rate matrix or Fokker-Planck operator, is not of the same type as the unbiased one since for instance it does not conserve probability. This leads to introduce the generalized Doob transform of these biased generators to define a suitable process called the driven process~\cite{Jack2010, Chetrite2013_vol111}. Using this intermediate process, the logarithmic equivalence of a conditioned path probability with a tilted path probability has been demonstrated~\cite{Chetrite2014}. In the case of jump processes, the Doob transform is technically a similarity transformation combined with a translation on the diagonal of the rate matrix. For the present work, it is fundamental to note first that a Doob transform is a gauge transformation~\cite{Garrahan2016_vol2016}, second that it relies on the Perron-Frobenius theorem to guarantee the existence of a unique left eigenvector of the biased generator. 

In the present work, we investigate the question of process conditioning when dealing with a large number of processes (independent or interacting) such that a reduced description emerges at large volume through intensive state variables (concentration, density, etc)~\cite{Ritort2004_vol2004,Book_Ross2008}. When looking at large deviations for a size-type scaling parameter \cite{wentzell1990} (e.g. volume, number of particles, etc.), the computation of a moment generating function yields a size and time extensive contribution given by an action~\cite{Phdthesis_Vroylandt2018,vroylandt2020efficiency}. This action describes the fluctuations at the corresponding large deviation scale. Its minimum provides the typical trajectories contributing the most to the moment generating function. For those trajectories, a Lagrangian--Hamiltonian description is possible and the Lagrangian can even be identified as a LDF for the state variables and their associated currents~\cite{Peletier2017, Lazarescu2019}. As usual when using reduced descriptions, the large size limit may lead to a nonlinear dynamics associated with critical phenomena \cite{Garrahan2007}. Letting aside such interesting questions, we aim in this paper to generalize the problem of conditioning, biasing and rectification in the Lagrangian--Hamiltonian formalism {emerging at large volume, i.e. when there exists a large deviation principle associated to a size-type scaling parameter.}

In section~\ref{Jump}, we consider a single Markov jump process and then $N$ independent copies of this process, and we review the problem of conditioning, biasing and rectification (generalized Doob transform) within the \formicro. The same calculations in the case of diffusion processes are done in Appendix.
{Treating $N$ independent processes does not add significant complexity and represents a first step for switching from the linear operator formalism to the Lagrangian/Hamiltonian formalism. However, the size of the linear operators increases making them less convenient to study their rectification. In addition, in the large-volume limit, coarse-grained observables such as empirical densities and currents become more interesting than their microscopic counterpart, and the Lagrangian/Hamiltonian formalism is particularly well adapted to their study.}

In Section~\ref{LHdescription}, we review large deviation theory in the Lagrangian--Hamiltonian formalism for nonlinear Markov processes~\cite{Lazarescu2019}. 
We explain how biased Lagrangians and Hamiltonians appear
in the large volume limit of the path integral used to compute probabilities or moment generating functions. Then, we recall the formulations of the dynamical problem (Euler-Lagrange, Hamilton and HJ equations) that must be solved to find the scaled cumulant generating function. Special solutions, called critical manifolds, of this dynamical problem provide dominant contributions to the path integral and thus take a central place in our discussions. 

In section~\ref{Perron-section}, we focus on understanding the structure of phase space and more precisely on the shape of the time-independent solutions of the HJ equation that go through a critical manifold. Those solutions called Hamilton's characteristic functions~\cite{Rax2020} play the same role as eigenvectors or eigenfunctions of the \formicro. For the rectification of nonlinear processes to exist, we need to guarantee the existence of such solutions. We do so by proposing a conjecture that generalizes the Perron-Frobenius theorem for nonlinear dynamics modeling stochastic processes, i.e. for statistical Hamiltonians.  

The rectification formula for nonlinear processes is finally given in section \ref{Sec_rectification}. The Doob transform of the \formicro\,becomes a canonical transformation corresponding to a gauge change, the gauge function being one of Hamilton's characteristic functions. Since process rectification relates different processes, we also investigate other equivalences between dynamics coming from fluctuation relation symmetry (dynamics at different affinities) or from the time reversal symmetry (dual dynamics). 

In section~\ref{application}, we apply our results to population processes: first for a general population process that includes the case of $N$ independent Markov processes as a sub-case, then for two specific models: a model of interacting machine called the Brownian donkey~\cite{Cleuren2001, vroylandt2019, vroylandt2020efficiency} and a simple model of chemical reaction \cite{Book_Ross2008}.

\section{Linear Markov jump processes: Exponential biasing and rectification} \label{Jump}

Before moving on to the Lagrangian--Hamiltonian framework, we review in this section the conditioning, biasing and rectification of Markov jump processes within the \formicro.
Conditioning (respectively biasing) a stochastic process on a specific value of an observable selects (respectively favors) trajectories leading to that value. Yet, the conditioned process is not Markovian, while the biased process is, although it does not conserve probability, i.e. it does not evolve a normalized initial state distribution into a normalized one. The rectification of the biased process enables to determine a norm-conserving Markov process that typically generates the trajectories of the conditioned process.

In this section, we focus on Markov jump processes starting with a single process for simplicity. We then consider $N$ independent processes as it will be the simplest application of the nonlinear framework of Section.~\ref{LHdescription}.
 
\subsection{Single process}

We consider a Markov jump process with a finite number $\Omega$ of states denoted by $l$, $m$ or $n$. The generator of this process is the time-independent square matrix $\tilde{\kk}$ whose non-negative off-diagonal component $\tilde{k}_{nm}$ is the transition rate from state $m$ to state $n$ \footnote{We use an over-tilde on the generator of the single process dynamics in order to distinguish it from the many-body process dynamics on which we focus in the remaining of this paper.}. The diagonal component $ \tilde{k}_{mm}\equiv -\sum_{n \neq m} \tilde{k}_{nm}$ corresponds to minus the escape rate from state $m$. We denote by $\pi_n \equiv \pi_n(t)$ the probability of occupying the state $n$ at time $t$. It satisfies the master equation
\begin{equation} \label{Eq_maitresse_standard}
\dot \pi_n = \sum_m \tilde{k}_{nm} \pi_m,
\end{equation}
where the over-dot stands for the time derivative. The master equation conserves the normalization of the probability since by construction $\sum_n \tilde{k}_{nm} = 0, ~ \forall m$. Combined with the normalized initial probability, this ensures the normalization of the probability at all times $\sum_n \pi_n(t) =1$. We assume that $\bm{\pi}$ reaches a stationary solution of the master equation after a sufficiently long time. We denote by $[n]$ a path consisting of the succession of states visited by the system during a interval of time $[0,t]$ and the times $t_\textsc{i}$ ($\textsc{i}=1,\dots,M$) at which the system jumps. In other words, $[n]$ includes all the information to build the piecewise constant function giving the state occupied by the system at all times:
\begin{eqnarray}
n(\t) = n_\textsc{i} \qquad \mathrm{for} \qquad  t_\textsc{i} \leq \t < t_{\textsc{i}+1}.
\end{eqnarray}
For a path with $M$ jumps, $t_0=0$ is the initial time and $t_M$ is the last jump time before the final time $t$. The path probability $\uPk$ of path $[n]$ is given by
\begin{equation} \label{path_proba}
\uPk = \pi_{n_0}(0) \exp\left[ \sum_{\textsc{i}=0}^{M-1} \ln\left(\tilde{k}_{n_{\textsc{i}+1},n_\textsc{i}} \right)  - \int_0^{t} \sum_{m \neq n(t')} \tilde{k}_{m, n(t')} \ud t' \right], 
\end{equation}
where $\pi_{n_0}(0)$ is the initial probability to occupy the initial state $n_0$. 

Most thermodynamic observables (heat, matter currents, work, entropy production, energy, etc.) write as linear combinations of empirical transition current and empirical occupancy.
The empirical transition current $\tilde{\om}_t[n]$ is a $\Omega \times \Omega$ matrix and its component $\tilde{\omega}_{lm}[n]$ counts the number of transitions $m \rightarrow l$ per unit of time along the trajectory~$[n]$:
\begin{eqnarray} \label{emp_current_1process}
    \tilde{\omega}_{lm} [n] &=& \frac{1}{t} \sum_{\textsc{i}=0}^M \delta_{n_{\textsc{i}+1},l} \delta_{n_{\textsc{i}},m}.
\end{eqnarray}
The empirical occupancy $\tilde{\bmu}_t[n] $ is a vector of dimension $\Omega$ and its component $\tilde{\bmu}_l[n]$ counts the rate of occupancy of each state along the trajectory $[n]$:
\begin{eqnarray} \label{emp_occupancy_1process}
    \tilde{\mu}_{l} [n] &=& \frac{1}{t} \int_{0}^{t} dt' \delta_{n(t'),l}.
\end{eqnarray}
Let us consider the observable $\tilde \A_t = (\tilde{\om}_t,\tilde{\bmu}_t)$. 
We want to condition our original Markov process of generator $\tilde{\kk}$ by filtering the ensemble of paths to select those leading to a chosen value $\ba$ of the observable $\tilde\A_t$. This defines a new process called \textit{the conditioned process} for which we aim to find an equivalent Markov process in the long-time limit~\cite{ChetriteHDR}.
This process is described by the \textit{microcanonical path probability}~\cite{Chetrite2014}
\begin{equation} \label{pathprobamicro}
\uPmicro = \uPkcond.
\end{equation}
In general, there is no Markov generator associated with this path probability. Yet, one can build a norm-conserving Markov process called \textit{the driven process} which enforces $\tilde\A_t$ to have the value $\ba$ as a typical value~\cite{Chetrite2014, Touchette2015}. To find the generator of this driven process, we need to introduce an intermediate process --- \textit{the biased process} --- which arises from the exponential bias of the path probability, and allows the calculation of the moments of $\tilde\A_t$. The moment generating function for the observable $\tilde\A_t$ imposing a final state $m$ reads
\begin{equation}
G_m(t{,\g}) \equiv \left\langle \e^{t \g \cdot \tilde\A_t[n]} \delta_{n(t),m} \right\rangle_{\tilde{\kk},\bpi(0)},
\end{equation}
where $\langle \ldots \rangle_{\tilde{\kk},\bpi(0)}$ is the path average with respect to the path probability~\eqref{path_proba}. For clarity, we made implicit the dependence of $G_m(t{,\g}) $ on the conjugated variable vector $\g \equiv \left(  \g^1 ~ \g^2 \right)$, with $\g^1$ the matrix with components $\gamma^1_{lm}$ conjugated to $\tilde{\omega}_{lm}$ and $\g^2$ the vector with components $\gamma^2_{l}$ conjugated to $\tilde{\bmu}_l$. 
This generating function evolves according to
\begin{equation}
\dot{\bm{G}} = \tilde{\bkappa} \bm{G},
\end{equation} 
where we defined the biased matrix $\tilde{\bkappa} \equiv \tilde{\bkappa}(\g)$ with components
\begin{equation} \label{biasedmatrix}
\tilde{\kappa}_{nm}(\g) \equiv \left\{
      \begin{aligned}
        & \tilde{k}_{nm} \e^{\gamma^1_{nm}}   & & \mathrm{if} ~ n \neq m, \\
         - & \sum_{n \neq m} \tilde{k}_{nm} + \gamma^2_{m} & & \mathrm{if} ~ n = m.
      \end{aligned}
    \right.
\end{equation}
This biased matrix is not norm-conserving: $\sum_n \kappa_{nm} \neq 0$, $\forall m$. Then, the generator of the driven process follows from applying to the biased matrix a generalized Doob transform which allows to build norm-conserving generators out of arbitrary ones~\cite{Chetrite2013, Chetrite2014}. Mathematically, the Doob transform $\bm M^{\bm v}$ of a matrix $\bm M$ using a vector $\bm v$ reads component-wise
\begin{equation}
M^{\bm v}_{nm} = v_n M_{nm} v_m^{-1} - v_n^{-1} (\bm v \bm M)_n \delta_{nm}, \label{general_doob}
\end{equation}
with $\bm M$ an arbitrary Metzler matrix~\cite{Farina2000} and $\bm v$ a vector whose elements are positive. The generator $\tilde{\K}$ of the driven process follows from the Doob transform of the biased matrix $\tilde{\bkappa}$ using its left eigenvector $\e^{\u}$ associated with its dominant eigenvalue $\scgfn \equiv \scgfn(\bm \gamma)$:
\begin{equation} \label{drive_generator}
\tilde{K}_{nm}(\g) \equiv \left\{
      \begin{aligned}
        & \tilde{\kappa}_{nm} \e^{u_{n} - u_{m}} = \tilde{k}_{nm} \e^{\gamma^1_{nm}} \e^{u_{n} - u_{m}}  & & \mathrm{if} ~ n \neq m, \\
        & - \sum_{n \neq m} \tilde{k}_{nm} + \gamma^2_{m} - \scgfn & & \mathrm{if} ~ n = m.
      \end{aligned}
    \right.
\end{equation}
Note that the positivity of $\e^{\u}$ is ensured by the Perron-Frobenius theorem~\cite{Johnson2012, Bokharaie2012}. The fact that $e^{\u}$ is the left eigenvector of $\tilde{\bkappa}$ associated with $\scgfn$ ensures that
\begin{equation} \label{Knn}
\tilde{K}_{mm} = - \sum_{n \neq m} \tilde{K}_{nm}.
\end{equation}
One can show that $\scgfn$ is the scaled cumulant generating function (SCGF) defined by
\begin{equation}  \label{SCGF}
\scgfn \equiv \lim_{t \rightarrow \infty} \frac{1}{t} \ln \left \langle \e^{t \g \cdot \tilde\A_t[n]} \right \rangle_{\tilde{\kk}, \bpi(0)}.
\end{equation} 
With such a definition of the driven process, it was established that for a specific value of $\g$, the dynamics generated by $\tilde{\K}$ has typical trajectories for which $\tilde\A[n]$ converges to the imposed value $\ba$ used to condition our original process (given the convexity of its large deviation function)~\cite{Chetrite2014, Chetrite2015}.

\subsection{$N$ independent processes} \label{SecNjumps}
We now consider $\un$ independent and identical systems, each one modeled by a Markov jump process described by Eq.~\eqref{Eq_maitresse_standard}. We label by $\nu \in \{1, 2, \dots N\}$ the $\nu^\text{th}$ system and by $n^{\nu} \in \{1, \dots  \Omega \}$ the associated states. The microstate vector $\n \equiv \left( \begin{array}{ccccc} n^1 &\dots & n^{\nu} & \dots &  n^N \end{array} \right)^T$ denotes the state of the global system and informs on the state of each system.  The probability $p_{\n}$ that the global system is in state $\n$ satisfies the master equation
\begin{equation}
\dot{p}_{\n} = \sum_{\m} \tilde{k}_{\n\m} p_{\m},
\end{equation}
where we introduced (with a slight abuse of notation) the transition rate from $\m$ to $\n$:
\begin{equation}
\tilde{k}_{\n\m} \equiv \sum_{\nu=1}^N \tilde{k}_{n^\nu m^\nu} (1-\delta_{n^\nu,m^\nu}),
\end{equation}
meaning that the transition $\m \rightarrow \n$ at the level of the global system corresponds to only one transition $m^\nu \rightarrow n^\nu$ performed by the $\nu^\mathrm{th}$ system. We look for a more coarse-grained description of the global system. To do so, we introduce the mesostate vector $\N(\n) \equiv \left( \begin{array}{ccccc} N_1 & \ldots & N_n & \ldots & N_\Omega \end{array} \right)^T$ whose component $N_n \equiv \sum_{\nu} \delta_{n,n^{\nu}}$ gives the number of systems in state $n$ given the microstate $\n$. We are interested in the probability $P_{\N}$ that the global system is in state $\N$:
\begin{equation}
P_{\N} = \sum_{\n \mid \N(\n) = \N} p_{\n} \equiv  \sum_{\n_{\N}} p_{\n_{\N}},
\end{equation}
where the last sum is over the ensemble $\{ \n_{\N} \}$ of microstates compatible with a mesostate $\N$, implying that $p_{\n_{\N}}$ is the joint probability to be in $\n$ and $\N$. The master equation satisfied by $P_{{\N}}$ reads then
\begin{equation}
\dot{P}_{\N} = \sum_{\N'} k_{\N \N'} P_{\N'},
\end{equation}
where the transition matrix $\bm k$ reads in the Dirac notation:
\begin{equation} \label{rate_Npart}
\bm{k} \equiv \sum_{\N, \N' \neq \N} \sum_{n,m} \tilde{k}_{nm} N'_{m} \delta_{N_n, N'_n + 1} \delta_{N_m, N'_m - 1} \ket{\N} \bra{\N'} - \sum_{\N'} \sum_{n,m \neq n} \tilde{k}_{nm} N'_m \ket{\N'} \bra{\N'}.
\end{equation}
Note that $\sum_{\N} k_{\N \N'} = 0$ and that the sum $\sum_{\N, \N' \neq \N}$ is implied to run over the mesostates $\N$ and $\N'$ such that they differ by only one microscopic transition. Eq.~\eqref{rate_Npart} means that the transition probability to jump from $\N'$ to $\N$ is given by the probability of any microscopic transition $m \rightarrow n$ performed by any of the $N'_m$ systems occupying the state $m$.

We denote by $[\n]$ a path giving the succession of microstates $\n_t$ visited by the global system at any time $t$. We assume that the typical time scale during which a single process performs one transition is $\delta t$. Hence during $\delta t$ the global system undergoes typically $N$ transitions.
Similarly to the case of a single process, we are interested in two empirical observables from which one can define many thermodynamic observables: the empirical transition current $\om[\n]$ and the empirical density $\bmu[\n]$. The component $\un \ud t \, \omega_{lm}[\n](t)$ counts the number of systems performing the transition $m \rightarrow l$ between times $t$ and $t+\delta t$ along the path~$[\n]$:
\begin{equation} \label{emp_current_Npart}
\omega_{lm}[\n](t) = \frac{1}{\un \ud t} \sum_{\nu=1}^N \sum_{s \in [t,t + \ud t[} \delta_{l,n^\nu(s^+)} \delta_{m,n^\nu(s^-)} = \frac{1}{\un \ud t} \sum_{s \in [t,t + \ud t[} \left[\delta_{N_l(s^+),N_l(s^-)+1} \delta_{N_m(s^+),N_m(s^-) - 1} \right],
\end{equation}
where the sum on $s$ runs over the transition times in $[t,t+\ud t[$ with $s^+$ (respectively $s^-$) the time right after (respectively before) the transition.
The component $\mu_m[\n](t)$ counts the fraction of systems being at state~$m$ at time $t$ along the path~$[\n]$:
\begin{equation} \label{emp_density_Npart}
\mu_m[\n](t) \equiv \frac{1}{N} \sum_{\nu = 1}^N \delta_{n^\nu(t),m} = \frac{N_m(\n(t))}{N}.
\end{equation}
Note that these observables are related to the empirical transition current~\eqref{emp_current_1process} and occupancy~\eqref{emp_occupancy_1process} for a single process through
\begin{eqnarray}
\frac{1}{t} \int_0^t \ud t' \, \om[\n](t') &=& \frac{1}{\un} \sum_{\nu = 1}^N \tilde{\om}[n^\nu], \\
\frac{1}{t} \int_0^t \ud t' \,  \bmu[\n](t') &=& \frac{1}{\un} \sum_{\nu = 1}^N \tilde{\bmu}[n^\nu].
\end{eqnarray}
We would like to condition our original Markov process by filtering the ensemble of paths to select those leading to a chosen value of $\bm A_t = \un \left( \frac{1}{t} \int_0^t \ud t' \bm \omega(t'),\frac{1}{t} \int_0^t \ud t' \bm \bmu(t')\right)$. The generating function $G_{\N}(t{,\g}) \equiv \left\langle\e^{t \g \cdot \A_t[\n]} \delta_{\N(\n_t),\N}\right\rangle$ evolves according to
\begin{equation}
\dot{G}_{\N} = \sum_{\N'}  \bkappa_{\N \N'} G_{\N'},
\end{equation}
where we used the conjugated variable vector $\g \equiv \left(  \g^1 ,~ \g^2 \right)$ with components $\gamma^1_{lm}$ for the first one and $\gamma^2_{l}$ for the second one, and where the biased matrix $\bkappa \equiv \bkappa(\g)$ is given by
\begin{equation}
\bkappa(\g) = \sum_{\N, \N' \neq \N} \sum_{n,m \neq n} N'_m \tilde{\kappa}_{nm}(\g) \delta_{N_n, N'_n + 1} \delta_{N_m, N'_m - 1} \ket{\N} \bra{\N'} + \sum_{\N'} \sum_{m} N'_m \tilde{\kappa}_{mm}(\g) \ket{\N'} \bra{\N'},
\end{equation}
where $\tilde{\bkappa}$ is the biased matrix for a single system defined in Eq.~\eqref{biasedmatrix}. Again, we define the generator of the driven process $\K$ by taking the Doob transform of the biased matrix using its dominant left eigenvector. Keeping in mind that $\scgfn$ is the dominant eigenvalue of the single-process biased matrix $\tilde{\bkappa}$ and that $\e^{\u}$ is its associated left eigenvector, one can show that the dominant eigenvalue of $\bkappa$ is $N \scgfn$ and that the associated left eigenvalue reads in the Dirac notation: $\bra{\bm L} \equiv \sum_{\N} \e^{\N \cdot \u} \bra{\N} = \sum_{\N} \e^{\sum_{m=1}^\Omega N_m u_{m}}  \bra{\N}$. Computing the Doob transform, we obtain that the generator of the driven process $\K$ is related to the driven generator of a single process $\tilde{\K}$ through
\begin{equation}
\K = \sum_{\N, \N' \neq \N} \sum_{n,m \neq n} N'_m \tilde{K}_{nm} \delta_{N_n, N'_n + 1} \delta_{N_m, N'_m - 1} \ket{\N} \bra{\N'} + \sum_{\N'} \sum_{m} N'_m  \tilde{K}_{mm}  \ket{\N'} \bra{\N'}.
\end{equation}
In this way, we \textit{rectify} the biased process yielding a norm-conserving Markov jump process for our global process made up of $N$ independent and identical Markov processes. This linear process is a particular case of the more general class of nonlinear population processes that we study in Section~\ref{poprocesses}.

\section{Nonlinear Markov processes: Exponential biasing and spectral problem} \label{LHdescription}

In this section, we review the Lagrangian--Hamiltonian formalism used to describe time-independent stochastic processes in the framework of {low-noise} large deviation theory \cite{wentzell1990}. {Such systems can for instance be obtained as a large $N$ limit of interacting microscopic processes, of which examples can be found in section \ref{application}, though in this section we will study their properties irrespective of their microscopic origin, in as much generality as possible.} {This formalism applies in particular to the $N$ independent processes of the previous section in the limit $N \rightarrow \infty$ as explained in chapter 3 of~\cite{Book_Ross2008}. In those simple cases, the Hamiltonian will be linear in the scaled mesostate vector $\N / N$, which is not always true: most Hamiltonians are nonlinear in that sense, hence the title of the section.} 

\subsection{Lagrangian and Hamiltonian for Markov processes}

We consider time-homogeneous Markov processes characterized by a large size-type parameter $N$ (number of particles, volume, etc) and focus on two empirical observables: a current variable $\bl$ and a state variable $\z$. These variables will have precise definitions in specific contexts. {For instance, the variable $\bl$ may represent the empirical transition current $\bm \omega$ of Section~\ref{SecNjumps}, a matter current or a chemical current, while the variable $\z$ may represent the empirical density $\bm \mu$ of Section~\ref{SecNjumps} or a concentration. The domain of definition of $\z$, which we will call $\z$-space or state-space, is only assumed to be connected and open, and can vary depending on the context: if $\z=\{z_i\}$ is a concentration vector of size $n$, for instance, it can be any connected open subset of $\R^n$, whereas if $\z=z(x)$ is a density on $x \in \R^n$, it will be a connected open subset of some function space, e.g. $L^1\left({\R^{+}}^{n}\right)$. The nature of the current variable $\bl$ also varies accordingly.} 
The dynamics of $\z$ is determined by the currents $\bl$ through a conservation law:
\begin{equation} \label{EqConserv_observablesEmpiriques}
\dot{\z} = \D \bl,
\end{equation}
where $\D$ stands for a generalized divergence operator that will have precise definitions in specific contexts. We are interested in the transition probability $P(\z', t+\delta t \mid \z, t)$ of observing $\z(t~+~\delta t) = \z'$ at time $t + \delta t$ given $\z(t) = \z$ at time $t$, with $\delta t$ an infinitesimal time. Since we consider time-homogeneous Markov processes, the transition probability depends only on the difference $\delta t$ between final and initial times and we write $P_{\delta t}(\z' \mid \z)$ the conditional probability to observe $\z'$ after a time $\delta t$ given that the system was in $\z$. From Eq.~\eqref{EqConserv_observablesEmpiriques}, observing $\z'$ after $\delta t$ given $\z$ is entirely determined by knowing $\bl$ and $\z$ since $\z(t+\delta t) = \z(t) + \delta t \D \bl$. We can thus equivalently consider $P_{\delta t}(\bl \mid \z)$ the conditional probability of the current variable $\bl$ given the state variable $\z$ during the infinitesimal time interval $\delta t$. We assume that this probability satisfies a LDP and we call the associated LDF the \textit{detailed Lagrangian} $\L(\bl, \z)$ defined by
\begin{equation} \label{L_det}
\L(\bl,\z) \equiv - \underset{\delta t \rightarrow 0}{\lim_{\un \delta t \rightarrow \infty}} \; \frac{1}{\delta t \un} \ln P_{\delta t}(\bl \mid \z),
\end{equation}
{where the limit $\delta t N \rightarrow \infty$ and $\delta t \rightarrow 0$ guarantees that a large number of transitions occurs during the infinitesimal time $\delta t$. Then, we write for an infinitesimal $\delta t$}
\begin{equation}\label{lagr}
P_{\delta t}(\bl \mid \z) \underset{\un \delta t \rightarrow \infty}{\asymp}  \e^{-\delta t N \L(\bl,\z)}.
\end{equation}
The fact that those Lagrangians come from distributions $P_{\delta t}(\bl \mid \z)$ normalised with respect to $\bl$, which we will call {\it proper Lagrangians} in the rest of the paper, implies that for all $\z$ we have
\begin{equation} \label{cond_L}
\begin{cases}
\begin{aligned}
& \forall \bl, ~ \L(\bl,\z) \geq 0  , \\
& \exists \, \bl_*(\z), ~  \L(\bl_*(\z),\z) = 0.
\end{aligned}
\end{cases} 
\end{equation}
We can also consider a less detailed level of description by introducing a new Lagrangian $L(\dot{\z},\z)$ --- the \textit{standard Lagrangian} ---  defined by contracting the detailed Lagrangian~\eqref{L_det} over $\bl$ under the constraint~\eqref{EqConserv_observablesEmpiriques}:
\begin{equation} \label{L_stand}
L(\dot{\z},\z) \equiv \inf_{\bl | \dot{\z} = \D \bl} \L(\bl,\z).
\end{equation}
This is the usual Lagrangian of analytical mechanics (hence \textit{standard}). This Lagrangian is usually difficult to obtain explicitly while the detailed Lagrangian has an explicit formula for a wide number of systems, such as systems modeled by diffusive processes or by Markov jump processes.

Since the detailed Lagrangian corresponds to a large deviation function, one can also define the \textit{detailed Hamiltonian} $\H(\f,\z)$ corresponding to the scaled cumulant generating function for $\bl$ obtained from the Legendre-Fenchel transform of $\L(\bl,\z)$:
\begin{equation} \label{Hdet}
\H(\f,\z) \equiv \sup_{\bl} \{ \f \cdot \bl - \L(\bl,\z) \},
\end{equation}
where the central dot $\cdot$ denotes the scalar product (here in current space) and $\f$ is conjugated to $\bl$. Note that $\H$ is convex in $\f$ since it follows from a Legendre-Fenchel transform with respect to $\bl$. {In general, the Hamiltonian is nonlinear in $\z$, hence the name of "nonlinear process", but in the case where $\z$ is a density of independent particles (e.g. as in section \ref{SecNjumps}), it will be linear.}
{\it Proper Hamiltonians} associated with stochastic processes must satisfy, $\forall \z$,
\begin{equation} \label{cond_H}
\H(\f=0,\z) = 0
\end{equation}
to ensure that $P_{\delta t}(\bl \mid \z)$ is a propability.
Indeed, condition~\eqref{cond_L} and 
\begin{equation}
\H(\f=0,\z) = \sup_{\bl} \{- \L(\bl,\z) \} = - \inf_{\bl} \{\L(\bl,\z) \}
\end{equation}
imply Eq.~\eqref{cond_H}.
Similarly, we can define a standard Hamiltonian $H(\p,\z)$ by taking the Legendre-Fenchel transform of the standard Lagrangian $L(\dot{\z},\z)$:
\begin{equation} \label{H_stand}
H(\p,\z) \equiv \sup_{\dot{\z}} \{ \p \cdot \dot{\z} - L(\dot{\z},\z) \}.
\end{equation}
Standard and detailed Hamiltonians are simply related by:
\begin{eqnarray} \label{standard_first}
\H(\f=\D^\dagger \p,\z) &=& \sup_{\bl} \left\{(\D^\dagger \p) \cdot \bl - \L(\bl,\z)  \right\} \\
&=& \sup_{\bl} \left\{\p \cdot (\D \bl) - \L(\bl,\z)  \right\} \\
&=& \sup_{\dot{\z}} \left\{ \sup_{\bl \mid \dot{z} = \D \bl} \left[ \p \cdot \dot{\z} - \L(\bl,\z) \right]  \right\} \\
&=& \sup_{\dot{\z}} \left\{ \p \cdot \dot{\z} - \inf_{\bl \mid \dot{z} = \D \bl} \left[  \L(\bl,\z) \right]  \right\} \\
&=& \sup_{\dot{\z}} \left\{ \p \cdot \dot{\z} - L(\dot{\z},\z)  \right\} \\
&=& H(\p, \z) \label{standard_H},
\end{eqnarray}
where $\D^\dagger$ is the adjoint of $\D$ and where we used Eqs.~\eqref{EqConserv_observablesEmpiriques}, \eqref{L_stand} and \eqref{H_stand} to obtain \eqref{standard_H}. Since standard Hamiltonians follow explicitly from evaluating the detailed ones in $\f=\D^\dagger \p$, we expect the Hamiltonian framework to be more convenient for analytical computations than the Lagrangian framework. Hence, we use mostly the (standard) Hamiltonian framework from now on.

\subsection{Biased Lagrangian and Hamiltonian} \label{Sec_biasedHL}

We are now interested in the fluctuations in the limit of large parameter $N$ of the two-component observable $\bA_t$ defined by
\begin{equation} \label{observable}
\bA_{t} \equiv \frac{1}{t} \left(
\begin{array}{c}
\int_0^t \ud t' \bl(t')  \\
\int_0^t \ud t' \z(t') 
\end{array} \right).
\end{equation}
We use an overbar to emphasize that the observable is rescaled by $\un$ such that $\un \bA$ is an extensive observable. The SCGF $\scgfn \equiv \scgfn(\g)$ defined by
\begin{equation} \label{scgf_N_nonlin}
\scgfn(\g) \equiv \lim_{t \rightarrow \infty} \frac{1}{t \un} \ln G_{\g}(t)
\end{equation}
describes the fluctuations of $\bA$ as its successive partial derivatives at $\g = 0$ give the (scaled) cumulants of the observable, with $G_{\g}(t)$ the moment generating function $G_{\g}(t)$ defined by
\begin{equation}  \label{Gen_func_nonlin}
G_{\g}(t) \equiv \left\langle \e^{\un t \g \cdot \bA_t} \right\rangle_{{\z_\mathrm{i}}} = \int  \df[\bl, \z] \e^{\un t \g \cdot \bA_t} \uP_t[\z \mid \z_\mathrm{i}] \delta[\dot \z - \D \bl ],
\end{equation}
where $\left\langle \dots \right\rangle_{{\z_\mathrm{i}}}$ is the average with respect to $\uP_t[\z \mid \z_\mathrm{i}]$ the path probability of trajectory $[\z]$ up to time $t$ given the initial state $\z_\mathrm{i}$ at time $t=0$, $\df[\bl, \z]$ is the path measure and $\delta[\dot \z - \D \bl ]$ is a Dirac delta ensuring Eq.~\eqref{EqConserv_observablesEmpiriques} at all times. 
Using the fact that
\begin{equation}
\uP_t[\z \mid \z_\mathrm{i}] = \prod_{\ell = 0}^M P_{\delta t}(\bl_{\t_\ell}\mid \z_{\t_\ell}),
\end{equation}
where the product runs over times $\t_\ell \equiv \ell \delta t$ with initial time $\t_0 = 0$ and final time $\t_M = t$, it follows from Eq.~\eqref{Gen_func_nonlin} that
\begin{equation}
G_{\g}(t) = \int \prod_{\ell = 0}^M \ud \bl_{\t_\ell} \ud \z_{\t_\ell} G_{\delta t}(\bl_{\t_\ell} \mid \z_{\t_\ell}) \delta(\z_{\t_{\ell+1}} - \z_{\t_\ell} - \delta t \D \bl_{\t_\ell} ),
\end{equation}
where we introduced the \textit{biased transition probability} $G_{\delta t}(\bl \mid \z)$ during the infinitesimal time $\delta t$
\begin{equation} \label{gen_func}
G_{\delta t}(\bl \mid \z) \equiv P_{\delta t}(\bl \mid \z)  \e^{\un \delta t (\g_1 \cdot \bl + \g_2 \cdot \z)}.
\end{equation}
{Note that the dependence of $G_{\delta t}(\bl \mid \z)$ on $\g$ is implicit}. From Eqs.~(\ref{lagr}, \ref{gen_func}), we find that the biased transition probability is associated with the \textit{detailed biased Lagrangian} $\L_{\g}(\bl,\z)$:
\begin{equation}
G_{\delta t}(\bl \mid \z) \underset{\un \rightarrow \infty}{\asymp} \e^{- \un \delta t \L_{\g}(\bl,\z)}, \label{generating_function_action}    
\end{equation}
with
\begin{equation} \label{Biased_det_lagrangian}
\L_{\g}(\bl,\z) \equiv \L(\bl,\z) - \g_1  \cdot \bl - \g_2 \cdot \z.
\end{equation}
We define the \textit{detailed biased Hamiltonian} $\H_{\g}(\f,\z)$ as the Legendre-Fenchel transform of the detailed biased Lagrangian
\begin{equation}   \label{def_Hbiais_det}
\H_{\g}(\f,\z) \equiv \sup_{\bl} \{ \f \cdot \bl - \L_{\g}(\bl,\z) \} = \H(\f+\g_1,\z) + \g_2 \cdot \z,
\end{equation}
where we used in the second equality  Eqs.~(\ref{Hdet}, \ref{Biased_det_lagrangian}) and the fact that $\bl$ and $\z$ are independent. Note that the standard biased Lagrangian $L_{\g}(\dot{\z},\z)$ and Hamiltonian $H_{\g}(\f,\z)$ follow from the detailed ones :
\begin{eqnarray} \label{biased_stand_lagrangian}
L_{\g}(\dot{\z},\z) & \equiv & \inf_{\bl | \dot{\z} = \D\cdot\bl} \L_{\g}(\bl,\z), \\
H_{\g}(\p,\z) & \equiv & \sup_{\dot{\z}} \{ \p \cdot \dot{\z} - L_{\g}(\dot{\z},\z) \} = \H_{\g}(\f = \D^\dagger \p, \z),  \label{Hb_lien_det_stand}
\end{eqnarray}
as in the non-biased case.
The biased Lagrangian and Hamiltonian are not associated with a norm-conserving Markov process as they do not satisfy conditions~\eqref{cond_L} and~\eqref{cond_H} respectively. 
The transformation of these Lagrangian and Hamiltonian that restores these conditions will be called \textit{rectification}. This rectification plays the role of the generalized Doob transform that, in the linear operator formalism, produces the driven generator from the biased generator.
However, since the construction of the driven generator relies on the Perron-Frobenius theorem constraining the spectral properties of the bias generator, we study in the following the spectral problem associated with the biased Lagrangian and Hamiltonian. In the next section, we provide a conjecture translating the Perron-Frobenius theorem in the context of this spectral problem.

\subsection{Equations of motion}

We are interested in the typical behavior of our system during the time-interval $[0,\tt]$. We saw that the biased transition probability of observing $\z_\mathrm{f} = \z_\mathrm{i} + \int_0^\tt \dot{\z}_t \ud t = \z_\mathrm{i} + \int_0^\tt \D \bl_t \ud t$ at $t=\tt$ given an initial state $\z_\mathrm{i}$ at $t=0$ satisfies the LDP
\begin{equation}
G_\tt[\z_\mathrm{f} \mid \z_\mathrm{i}] \apeupn \int \df[\bl,\z]\delta[\dot \z - \D \bl ]\e^{- \un \int_0^{\tt} \L_{\g}(\bl_t,\z_t) \ud t} \apeupn \int \df[\dot{\z},\z]\e^{- \un \int_0^{\tt} L_{\g}(\dot{\z}_t,\z_t) \ud t},
\end{equation}
where we used Laplace's approximation and the definition of the standard biased Lagrangian~\eqref{biased_stand_lagrangian} in the last equality. There is a family of trajectories $\{(\z_t)^\varepsilon\}$ indexed by $\varepsilon$ connecting $\z_\mathrm{i}$ to $\z_\mathrm{f}$ during a time $\tt$. The typical trajectory followed by the system is the one minimizing the action $S[\dot{\z},\z]_0^\tt \equiv \int_0^\tt \ud t L_{\g}(\dot{\z_t},\z_t)$ and that hence solves the Euler--Lagrange equation
\begin{equation} \label{Euler_Lagrange}
\frac{\partial L_{\g}}{\partial \z} - \frac{\ud}{\ud t}\left( \frac{\partial  L_{\g}}{\partial \dot{\z}} \right) = 0,
\end{equation}
with initial and final conditions $\z_\mathrm{i}$ and $\z_\mathrm{f}$, respectively. 
Alternatively, using the extremum action principle on the action \begin{equation}
S[\p,\z]_0^\tt = \int_0^{\tt} \left[ \p_t \cdot \dot{\z}_t - H_{\g}(\p_t,\z_t) \right] \ud t    
\end{equation}
written in term of the Hamiltonian leads to Hamilton's equations
\begin{equation}  \label{Hamilton_eq}
\begin{cases}
\dot{\z}_t = \partial_{\p} H_{\g}(\p_t,\z_t), \\
\dot{\p}_t = - \partial_{\z} H_{\g}(\p_t,\z_t),
\end{cases}
\end{equation}
with the same boundary conditions $\z_\mathrm{i}$ and $\z_\mathrm{f}$. {Since the Hamiltonian is a constant of motion along the solutions $\p_t^*$ and $\z_t^*$ of Hamilton's equations Eq.~\eqref{Hamilton_eq}, i.e.   $\Hg(\p_t^*,\z_t^*) = E$ for all $t$, the biased transition probability reads
\begin{equation} \label{scgf_nonlinbis}
G_\tt[\z_\mathrm{f} \mid \z_\mathrm{i}] \apeupn \e^{\un \left[\tt H_{\g}(\p^*_t,\z_t^*)  - \int_0^{\tt} \p_t^* \cdot \dot{\z}_t^* \ud t \right]}.
\end{equation}}
We recognize in the second term of the exponential the so-called \textit{reduced action} {which} we write $\Sr(\tt) \equiv \int_0^\tt \p^*_t \cdot \dot{\z}^*_t \ud t$.
{We will focus in the following on cases where this reduced action is not extensive in time, with an exponential contribution of order $N$ to the generating function, negligible compared to the Hamiltonian term of order $NT$.}

\subsection{Hamilton-Jacobi equation}
\label{3.4}

An alternative description of the dynamics can be obtained by considering \textit{Hamilton's principal function} --- also called \textit{Jacobi's action} --- defined as the action evaluated along the solutions of Hamilton's equations (or equivalently Euler-Lagrange equation) with initial state $ \z_\mathrm{i}$ and arrival state $\z_t = \z$:
\begin{equation}
S(\z,\z_\mathrm{i},t) \equiv  \int_0^t \ud \tau  L_{\g}(\dot{\z}_\tau^*,\z_\tau^*)  .
\end{equation}
The action $S$ contains all the information on the dynamics of the system (see Ref.~\cite{Rax2020} for more details). It can be obtained by solving a partial differential equation called \textit{Hamilton-Jacobi equation}:
\begin{equation}
\frac{\partial S}{\partial t} + \Hg(\partial_{\z} S, \z) = 0.
\end{equation}
When the Hamiltonian is time-independent and hence a constant of motion $\Hg(\p_t^*,\z_t^*) = E$, it is convenient to consider \textit{Hamilton's characteristic function} defined as the Legendre transform of $S$ with respect to time:
\begin{equation}
W(\z,\z_\mathrm{i},E) = E t + S(\z,\z_\mathrm{i},t),
\end{equation}
where the \textit{eigenrate} $E$ has the dimension of an inverse time and replaces "energy" in the HJ equation
\begin{equation} \label{HJeq}
\Hg(\partial_{\z} W,\z) = E,
\end{equation}
where the momentum $\p =\partial_{\z} W $ is the gradient of Hamilton's characteristic function. In the following, the terminology "HJ equation" always refers to Eq.~\eqref{HJeq}. Solving this equation is the nonlinear equivalent of a spectral problem for linear operators.

We say that a solution $W(\z,E)$ of this equation is \textit{global} if it is defined and analytic for all $\z$. We define the corresponding \textit{reduced dynamics} describing the evolution of the state variable only by the equation
\begin{equation}
    \dot{\z} = \left. \frac{\partial H_{\g}}{\partial \p} \right|_{\p= \partial_{\z} W,\z}. \label{DefReducedDynamics}
\end{equation}
This dynamics is said to be \textit{globally stable} (respectively \textit{globally unstable}) if there exists a compact set $C$ in $\z$-space such that all trajectories of the reduced dynamics converge to (respectively exit from) $C$, i.e.
{\begin{eqnarray}
    &\forall \z_\mathrm{i}, \exists t^\star \in \R, \forall t > t^\star, \z_t\in C &\qquad \text{(globally stable)}, \\
    &\forall \z_\mathrm{i}, \exists t^\star \in \R, \forall t < t^\star, \z_t\in C  &\qquad \text{(globally unstable)}.
\end{eqnarray}}
We will see that such stability conditions can guarantee the existence of special solutions of the HJ equation called \textit{critical manifolds}, which are described in the next section.

\subsection{Critical manifolds}

Given a solution of Hamilton's equations, one can draw a line in the phase space, called \textit{orbit}, that is parametrized by the time dependence of $(\p_t, \z_t)$. Orbits hence belong to a subspace of the phase space where the Hamiltonian remains constant. There is a particular class of orbits that we call \textit{critical manifolds}\footnote{A manifold is informally defined as a geometrical space generalizing the notion of curve or surface to arbitrary dimensions. For instance, a one-dimensional manifold is a curve and includes lines and circles. A two-dimensional manifold is a surface and includes planes, spheres and tori.} defined as an ensemble of compact trajectories (in the sense that they are entirely included in some compact set of phase space) and such that at least one other trajectory converges towards it forward or backward in time. \textit{Fixed points} are the simplest critical manifolds: their phase space coordinates solve the stationary Hamilton's equations
\begin{equation} \label{def_fixedpoint}
\begin{cases}
\partial_\p H_{\g} = 0, \\
\partial_\z H_{\g} = 0.
\end{cases}
\end{equation}
\textit{Limit cycles} are also critical manifolds of dimension~$1$ that are periodic solutions of Hamilton's equations. Limit cycles arise for nonlinear dynamics and cannot occur in linear systems. Note that neither centers (see the magenta point in Fig.~\ref{fig_espace_des_phases_BD_3valeurs}) nor the periodic orbits surrounding them are critical manifolds, as no trajectory converges to them. Other examples of critical manifolds include tori or complex geometric structures called \textit{strange attractors}\footnote{We made here an abuse of language as \textit{attractor} means that all trajectories converge toward it forward in time. Here, the strange attractor may be stable for some trajectories (attractor) and unstable for others (repeller).}. When the system is at a critical manifold, it will take an infinite time to leave it. {Conversely, the system cannot reach a critical manifold in finite time.}

\section{Spectral properties of statistical Hamiltonians}
\label{Perron-section}
 
The rectification procedure in the linear operator formalism relies heavily on the Perron-Frobenius theorem since it ensures the non-degeneracy of the largest eigenvalue of the biased generator and the positivity of its dominant left eigenvector, both used in the definition of the driven generator. In order to extend the rectification to nonlinear processes, one needs to translate the Perron-Frobenius theorem in the nonlinear framework in which the spectral problem is expressed by a HJ equation. Given the difficulty of such a generalization, we instead propose a conjecture based on physically reasonable assumptions on the structure of the Hamiltonian under consideration.

\subsection{Assumptions on statistical Hamiltonians} \label{Sec_assumptions}

In the following, we make a series of assumptions on the properties of the Hamiltonians we consider. We assume these properties to be generically preserved under biasing, if not we restrain $\g$ to the values for which it is the case. We call the class of Hamiltonians satisfying the following properties \textit{Statistical Hamiltonians}. Without loss of generality, we focus on the biased Hamiltonian $\Hg$ (the non-biased case follows from $\g = 0$) and illustrate numerically each assumption on the nonlinear model called "Brownian Donkey" that will be studied in Sec.~\ref{Sec_BK}. In the remaining of this paper, we assume $\z$ and $\p$ to be defined on $\R^n$ or open sets of $\R^n$ (with $n$ an \textbf{}integer).

~~

Our assumptions are the following.

First, $\Hg$ is convex in $\p$ for any $\z$ since it follows from a Legendre-Fenchel transform. We assume in addition that it is {\it strictly convex} as well as {\it coercive}, i.e. for any $\z$, $\Hg(\p,\z)\rightarrow\infty$ when $|\p|\rightarrow\infty$, where $|\p|$ is the Euclidean norm of $\p$. Given these assumptions, there is for any $\z$ a unique value $\p=\p_{\min}(\z)$ that minimizes $H_{\g}(\p,\z)$:
\begin{equation} \label{pmin_assump_Hg}
\partial_{\p} H_{\g}(\p_{\min}(\z),\z)=0~~~~\mathrm{and}~~~~\partial_{\p}^2 H_{\g}(\p_{\min}(\z),\z)>0,
\end{equation}
see Figure~\ref{fig_colormap_BD_H(z,p)}. From the first equation of Eq.~\eqref{Hamilton_eq}, the minimizer $\p_{\min}(\z)$ is associated with a stopping point for $\z$, i.e. $\dot{\z} = 0$. We define, for future use, the minimal value of $\Hg$ for each $\z$:
\begin{equation}  \label{Hmin_PF}
   H_{\min}(\z)\equiv \Hg(\p_{\min}(\z),\z).
\end{equation}
\begin{figure}
\begin{center}
\includegraphics[scale=0.5]{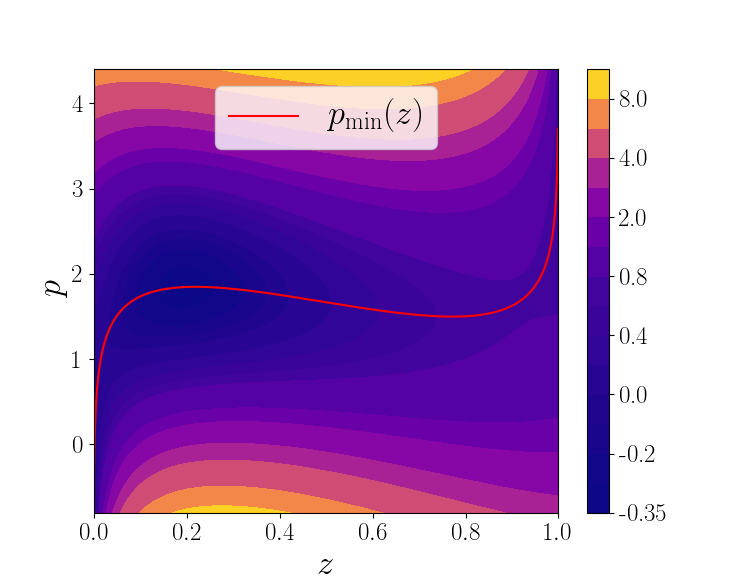}
\end{center}  
\caption{$\Hg(p,z)$ and $p_{\min}(z)$ for the model of the Brownian Donkey. $\Hg(p,z)$ is strictly convex in $p$ and diverges for infinite values of $p$ for any $z$ (except near the edges of $z$ where the assumptions of the model are not verified anymore), implying the existence of a unique minimum $p_{\min}(z)$ as illustrated on the figure.  \label{fig_colormap_BD_H(z,p)} }
\end{figure}  
{Second, the vector field $\mathbf{v}=\partial_z H_{\min}(\z)$ is globally stable, i.e. there exists a compact set $B$ in $\z$-space (e.g. a large enough ball) such that all trajectories flowing along $\mathbf{v}$ \textbf{outside} of $B$ reach $B$ eventually. This is equivalent to requiring that $H_{\min}(\z)$ be strictly decreasing outside of $B$ towards the boundaries of state-space, in the sense that its level sections should be nested closed sets, with decreasing levels. For $\z$ in $\R^n$ for instance, the level sections will have to be isomorphic to nested balls. This is physically reasonable for the following reason: $H_{\min}(\z)$ being the value of the Hamiltonian at $\z$ when $\partial_{\p} H_{\g}=\dot\z=0$, we have $H_{\min}(\z)=-L_{\g}(0,\z)$, so that this condition is equivalent to requiring that the probability of observing a zero velocity should decrease towards the boundaries of state-space, and hence reaching those boundaries should be increasingly difficult, consistently with the system being globally stable.} 

{This property implies that $H_{\min}(\z)$ admits at least one maximum inside $B$ and no extrema outside.} Note that the extrema $\{\z^\star\}$ of $H_{\min}(\z)$ are the positions of the fixed points $\{(\p^\star = \p_{\min}(\z^\star),\z^\star)\}$ of the Hamiltonian dynamics since at an extremum $\z^\star$, we have
\begin{equation}
\dot{\p}=\partial_{\z} H_{\g}(\p_{\min}(\z^\star),\z^\star) = 0.
\end{equation}
We label $\z_\alpha^\star$ ($\alpha=0,1,\dots)$ the positions of the maxima of $H_{\g}$ on the manifold $\p=\p_{\min}(\z)$ (which we assume to be countable for the sake of simplicity) and we introduce $\p^\star_\alpha \equiv \p_{\min}(\z^\star_\alpha)$. We define $H_\alpha^\star(\g)\equiv H_{\min}(\z_\alpha^\star) = H(\p_\alpha^\star,\z_\alpha^\star)$ and choose the indices of the $\z_\alpha^\star$ such that $H_0^\star(\g)\ge H_1^\star(\g) \ge H_2^\star(\g)\ge \dots $, so that $H_0^\star(\g)=\max_\alpha H_\alpha^\star(\g)$, see Figure~\ref{fig_Hg(p0(z),z)}. The corresponding fixed point $(\p_0^\star,\z_0^\star)$ is particularly important and will be called the \textit{dominant fixed point} for reasons that will be explained in section \ref{Sec_cv_trajectoires}.
\begin{figure} 
\begin{center} 
\includegraphics[scale=0.5]{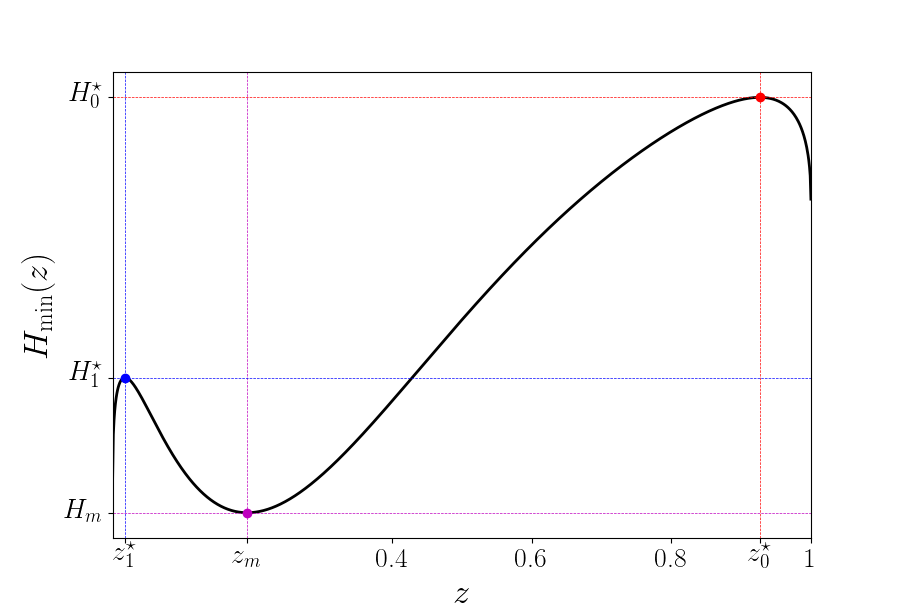} 
\end{center}
\caption{$H_{\min}(z)$ for the model of the Brownian Donkey. It admits three extrema corresponding to the positions of the fixed points: two maxima $z_0^\star$ and $z_1^\star$ and one minimum $z_m$. {$H_{\min}$ is strictly decreasing away from $B = [z_1^\star,z_0^\star]$.} 
\label{fig_Hg(p0(z),z)}}
\end{figure}

Finally, we assume that the absolute maximum of $H_{\min}(\z)$ is non-degenerate, i.e. $H_0^\star(\g)> H_1^\star(\g)$, in order to avoid first-order phase transition points, where the Perron-Frobenius theorem is expected to fail due to the breaking of ergodicity~\cite{vroylandt2019, Lazarescu2019}. {This implies that
\begin{equation}
     H_0^\star(\g) = \max_{\z} \min_{\p} \Hg(\p,\z),
\end{equation}
which will be an important object in the following conjecture.}

{Note that these assumptions are only sufficient for what follows, but perhaps not necessary. In particular, the second condition may be too restrictive, but we leave such questions for later.}

\subsection{Conjecture for a nonlinear generalization of the Perron-Frobenius theorem} \label{Sec_conj_PF}

Under the assumptions of the previous section, we make the following conjecture concerning the solutions of the HJ equation~\eqref{HJeq}:

\begin{conjecture} \label{PFconj}
There exists a value $E^\star(\g)$ of $\Hg(\p,\z)$ such that
\begin{enumerate}

\item For $E > E^\star(\g)$, all orbits tend towards the boundaries of the system forward and backward in time, so that none of them contain or reach a critical manifold (fixed points, limit cycles, strange attractors, etc.).
\label{E>Estar}

\item For $E < E^\star(\g)$, there is no global solution to the HJ equation, and the reduced action of any solution $W$ along any bounded orbit (such as closed orbits or strange attractors) is non-negative: $\int \partial_{\z} W \cdot \ud \z \geq 0$. \label{E<Estar}

\item For $E = E^\star(\g)$, the HJ equation admits at least two global solutions (up to an additive constant). Among these solutions, there is exactly one globally stable solution $W_{\sm}(\z,\g)$, and one globally unstable solution $W_{\um}(\z,\g)$. These two solutions $W_{\sm}(\z,\g)$ and $W_{\um}(\z,\g)$ coincide on each of their critical manifolds.\label{E=Estar}

\item The dominant fixed point $(\p_0^\star,\z_0^\star)$ is contained in both the globally stable solution $W_{\sm}(\z,\g)$ and the globally unstable solution $W_{\um}(\z,\g)$. The critical value $E^\star(\g)$ of the eigenrate can therefore be obtained by a max-min formula: 
\begin{equation}\label{minmaxE}
    E^\star(\g) =  \max_{\z} \min_{\p} \Hg(\p,\z)\equiv H_0^\star(\g).
\end{equation} \label{minmax}
\end{enumerate}
\end{conjecture}

\begin{figure} 
\begin{center}
\includegraphics[scale=0.5]{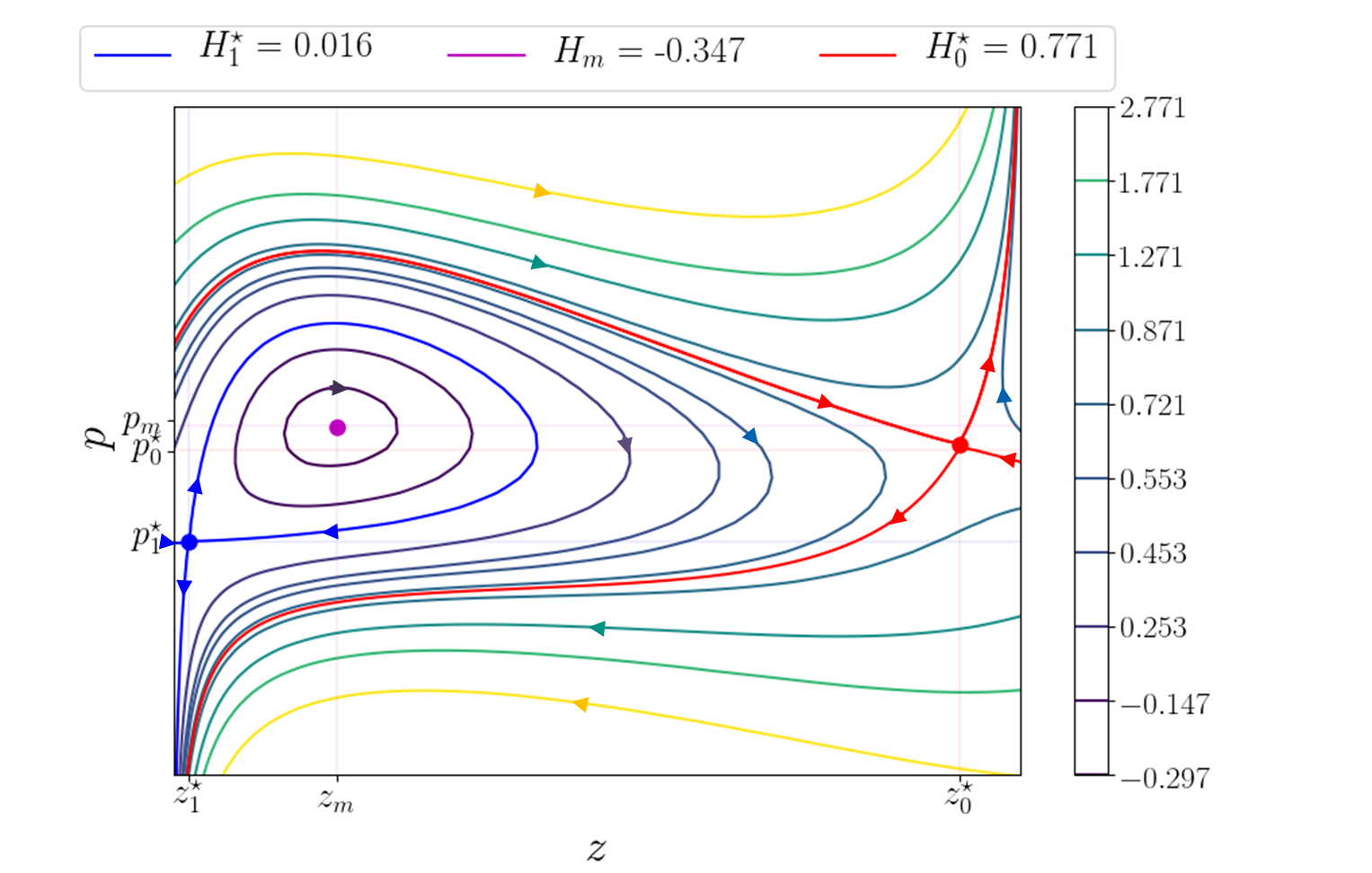}
\caption{Trajectories in the phase space associated with $H_{\g}(p,z)$ for the model of the Brownian donkey. The red, magenta and blue points correspond respectively to the fixed points of positions $z_0^\star$, $z_m$ and $z_1^\star$ given by the extrema of $H_{\min}(z)$ as illustrated in Figure~\ref{fig_Hg(p0(z),z)}. In red, the trajectory of eigenrate $E = H_0^\star$. \label{fig_espace_des_phases_BD_3valeurs}}
\end{center}
\end{figure} 
By analogy with the Perron-Frobenius theorem, $E^\star(\g)$ corresponds to the dominant eigenvalue, $W_{\sm}(\z,\g)$ corresponds to the dominant left eigenvector, which is the solution that vanishes when $\g = 0$, whereas $W_{\um}(\z,\g)$ corresponds to the dominant right eigenvector, and determines the stationary distribution/quasipotential of $\z$ when $\g=0$.  Fig.~\eqref{fig_espace_des_phases_BD_3valeurs} provides an illustration of this conjecture. We see that for values $E$ of the Hamiltonian smaller than $H_0^\star$ (trajectories between the magenta point and the red trajectory), there are intervals of $\z$ for which the equation $H_{\g}(p,z) = E$ does not admit solutions for $\p$. Starting from $E = H_0^\star$, we see that $H_{\g}(p,z) = E$ admits two solutions for $p$ for any $z$. For $E > H_0^\star$, all orbits tend to the boundaries of state-space forward and backward in time.

\begin{figure} 
\begin{center} 
\includegraphics[scale=1.2]{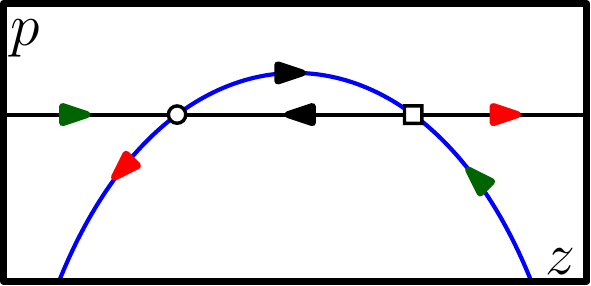} 
\end{center}
\caption{Schematic example of a system sitting at a first order transition with two fixed points (circle and square), both included in the same two manifolds (black and blue lines). As can be seen, each of the manifolds has one stable and one unstable fixed point, so that neither can be globally stable or unstable. \label{fig_1Dunst}}
\end{figure}

Additional remarks can be made about this conjecture, including some elements of proof:
\begin{itemize}
    \item Point \ref{minmax} of the conjecture is a consequence of points \ref{E>Estar} and \ref{E=Estar}, and of our state-space having Euler characteristic $1$. Consider the vector field defined by the reduced dynamics of $W_{\sm}(\z,\g)$ restricted to the compact set $C$ involved in the global stability of $W_{\sm}$, as defined in Section~\ref{3.4}, and which is such that all vectors on the boundary of the compact set point inwards. By virtue of the Poincaré-Hopf theorem~\cite{Poincare1881, Poincare1882, Hopf1927}, the total topological index of the field inside $C$ must be $1$, which implies that it contains at least one fixed point. Since point~\ref{E>Estar} excludes fixed points with an eigenrate higher than $E^\star(\g)$, the fixed points contained in $W_{\sm}(\z,\g)$ are those with the maximum value of $H$. Moreover, we have assumed that value to be non-degenerate, so that the fixed point is in fact unique and must be $(\p_0^\star,\z_0^\star)$. Finally, we recall that the value of the Hamiltonian at that point is $H_0^\star(\g)$, obtained as the max over state-space of the min over momentum-space of $\Hg$, which leads to Eq.~\eqref{minmaxE}.
    \item  The uniqueness of the stable solution $W_{\sm}(\z,\g)$ can be understood more easily when considering the dynamics close to the fixed point: due to the convexity/concavity of the Hamiltonian, for a dynamics on $\R^{2n}$, the fixed point will have $n$ independent stable directions, and $n$ independent unstable directions (i.e. $n$ positive/negative Lyapunov coefficients, being the eigenvalues of the Hessian matrix of $\Hg$ at the fixed point), defining two $n$-dimensional tangent spaces with the corresponding stability. It is however not trivial that those two spaces extend into complete solutions of the HJ equation when following each of their orbits (e.g. that those orbits cannot cross). A proof for quadratic Hamiltonians with a single fixed point can be found in~\cite{richardson1986positive}.
    \item The globally unstable solution $W_{\um}(\z,\g)$ is related to the globally stable solution of the dual dynamics (see section \ref{DualDyn}) by the fluctuation symmetry~\eqref{Sym_H_biased}. Its existence and uniqueness is therefore a consequence of that of the stable solution $W_{\sm}(\z,\g)$.
    \item In the case where other critical manifolds exist at $E^\star(\g)$ (and are therefore not fixed points, as per our assumptions), the reduced action $\Sr = \int_0^{\tt} \p_t \cdot \dot{\z}_t \ud t=W_{\sm}(\z_\mathrm{f},\g)-W_{\sm}(\z_\mathrm{i},\g)$ accumulated along any one of those manifolds cannot be time-extensive, which implies that it is negligible in the action.
    \item The properties conjectured above rely heavily on the topology of state-space and are for instance not true for processes defined on compact manifolds (e.g. on the unit circle ; see ex.~\ref{exDiffS1}).
    \item The max-min formula \eqref{minmaxE} is continuous in $\g$, and therefore remains valid even if the dominant fixed point of $\Hg$ is degenerate, i.e. at first-order {dynamical} transition points. The existence of globally stable/unstable solutions of the HJ equation is however no longer expected, since simple counter-examples can be found (e.g. a one-dimensional system with one stable and one unstable fixed points on the same characteristic manifold, as illustrated on Fig.~\ref{fig_1Dunst}). {Moreover, when approaching such a transition, the values of $\Hg$ at each fixed point cross each-other, as can be seen in \cite{Lazarescu2019}, section V. Other works describing dynamical phase transitions include \cite{vroylandt2019,Lecomte2007_vol,Garrahan2007vol98,Lecomte2007_vol127,Garrahan2009_vol42, lazarescu2017generic, bodineau2008long, Baek2017, bunin2012non} ; in each of those systems, we expect our conjecture to hold on either side of the transition, and to play the part of the large-$N$ equivalent to the Perron-Frobenius theorem.}
\end{itemize}

\subsection{Long-time limit and SCGF}  \label{Sec_cv_trajectoires}

As in the linear operator formalism, we are interested in finding an equivalent process to the conditioned process in the long-time limit. Finding this process relies on the Perron-Frobenius theorem in the linear operator formalism. In the Lagrangian--Hamiltonian framework, we use instead our conjecture to obtain similar information.  More specifically, the globally stable and unstable solutions mentioned in points~\ref{E=Estar} and \ref{minmax} contain the long-time dynamics of the system in the sense that, for any choice of boundary conditions, the orbits that dominate the action in the long-time limit are included in those two manifolds. 
Under the assumptions of Sec.~\ref{Sec_assumptions} and using the previous conjecture, we have the following result:
\begin{itshape}
In the long-time limit $\tt \rightarrow \infty$, the orbit $(\p^*_t,\z^*_t)_{t \in [0,\tt]}$ dominating the path integral is such that $\Hg(\p_t,\z_t) \rightarrow E^\star(\g)$ and collapses onto a trajectory with the following structure:
\begin{itemize}

\item A relaxation phase in which the system follows an orbit of the stable manifold (corresponding to $\p = \partial_{\z} W_{\sm}$) from $\z_\mathrm{i}$ to the associated critical manifold.

\item A stationary phase (or switching phase) in which the system remains at the critical manifold (or alternates between multiple critical manifolds through the heteroclinic orbits connecting them).

\item A fluctuation phase in which the system leaves the critical manifold to reach $\z_\mathrm{f}$ via an orbit of the unstable manifold (corresponding to $\p = \partial_{\z}W_{\um}$).

\end{itemize}

Moreover, the SCGF, i.e. the dominant value of the scaled action, is given by
\begin{equation}   \label{SCGF_H0_PF}
\scgfn(\g) = H_0^\star(\g)=E^\star(\g),
\end{equation}
which is to say that the contribution of the reduced action is negligible.
\end{itshape} 

This result can be proven when there is a single critical manifold at $\Hg(\p_t,\z_t)=E^\star(\g)$ (i.e. the fixed point $(\p_0^\star,\z_0^\star)$). The proof, which we present in the following, relies on first showing that such a trajectory exists, and then that any orbit of higher or lower eigenrate necessarily has a lower Jacobi's action. For complex cases with more that one dominant critical manifold, the statement above is presented as a conjecture. The reader can lean on Fig.~\ref{fig_espace_des_phases_BD_3valeurs} to illustrate each argument. \vspace{0.2cm}

~~

Our first task is to show that the initial and final conditions can be connected continuously through a trajectory fitting the description above. This relies on point~\ref{E=Estar} of our conjecture: for $\Hg(\p_t,\z_t) = E^\star(\g)$, the HJ equation admits a globally stable solution $ W_{\sm}$ and a globally unstable solution $ W_{\um}$. Considering first $ W_{\sm}$, we know that for any initial condition $\z_\mathrm{i}$, the stable reduced dynamics converges towards an attractor inside the compact set $C$ along an orbit such that $\p = \partial_{\z} W_{\sm}$. Similarly, considering now $ W_{\um}$, and given any final condition $\z_\mathrm{f}$, the unstable reduced dynamics also converges to an attractor of $C$ with $\p = \partial_{\z} W_{\um}$, though this time backwards in time. Under our assumption about both solutions containing a unique critical manifold, both attractors must be $\z_0^\star$, and the boundary conditions are therefore connected through a trajectory with the correct eigenrate, passing through the fixed point $(\p_0^\star,\z_0^\star)$ (so that it is of infinite duration). We also note that any trajectory with $\Hg(\p_t,\z_t) = E^\star(\g)$ that might connect the boundary conditions without passing through $\z_0^\star$ must then be of finite duration and is therefore not a candidate for the infinite time limit.

Having identified this trajectory, we can look at the opposite of the corresponding scaled action:
\begin{equation}
-\tt^{-1}S[\p,\z]_0^\tt = H_{\g}(\p_t,\z_t) - {\tt^{-1}}\int_0^{\tt} \p_t \cdot \dot{\z}_t\ud t .
\end{equation}
The reduced action can be computed in terms of the global solutions $W_{\sm,\um}$ according to:
\begin{align}\label{RedAct}
\int_0^{\tt}\p_t \cdot \dot{\z}_t\ud t \underset{T \rightarrow \infty}{\simeq}& \int_{\z_\mathrm{i}}^{\z_0^\star}\partial_{\z} W_{\sm} \cdot \ud\z +\int_{\z_0^\star}^{\z_\mathrm{f}}\partial_{\z} W_{\um} \cdot \ud\z\nonumber\\=~~&W_{\um}(\z_\mathrm{f},\g)-W_{\um}(\z_0^\star,\g)+W_{\sm}(\z_0^\star,\g)-W_{\sm}(\z_\mathrm{i},\g),
\end{align}
and turns out to not be time-extensive, so that the scaled reduced action vanishes for $\tt\rightarrow\infty$, leaving us with
\begin{equation} 
-\tt^{-1}S[\p,\z]_0^\tt \xrightarrow[\tt \rightarrow \infty]{}  H_{\g}(\p_t,\z_t)=E^\star(\g).
\end{equation}

We now need to exclude possible trajectories at different values of $\Hg$. Let us first look at the case $\Hg>E^\star(\g)$. From point~\ref{E>Estar} of our conjecture, all the corresponding trajectories tend towards the boundaries of state-space when $\tt \rightarrow \pm \infty$ and away from any specific point, so that any orbit connecting $\z_\mathrm{i}$ and $\z_\mathrm{f}$ is necessarily of finite duration.

Finally, let us look at the case $\Hg<E^\star(\g)$, which is the most complex. Point~\ref{E<Estar} of our conjecture tells us that there are no global solutions to the HJ equation so that some boundary conditions $\z_\mathrm{i}$ to $\z_\mathrm{f}$ are not connected by orbits, but some may be. Given $\z_\mathrm{i}$ and $\z_\mathrm{f}$, we distinguish three cases:
\begin{itemize}
    \item There is no orbit connecting $\z_\mathrm{i}$ and $\z_\mathrm{f}$.
    \item There is an orbit connecting $\z_\mathrm{i}$ and $\z_\mathrm{f}$ which is neither periodic nor leads to a critical manifold, so that it is necessarily of finite duration.
    \item There is an orbit connecting $\z_\mathrm{i}$ and $\z_\mathrm{f}$ that is either periodic or leads to a critical manifold. It is then necessary to compare %its scaled action 
    $-\tt^{-1}S[\p,\z]_0^\tt = H_{\g}(\p^*_t,\z_t^*) - \tt^{-1}\int_0^{\tt} \p_t^* \cdot \dot{\z}_t^* \ud t$ along that orbit with the value $E^\star(\g)$ found above. Given the case we are considering, the Hamiltonian term is smaller than $E^\star(\g)$. Moreover, we have conjectured that the reduced action is nonnegative, so that it reduces the value of $-\tt^{-1}S[\p,\z]_0^\tt$ %of the action 
    even more. This implies that such a trajectory will be exponentially less likely than the one found at $\Hg(\p_t,\z_t) = E^\star(\g)$.
\end{itemize}

We conclude that the dominant trajectory in the long-time limit is the one corresponding to $\Hg(\p_t,\z_t) = E^\star(\g)$. It follows from Eq.~\eqref{scgf_nonlinbis} and point~\ref{minmax} of the conjecture that, for any initial and final conditions,
\begin{equation}
G_\tt(\z_\mathrm{f} \mid \z_\mathrm{i})  \underset{N,\tt \rightarrow \infty}{\asymp}
 \e^{\un \tt H_0^\star(\g) }. \label{biasedpropasymp}
\end{equation}
The SCGF can then be identified as
\begin{equation}
\scgfn(\g) = H_0^\star = E^\star(\g),
\end{equation}
that is to say the critical value of the Hamiltonian described in our conjecture, which is the nonlinear equivalent to the famous result by Donsker and Varadhan relating the SCGF of a Markov process to the largest eigenvalue of the biased generator~\cite{donsker1975asymptotic, donsker1975asymptotic2, donsker1976asymptotic, donsker1983asymptotic}. {This generalises the result found in \cite{Bertin2019} for linear diffusions, and in \cite{Lazarescu2019} for chemical reaction processes.} For the Brownian Donkey, the orbit at this value $E^\star(\g)$ corresponds to the red trajectory on Figure~\ref{fig_espace_des_phases_BD_3valeurs}.

{In the simple case of a single dominant critical manifold, we can also evaluate the first correction to the leading order of Eq.~\eqref{biasedpropasymp}, yielding an expression depending on the two states $\z_\mathrm{i}$ and $\z_\mathrm{f}$ and scaling only with $N$: knowing the value of the reduced action \eqref{RedAct}, which we can plug in \eqref{scgf_nonlinbis}, we get
\begin{equation}\label{TransProb}
\frac{G_\tt(\z_\mathrm{f} \mid \z_\mathrm{i})}{\e^{\un \tt H_0^\star(\g)}}\underset{N,\tt \rightarrow \infty}{\asymp}\e^{\un \left[-\int_0^{\tt} \p_t^* \cdot \dot{\z}_t^* \ud t \right]}\propto\e^{-\un (W_{\um}(\z_\mathrm{f},\g)-W_{\sm}(\z_\mathrm{i},\g) )}.
\end{equation}
This implies that the biased long-time propagator is a projector, much like its linear counterpart as a result of the Perron-Frobenius theorem. This guarantees the uniqueness of stationary distributions both forward and backward in time.}

{Note that, although our purpose will be to use the special trajectories characterised here as a tool to build a rectified process (i.e. an effective proper stochastic process), they also carry some significance in themselves: the value of $\p$ along each trajectory corresponds to the value of the random force necessary to realise that trajectory (and which is space-dependent for nonlinear processes), so that $\p=\partial_{\z} W_{\um,\sm}$ solves the optimal control problem of finding the cheapest force (in the sense of probability cost) to take the system from its initial to its final position in the prescribed time \cite{Chetrite2015}}.

~~

In the case where there is more than one critical manifold at $\Hg(\p_t,\z_t) = E^\star(\g)$, we conjecture that it is always possible to connect one critical manifold to another via orbits of the stable and unstable manifolds. Combined with point $\ref{E=Estar}$, it follows that there exists an orbit of the stable manifold connecting $\z_\mathrm{i}$ to a first critical manifold, an orbit of the unstable manifold connecting a second (or possibly the same) critical manifold to $\z_\mathrm{f}$, and in between there exists orbits connecting the first and second critical manifolds (switching phase). Along any of these critical manifolds, the reduced action is non-extensive in time as discussed in Sec.~\ref{Sec_conj_PF}. Subsequently, the rest of the proof done for the case of a single fixed point holds, concluding our reasoning. {In this case, however, the spatial dependence of the biased transition probability \eqref{scgf_nonlinbis} is not guaranteed to be as simple, since it will contain extra terms depending on which attractors are visited.}

\subsection{Illustrative examples}
\label{exDiff}

In this section, we present  few examples in order to illustrate and justify our conjecture, give an idea of the type of proofs that might apply, and point to a subtlety relating to the topology of state-space. To make things simpler, we will only consider \textit{non-biased} low-dimensional diffusion Hamiltonians here, given that the principle of process rectification is precisely that biased processes are not qualitatively different. Examples of biased one-dimensional processes will be examined in sections~\ref{Sec_BK} and \ref{Sec_chimie}.

\subsubsection{Diffusion in $\R$}
\label{exDiff1}

For the first example, let us consider a one dimensional diffusive Hamiltonian (i.e. quadratic in $p$):
\begin{equation}
    H(p,z)=p k(z)\left(\frac{p}{2}+f(z)\right)
\end{equation}
where the variance $k(z)$ and the deterministic force $f(z)$ are two real functions of the state variable $z \in \R$, and $k(z)$ is strictly positive. For every $z$, the minimal value of $H$ reads
    \begin{equation}
       H_{\min}(z)=H(p_{\min}(z),z)=-\frac{k(z) f(z)^2}{2} \le 0
    \end{equation}
and is negative. Our assumptions on $H$ are:
\begin{itemize}
    \item {\it Strict convexity and coercivity in $p$}: The function $k(z)$ is strictly positive for all $z \in \R$ ensuring the convexity of $H$ in $p$ and that $H(p,z) \xrightarrow[|p|\rightarrow \infty]{} \infty$. 
    \item {\it {Stability} in $z$}: There is an interval $B=[z_1,z_2]$ such that $H_{\min}'(z)$ is {positive for $z<z_1$ and negative for $z>z_2$}.
    \item {\it uniqueness of the fixed point}: $\exists!~z_0^\star$, $f(z_0^\star)=0$.
\end{itemize}
The second assumption is satisfied for instance if $f(z)\sim-z^{n}$ when $|z|\rightarrow\infty$. The third assumption then requires $n$ odd and $k(z)$  bounded from below by a positive constant. 

Under these assumptions, since the absolute maximum of $H_{\min}(z)$ is $0$ and is reached at $z_0^\star$ as we have assumed, the dominant solutions of the HJ equation are obtained for $E^\star=0$ and read
\begin{eqnarray}
	W_{\sm}(z,0) &=& 0, \\
	W_{\um}(z,0) &=& -2\int^z  f(z')\mathrm{d}z'.
\end{eqnarray}
We emphasize that there are two solutions and that they are both defined for all $z$ as required by point \ref{E=Estar} of our conjecture.
We check the stability of these solutions by studying the reduced dynamics:
\begin{equation}
\dot z = \frac{\partial H }{\partial p} = p k(z)+k(z) f(z),
\end{equation}
which leads to the two following dynamics:
\begin{eqnarray}
	\dot{z}_{s} &=& k(z)f(z), \\
    \dot{z}_{u} &=& -k(z)f(z),
\end{eqnarray}
corresponding respectively to $W_{\sm}$ or $W_{\um}$. As we see, given our assumptions on $f$, the first equation is globally stable, while the second is globally unstable.

For $E>E^\star=0$, we can solve the quadratic equation $H(p,z)=E$ for $p$ to obtain the following two solutions
\begin{equation}
    p_\pm(z)=-2f(z)\pm\sqrt{f(z)^2+2E/k(z)}, \label{momenta}
\end{equation}
that are global solutions since they are defined for all $z$. Notice that the discriminant is always strictly positive, implying that the two orbits in phase space do not cross. The corresponding reduced dynamics are
\begin{equation}
    \dot z_\pm=-k(z)f(z)\pm 2k(z)\sqrt{f(z)^2+2E/k(z)}.
\end{equation}
These equations have no fixed point since $\dot z_+ > 0$ and $\dot z_- < 0$, hence illustrating point~\ref{E>Estar} of our conjecture.

For $E<E^\star = 0$, the solutions of the HJ equation have the same expression as in Eq.~\eqref{momenta}, except that the discriminant may vanish since $E<0$. In this case, $p^\pm(z)$ are two branches of the same orbit in phase space. Those orbits are either closed and go around a center, or open orbits if they leave the interval $B$. The two branches $p_\pm(z)$ of a closed orbit meet for vanishing value of the discriminant, i.e. at solutions of $f(z)^2+2E/k(z) = 0$: $(z_{\min},p(z_{\min}))$ and $(z_{\max},p(z_{\max}))$ with $z_{\min}<z_{\max}$. These solutions are therefore not defined for all $z$, illustrating point~\ref{E<Estar} of our conjecture. 

\begin{figure}
\begin{center} 
\includegraphics[scale=.8]{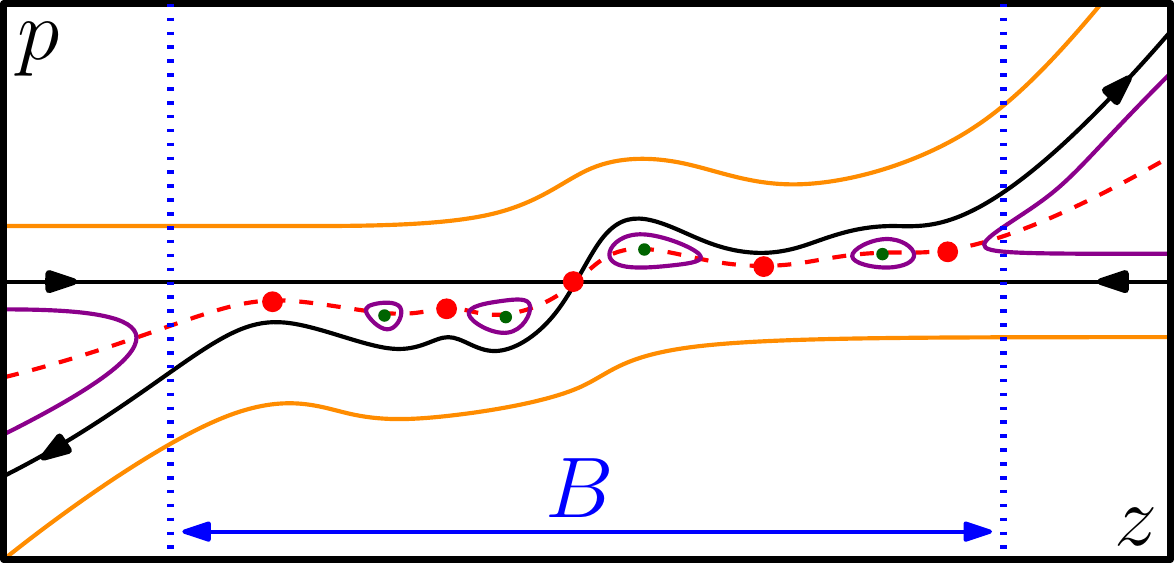} 
\end{center}
\caption{Schematic representation of the phase space of a proper 1D stochastic Hamiltonian, including a few orbits. The solutions $W_{\sm,\um}$ of the HJ equation are represented in black, and the (red) fixed point where they cross is the position of the absolute maximum of $H_{\min}(z)$. Trajectories with a larger Hamiltonian go to $\pm\infty$ on both sides (orange). Trajectories with a lower value of $H$ (purple) and that are periodic are all contained in $B$ (blue dotted lines). The red dashed line is $p_{\min} (z)$, the location of the minimum on $p$ of $H(p,z)$ for each $z$. \label{fig_1D}}
\end{figure}

All the observations of the previous paragraphs are summarised on Fig.~\ref{fig_1D}. We now illustrate that the dominant trajectories are obtained for $E=E^\star$. The reduced action reads
\begin{equation}
    \int p \dot z~\mathrm{d}t=4\int_{z_{\min}}^{z_{\max}}\sqrt{f(z)^2+2E/k(z)}\mathrm{d}z>0.
\end{equation}
For $E \le E^\star$, the scaled action is bounded by
\begin{equation}
    \tt^{-1}S=-\frac{1}{\tt} \int L \mathrm{d}t=E-\frac{1}{\tt}\int p \dot z~\mathrm{d}t \le E^\star.
\end{equation}
On the other hand, for $E = E^\star$, the scaled reduced action vanishes when $\tt\rightarrow \infty$, leaving only the value $E^\star$ which is optimal. Closed orbits are therefore sub-dominant compared to the orbit crossing the fixed point at $E=E^\star$.

\subsubsection{Diffusion in $\R^n$}
\label{exDiffn}

We now illustrate point~\ref{E>Estar} and point~\ref{E=Estar} of our conjecture for a more general Hamiltonian, associated with an $n$-dimensional diffusion, of the form
\begin{equation}
    H(\p,\z)=\p\cdot k(\z)\left(\frac{\p}{2}+\bm f(\z)\right)
\end{equation}
with $\z \in \R^n$. For every $\z$, the variance $k(\z)$ is now an $n$-dimensional positive definite matrix, and the force $\bm f(\z)$ is a vector field. We find for every $\z$ that the minimal value of $H$ is given by:
    \begin{equation}
        H_{\min}(\z)=H(\p_{\min}(\z),\z)=-\bm f(\z)\cdot\frac{k(\z)}{2}\bm f(\z),
    \end{equation}
which is again negative. The assumptions on $H$ are:
\begin{itemize}
    \item {\it Coercivity in $\p$}: $k(\z)$ is symmetric definite positive for all $\z$, and therefore invertible. 
    \item {\it {Stability} in $\z$}: there is a positive real number $R$ such that $H_{\min}(\z)$ is {decreasing away from $B=B_R$}, the ball of radius $R$ centred on the origin.
    \item {\it Uniqueness of the fixed point}: $\exists!~\z_0$, $\bm f(\z_0)=0$.
\end{itemize}
For a physically reasonable model, global stability needs to be guaranteed for $\p=0$ by constraining the direction of $\bm f$ outside of $B$, so that all deterministic trajectories for $|\z|>R$ converge towards $B$. A good way to enforce this constraint in practice is to assume that the vector field $\bm f(\z)$ admits a strictly orthogonal Helmholtz-Hodge decomposition~\cite{Suda2020} in metric $k(\z)$, expressed in terms of a rotational vector field $\bm V(\z)$ and a potential function $U(\z)$ such that $\forall \z$
\begin{equation}
    \bm f(\z)\equiv \bm V(\z)-\bm U'(\z)~~~~\mathrm{and}~~~~\bm U'(\z)\cdot k(\z)\bm V(\z) = 0,
\end{equation}
where we use $\bm U' = \partial_\z U$ to shorten notations. From this decomposition, we find
\begin{equation}
    H(2\bm U',\z)= \bm U'(\z)\cdot k(\z)\bm V(\z)=0,
\end{equation}
due to the strict orthogonality assumption. Since this equation is exactly the stationary HJ equation for an eigenrate $E^\star=0$, which is the absolute maximum of $H_{\min}$, the dominant solutions can immediately be found:
\begin{eqnarray}
	W_{\sm}(\z) &=& 0, \label{Wstable}\\
	W_{\um}(\z) &=& 2U(\z). \label{Wunstable}
\end{eqnarray}
The corresponding reduced dynamics are then
\begin{eqnarray}
	\dot{\z}_s &=& k(\z)(\bm V(\z)-\bm U'(\z)), \label{RDstable}\\
    \dot{\z}_u &=& k(\z)(\bm V(\z)+\bm U'(\z)).\label{RDunstable}
\end{eqnarray}
We can now justify our choice of stability subscripts for the two characteristic functions. The global stability condition can be expressed as $U(\z)\rightarrow\infty$ for $|\z|\rightarrow\infty$, so that $U$ (respectively $-U$) are Lyapunov functions for Eq.~\eqref{RDstable} (respectively Eq.~\eqref{RDunstable})~\cite{Suda2019}, meaning that the first equation is globally stable while the second is globally unstable, in agreement with point~\ref{E=Estar} of our conjecture.

Note that Eqs.~(\ref{RDstable}-\ref{RDunstable}) coincide whenever $\bm U'=0$, in which case we also have identical momenta $\p=0$, and $\dot \p=0$ for both the stable and unstable reduced dynamics. Such solutions include the fixed point $(\z_0^\star,\p^\star_0 =0)$, but also any limit cycles or strange attractors which may be contained in the reduced dynamics. For those more complex critical manifolds, the potential $U$ is extremal and constant along each whole manifold, and given that the deterministic flow $\dot \z=k\bm V$ is tangent to the level lines of $U$ by definition, the trajectory is then necessarily included in that manifold.

These intersections between $W_{\sm}$ and $W_{\um}$ allow for infinite-time trajectories that connect the initial condition to the final one by first relaxing along $W_{\sm}$, accumulating on one of those attractors, and then fluctuating along $W_{\um}$ towards the final condition, without discontinuity.

~~

Let us now look at the spectral properties of $H$ stated in point~\ref{E>Estar} of our conjecture. For simplicity, we assume from now on that $k$ is independent of $\z$. We also assume that for $|\z|>R$ (i.e. outside of $B$) the flow is potential, i.e. $\bm V(\z)=0$, with $U$ strictly convex and coercive which we guarantee by assuming that the spectrum of the Hessian matrix $U''$ is bounded from below by a strictly positive constant.

We will show that, under these assumptions, there is a value $E^\star$ such that all trajectories with $H>E^\star$ diverge to infinity at both ends. This will be done in two steps: 
\begin{enumerate}
    \item We first show that, outside of $B$ (i.e. for any $|\z| > R$), the value of $U(\z)$ is accelerating in time. This implies that a trajectory leaving $B$ forward or backward in time must go to $\infty$, and in particular cannot go back towards $B$ (i.e. cannot be internally tangent to any level line of $U$, as this would require $U(t)$ to have a maximum). Similarly, trajectories that do not pass through $B$ must start and end at infinity.
    \item We then show that there is a maximal value of $H$ that can sustain trajectories confined in $B$ (i.e. with $|\z| < R$), based on the fact that trajectories with a higher $H$ have a larger curvature radius. This forces those trajectories to exit $B$ and diverge to infinity due to the previous point.
\end{enumerate}
Note that, in most of the following, we will omit the argument $\z$ of all functions for clarity.

\textit{Step 1:} Let us show that $\ddot U>0$ for any $|\z|>R$, so that $U$ is at a minimum when $\dot U=0$ along a trajectory. We recall that $\bm V=0$ in this region by assumption. We have
\begin{equation}
    \ddot U=\dot \z\cdot U''\dot \z +\ddot \z \cdot \bm U'
\end{equation}
with
\begin{eqnarray}
    \dot \z &=& k(\p - \bm U') \label{velocity-nDim}\\
    \dot \p &=& U'' k \p\\
    \ddot \z &=& k \left(U'' k \p -U''k (\p - \bm U') \right)=kU''k\bm U',
\end{eqnarray}
which can be rewritten as a the quadratic form
\begin{equation}
    \ddot U = (\p-\bm U') \cdot kU'' k(\p - \bm U') + \bm U' \cdot k U'' k \bm U'>0.
\end{equation}
Moreover, both terms being positive, this quantity is in fact larger than the minimal value of the second term, which is strictly positive under the convexity assumptions made on $U$. This means that as long as $|\z|>R$, $U(\z)$ will accelerate towards $\infty$ forward and backward in time. Only three types of trajectories can then exist outside of $B$:
\begin{itemize}
    \item $\exists t_0$, $|\z(t_0)|=R$ and $|\z|\underset{t\rightarrow+\infty}{\longrightarrow}\infty$,
    \item $\exists t_0$, $|\z(t_0)|=R$ and $|\z|\underset{t\rightarrow-\infty}{\longrightarrow}\infty$,
    \item $|\z(t)|>R$ $\forall t$ with $|\z|\underset{t\rightarrow+\infty}{\longrightarrow}\infty$ and $|\z|\underset{t\rightarrow-\infty}{\longrightarrow}\infty$.
\end{itemize}

\textit{Step 2:} We now need to examine the trajectories reaching $|\z| < R$, and show that, for $H$ large enough, they cannot remain in that region. $H$ being convex and coercive in $\p$, choosing a large value of $H$ implies a large lower bound on $|\p|^2 = \p \cdot k \p$, so that we can neglect $\bm f$ within $|\z|<R$.

The curvature vector $\bm K$ of a trajectory (in $\z$ space and in metric $k$) is given by
\begin{equation}
    K=\frac{1}{\dot \z\cdot k^{-1}\dot \z}\left(\ddot \z-\frac{\ddot \z \cdot k^{-1} \dot \z}{\dot \z \cdot k^{-1}\dot \z}\dot \z\right),
\end{equation}
with
\begin{eqnarray}
    \dot \z &=& k(\p + \bm f),\\
    \dot \p &=& -(f')^t k \p,\\
    \ddot \z &=& -k(f')^t k \p + k f' k (\p + \bm f),
\end{eqnarray}
where $f'$ is the matrix of $(i,j)$ component $\partial_{z_j} f_i$. We find, by neglecting $f$ wherever appropriate, that
\begin{eqnarray}
    \dot \z \cdot k^{-1} \dot \z &=& (\p + f) \cdot k (\p + f) \sim |\p|^2,\\
    \ddot \z \cdot k^{-1} \dot \z &=& f\cdot k f' k (\p + f) \sim f \cdot k f' k \p, \label{ici}
\end{eqnarray}
so that the overall scaling in $|\p|$ of the curvature is
\begin{equation}
    K \sim |\p|^{-1}.
\end{equation}

This means that larger values of $p$ lead to smaller curvatures: we can find real numbers $E_m$ and $p_m$ such that
\begin{equation}
  H > E_m \Rightarrow  |\p| > p_m \Rightarrow |K| < R^{-1}.
\end{equation}
Any such trajectory cannot be contained in a region of radius $R$, which concludes this part of the proof.

\begin{figure} 
\begin{center} 
\includegraphics[scale=0.9]{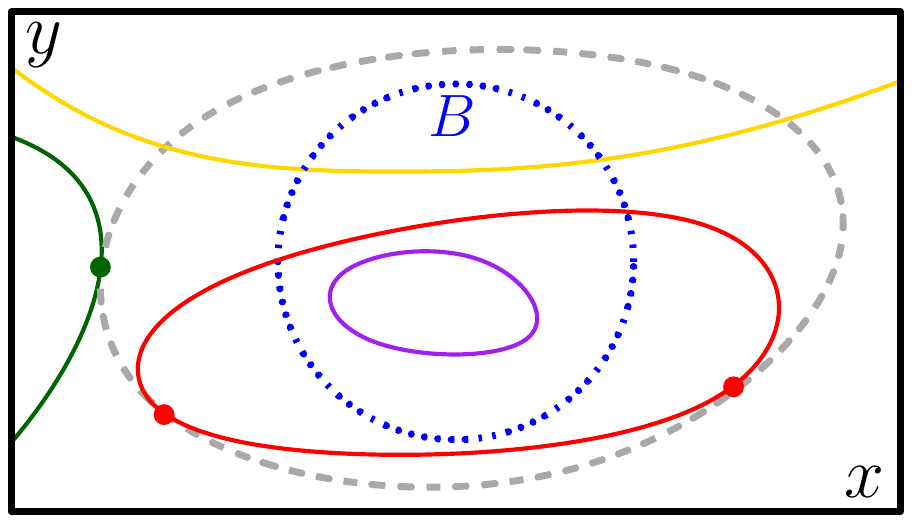} 
\end{center}
\caption{Schematic representation of Hamiltonian trajectories of a 2D diffusion projected on state space. Trajectories with a high value of $H$ must diverge on both sides (yellow and green), while periodic trajectories must be contained in $B$ (dotted circle) and have a low value of $H$ (purple). Periodic trajectories not contained in $B$ (red) would need to be internally tangent to a level line of $U$ (dashed line) outside of $B$ which is impossible.\label{fig_2D}}
\end{figure}

We can then combine those two steps in the following way: considering a trajectory such that $H>E_m$, either $|\z(t)| > R$ $\forall t$, or $\exists t_1 < t_2$ such that $|\z(t_1)| = |\z(t_2)| = R$ with on one side $|\z(t<t_1)| >R $ and on another side $ |\z(t>t_2)| > R$. In both cases, the trajectory diverges in both directions in agreement with the point~\ref{E>Estar} of our conjecture.

\subsubsection{Diffusion in $S^1$}
\label{exDiffS1}

Finally, we consider a very simple case which will highlight the importance of the topology of state-space. The Hamiltonian for a diffusion on the unit circle with constant force $f$ and constant noise variance $k$ is given by
\begin{equation}
    H(p,z)=p k\left(\frac{p}{2}+f\right).
\end{equation}
It is obviously coercive in $p$, but due to the compactness of state-space, the other two conditions cannot be satisfied. We have, for every $z$, $p_{\min}(z)=-f$, and
    \begin{equation}
       H_{\min}(z)=H(p_{\min}(z),z)=-\frac{kf^2}{2}<0.
    \end{equation}
Hamilton's equations are simply
\begin{align}
    \dot z &= k(p+f), \\
    \dot p &= 0
\end{align}
with solutions having a constant velocity, all of them being periodic except for a line of stationary points at $p=-f$. The HJ equation has only one global solution
\begin{equation}
	W_\sm(z) = 0,
\end{equation}
though nothing distinguishes it from other Hamiltonian trajectories in terms of their topology, and no max-min principle holds for $E^\star$. The dominance of this trajectory in the path integral comes in fact from the reduced action, which we have conjectured to be irrelevant in other cases:
\begin{equation}
    p \dot z = p k (p + f) = H + \frac{p k p}{2}
\end{equation}
such that
\begin{equation}
    p\dot z - H =\frac{p k p}{2}
\end{equation}
is positive and vanishes only for $p=0$. Moreover, the trajectory dominating the time-reversal of this process, which can be obtained for instance from a continuous limit of an asymmetric jump process on a cycle, corresponds to $p=-2f$, which cannot be integrated on $S^1$, and therefore does not derive from a characteristic function $W_\um$.

\section{Rectification of nonlinear Markov processes}
\label{Sec_rectification}

Relying on our conjecture replacing the Perron-Frobenius theorem for with statistical Hamiltonians, we now propose a procedure turning a biased dynamics into a \textit{rectified dynamics} by applying a gauge transformation to the biased Hamiltonian. This rectification procedure, which we define in Section~\ref{def-rectification}, is the nonlinear counterpart of the generalized Doob transform leading to the driven process in the linear operator formalism. In Section~\ref{Sec_equiv_micro_rectified_nonlin}, we comment on the equivalence between the rectified dynamics and the microcanonical process. We also look at the fluctuation relation, which relies on a symmetry of the  dynamics through time-reversal, and we show in Section~\ref{Sec_fluct_th} that this significant symmetry is inherited by the rectified dynamics, though with modified affinities. We also show in Section~\ref{DualDyn} that the dual dynamics, describing the evolution of the system backwards in time, can be obtained from the rectification of the Hamiltonian obtained by reversing the momenta. 

\subsection{Definition and properties of the rectified process}
\label{def-rectification}

For the purpose of this section, we denote by $\Z$ and $\P$ the variables of the biased Hamiltonian $\Hg$. We aim to introduce a \textit{rectified Hamiltonian} which satisfies \eqref{cond_H} and which preserves the Hamiltonian structure, i.e. such that the transformed dynamics is given by Hamilton's equations. To guarantee the latter condition, we define the rectified Hamiltonian through a canonical transformation $(\Z,\P) \to (\bm{z},\bm{p})$ associated with the generating function of the second type
\begin{equation}
	F_{2}(\bm{p}, \bm{Z},t) = \bm{Z} \cdot \bm{p} +  W_{\sm}(\Z,\g) - \scgfn t \label{generating-function}
\end{equation}
where appears the characteristic function $W_{\sm}$. The new variables $\bm z$ and $\bm p$ are obtained from the transformation rules:
\begin{eqnarray}
	\bm{z} &=& \partial_{\bm{p}} F_{2} = \Z \\
	\P &=& \partial_{\Z} F_{2} = \bm{p} + \partial_{\Z} W_{\sm},
\end{eqnarray}
with $\partial_{\bm{p}}$ being the gradient operator with respect to $\bm{p}$. The new Hamiltonian follows from
\begin{equation}
H^{\ur}(\bm{p},\bm z;\g) = H_{\g}(\P,\Z) + \partial_t F_2 = H_{\g}(\bm{p}+\partial_{\bm{z}} W_{\sm},\bm{z}) - \scgfn,
\end{equation}
leading to
\begin{equation} \label{Hr_stand}
H^{\ur}(\bm{p},\bm z;\g) = H_{\g}(\bm{p}+\partial_{\bm{z}} W_{\sm},\bm{z})- H_{\g}(\partial_{\bm{z}} W_{\sm},\bm{z}),
\end{equation}
where we used the fact that $W_{\sm}$ is solution of HJ equation~\eqref{HJeq} with $E = \scgfn$, and where the superscript ``$\ur$'' refers to $rectified$. 
The detailed rectified Hamiltonian follows using~\eqref{Hb_lien_det_stand}:
\begin{equation} \label{Hr_det}
\H^{\ur}(\f,\z; \g) =  \H_{\g}(\f + \D^\dagger \partial_{\z}W_{\sm}, \z) - \H_{\g}(\D^\dagger \partial_{\z}W_{\sm}, \z).
\end{equation}
Assuming $\H^{\ur}$ is everywhere differentiable in $\f$, the Legendre-Fenchel transform is involutive and the rectified detailed Lagrangian is given by
\begin{equation}  \label{Lr_det}
\L^{\ur}(\bl,\z;\g) = \L_{\g}(\bl,\z) - \bl \cdot \D^\dagger \partial_{\z}W_{\sm} + \scgfn(\g).
\end{equation}
The rectified standard Lagrangian follows immediately from
\begin{equation}
    L^\ur(\dot{\z},\z;\g) = \inf_{\bl | \dot{\z} = \D \bl} \L^{\ur}(\bl,\z;\g).
\end{equation}

We have introduced the rectified Hamiltonian and Lagrangian at standard and detailed levels. Without loss of generality, we focus again on the rectified standard Hamiltonian and show that it is proper, i.e. it has the properties of a statistical Hamiltonian (see Section \ref{Sec_assumptions}) on the one hand, and the properties of a non-biased Hamiltonian on the other hand.

We first show that the rectified Hamiltonian $H^{\ur}(\g)$ is a statistical Hamiltonian given that the biased Hamiltonian $\Hg$ is a statistical Hamiltonian. First, the strict convexity in $\p$ of $H^{\ur}$ is inherited from the strict convexity in $\p$ of the biased Hamiltonian $\Hg$. Indeed, $\Hg$ is strictly convex if and only if for all $\p, \p'$, $\p \neq \p'$, 
\begin{equation} \label{def_conv_H}
\Hg(\p,\z) > \Hg(\p',\z) + \partial_{\p}\Hg(\p,\z) \cdot (\p - \p').
\end{equation}
It follows from Eqs.~\eqref{Hr_stand} and \eqref{def_conv_H}:
\begin{align}
H^{\ur}(\p,\z;\g) &> \left[ H_{\g}(\p'+\partial_{\z} W_{\sm},\z) - \scgfn \right] + \partial_{\p}\left[\Hg(\p+\partial_{\z} W_{\sm},\z) - \scgfn \right] \cdot (\p - \p') \\
&> H^{\ur}(\p',\z;\g) + \partial_{\p}H^{\ur}(\p,\z;\g) \cdot (\p - \p'),
\end{align}
proving the strict convexity of $H^{\ur}$. Note that the coercivity of $H^{\ur}$ follows immediately from the coercivity of $\Hg$. It follows that $H^{\ur}$ admits for each $\z$ a unique minimum reached for $\p = \p^{\ur}_{\min}(\z) = \p_{\min}(z) - \partial_{\z} W_{\sm}(\z)$ with $\p_{\min}$ the minimizer of $\Hg$:
\begin{align}
\partial_{\p} H^{\ur}(\p^{\ur}_{\min}(\z),\z;\g) &= \partial_{\p} \Hg(\p^{\ur}_{\min}(\z)+\partial_{\z} W_{\sm},\z) = \partial_{\p} \Hg(\p_{\min}(\z),\z) = 0, \\
\partial_{\p}^2 H^{\ur}(\p^{\ur}_{\min}(\z),\z;\g) &=  \partial_{\p}^2 H_{\g}(\p^{\ur}_{\min}(\z)+\partial_{\z} W_{\sm},\z) = \partial_{\p}^2 H_{\g}(\p_{\min}(\z),\z) > 0,
\end{align}
where we used Eq.~\eqref{pmin_assump_Hg}. The minimal value of $H^{\ur}$ is then related to the minimal value of $\Hg$ by:
\begin{equation} \label{Hmin_rectified}
H^{\ur}_{\min}(\z) \equiv H^{\ur}(\p^{\ur}_{\min}(\z),\z) = \Hg(\p_{\min}(\z),\z) - \scgfn = H_{\min}(\z) - \scgfn.
\end{equation}
Consequently, the extrema of $H^{\ur}_{\min}$ are given by the extrema of $ H_{\min}(\z)$ shifted by the constant $\scgfn$, both reached at the same positions $\z^\star_\alpha$. In particular, it implies the non-degeneracy of the absolute maximum of $H^{\ur}_{\min}$. Finally, we remind that the rectified dynamics satisfies Hamilton's equations since the rectified Hamiltonian derives from a canonical transformation.

We now show that the rectified Hamiltonian is a proper statistical Hamiltonian. First, we have by construction that $H^{\ur}(\p=0,\z) = 0$, as required for a non-biased Hamiltonian. Second, the absolute maximum of $H^{\ur}_{\min}$ is zero by virtue of Eqs.~\eqref{SCGF_H0_PF} and \eqref{Hmin_rectified}. It remains to show that the solution $\p = 0$ is the globally stable solution of the HJ equation for the eigenrate $E = 0$. From Eq.~\eqref{DefReducedDynamics} and \eqref{Hr_stand}, we have:
\begin{equation}
\dot{\z} = \left. \frac{\partial H^{\ur}}{\partial \p} \right|_{\p= 0,\z} = \left. \frac{\partial H_{\g}}{\partial \p} \right|_{\p= \partial_{\z} W_{\sm},\z},
\end{equation}
meaning that the reduced dynamics at $\p=0$ of the rectified Hamiltonian corresponds to the reduced dynamics of the biased Hamiltonian along its globally stable manifold. Hence, all the trajectories for the rectified dynamics at $\p=0$ converge to a compact set, showing the global stability of the manifold $\p = 0$ for the rectified dynamics. Notice that the corresponding globally unstable solution is given by $\p = \partial_{\z}\left(W_{\um} - W_{\sm} \right)$.

~~

Finally, let us comment on why it is necessary to perform the rectification with respect to $W_\sm$ rather than $W_\um$ or any other characteristic function. Among our assumptions on the unbiased Hamiltonian is the fact that $\p=0$ is a globally stable solution of the HJ equation. This follows from the fact that propagators between any two given states $\z_\mathrm{i}$ and $\z_\mathrm{f}$ should be normalized with respect to $\z_\mathrm{f}$. For the transition probability $P_\tt(\z_\mathrm{f} \mid \z_\mathrm{i})$ to be normalizable, the solution $\p=0$ has to correspond to the globally stable solution $W_{\sm} = 0$. Indeed, {considering for instance the case of a single dominant critical manifold and using \eqref{TransProb} in the nonbiased case, we see that the $\z_\mathrm{f}$-dependence} of $P_\tt(\z_\mathrm{f} \mid \z_\mathrm{i})$ comes from the globally unstable solution $W_{\um}$ in the reduced action term as $\z_\mathrm{f}$ is reached via the unstable manifold:
\begin{equation}
P_\tt(\z_\mathrm{f} \mid \z_\mathrm{i}) \apeupT \e^{-\un(W_{\um}(\z_\mathrm{f},0) - W_{\sm}(\z_\mathrm{i},0))}.
\end{equation}
If $W_{\um}$ was zero, rather than $W_\sm$, the transition probability would be a constant of $\z_\mathrm{f}$, hence non-normalizable. The solution $\p = 0$ must therefore correspond to the globally stable manifold. It is then necessary to construct the rectified Hamiltonian such that the new reference for the momentum $\p$ is the globally stable manifold, which justifies our choice of characteristic function when performing the canonical change of variable.

{In addition, we may remark that, in the single critical manifold case, our procedure also produces a simple expression for the stationary distribution of the rectified process. Indeed, after rectification, the two solutions to the HJ equation become $\p=0$ and $\p=\partial_{\z} \left(W_{\um}(\z,\g)-W_{\sm}(\z,\g)\right)$, so that the long-time rectified transition probabilities become
\begin{equation}\label{QuasiPot}
P^{\mathrm{r}}_\tt(\z_\mathrm{f} \mid \z_\mathrm{i}) \apeupT \e^{-\un(W_{\um}(\z_\mathrm{f},\g) - W_{\sm}(\z_\mathrm{f},\g))}.
\end{equation}
The quasipotential of the stationary state is therefore $W_{\um}(\z,\g) - W_{\sm}(\z,\g)$. Unfortunately, this simple result does not seem to extend to more complex cases with limit cycles or strange attractors.}

\subsection{Equivalence of microcanonical, rectified and canonical processes} \label{Sec_equiv_micro_rectified_nonlin}

In the linear operator formalism, the driven process (with the appropriate value of $\g$) is equivalent to the microcanonical process, i.e. the process conditioned on one value of the observable~$\A$. Similarly, the rectified Hamiltonian $H^{\ur}$ defines a new process and we expect the typical trajectory of the long-time limit dynamics to be such that $\A$ takes a new typical value according to the value of the biasing parameter $\g$. To support this assertion, let us show that the rectified path probability $\uP^{\ur}_{\tt}[\z \mid \z_\mathrm{i}]$ of the path $[\z]$ of duration $\tt$ given the initial state $\z_\mathrm{i}$ is asymptotically equivalent in the long-time limit to the canonical path probability
\begin{equation} \label{Pcano_pop}
\uP^{\cano}_{\g,\tt}[\z \mid \z_\mathrm{i}] \equiv \frac{\e^{\tt \g \cdot \A_\tt} \uP_{\tt}[\z \mid \z_\mathrm{i}]}{\left\langle\e^{\tt \g \cdot \A_\tt} \right\rangle_{\z_\mathrm{i}}},
\end{equation}
with $\A_\tt = \un \bA_\tt$. We know by construction that the rectified and biased Lagrangians satisfy
\begin{align}
\uP^{\ur}_{\tt}[\z \mid \z_\mathrm{i}] & \apeupn \e^{-\un \int_0^\tt \ud \t \L^{\ur}(\bl_\t,\z_\t)}, \label{lien_P_Lrectifie} \\
\uP_{\tt}[\z \mid \z_\mathrm{i}] \e^{\tt \g \cdot \A_\tt} & \apeupn \e^{-\un \int_0^\tt \ud \t \L_{\g}(\bl_\t,\z_\t)}, \label{lien_P_Lbiaise}
\end{align}
with $\bl$ and $\z$ related by Eq.~\eqref{EqConserv_observablesEmpiriques}. Combined with Eqs.~(\ref{scgf_N_nonlin}, \ref{Lr_det}), it follows
\begin{equation}
\frac{\uP^{\ur}_{\tt}[\z \mid \z_\mathrm{i}]}{\uP^{\cano}_{\g,\tt}[\z \mid \z_\mathrm{i}]} \apeupn \e^{-\un \int_0^\tt \ud \t \bl_\t \cdot \d^\dagger \partial_{\z} W_{\sm}(\z_\t)} \apeupn \e^{-\un \int_0^\tt \ud \t \dot{\z}_\t \cdot \partial_{\z} W_{\sm}(\z_\t)},
\end{equation}
where we used Eq.~\eqref{EqConserv_observablesEmpiriques} in the last equality. It follows
\begin{equation}
\frac{\uP^{\ur}_{\tt}[\z \mid \z_{\mathrm{i}}]}{\uP^{\cano}_{\g,\tt}[\z \mid \z_{\mathrm{i}}]} \apeupn \e^{-\un \left[W_{\sm}(\z_\tt) - W_{\sm}(\z_{\mathrm{i}}) \right]},
\end{equation}
leading to:
\begin{equation} \label{logequiv_rect_cano}
\lim_{\tt \rightarrow \infty} \frac{1}{\tt} \ln \frac{\uP^{\ur}_{\tt}[\z \mid \z_{\mathrm{i}}]}{\uP^{\cano}_{\g,\tt}[\z \mid \z_{\mathrm{i}}]} = 0.
\end{equation}
Hence, the rectified path probability and the canonical path probability are logarithmically equivalent:
\begin{equation}
\uP^{\ur}_\tt[\z \mid \z_{\mathrm{i}}] \apeupT \uP^{\cano}_{\g,\tt}[\z \mid \z_{\mathrm{i}}].
\end{equation}
Finally, the equivalence between the rectified path probability $\uP^{\ur}_\tt[\z \mid \z_{\mathrm{i}}]$ and the microcanonical path probability $\uP^{\text{micro}}_{\bm a,\tt}[\z \mid \z_{\mathrm{i}}] = \uP_{\tt}[\z \mid \z_{\mathrm{i}}, \A_\tt = \bm a]$ follows from the equivalence between the canonical and microcanonical path probabilities for $\g = \nabla I(\bm a)$, with $I$ the LDF (in time) of $\A_\tt$~\cite{Touchette2015}.

\subsection{Fluctuation relations} \label{Sec_fluct_th}

We say that the Hamiltonian $\H$ satisfies a fluctuation relation if there exists a quantity $\S$, called \textit{affinity}, such that
\begin{equation} \label{Sym_H}
\H(\f,\z)=\H(\S+\theta \f,\z), 
\end{equation}
where the current-reversal operator $\theta$ is an involutive linear operator acting on $\f$, i.e. $\theta^2\f=\f$. For instance, for overdamped diffusion processes, $\theta$ is equal to minus the identity: $\theta \f = - \f$, while for jump processes $\theta$ is the operator that exchanges initial and final states of a jump: $\theta f_{nm} = f_{mn}$. The affinity $\S$ is such that $\theta \S = -\S$ and may depend on $\z$. For example, we have $F_{nm} = \ln \left( \tilde{k}_{nm} \mu_m/\tilde{k}_{mn} \mu_n \right)$ for independent many-body jump processes (Section~\ref{Sec_many_indep_jump}) and $F(x) = - \frac{2 J^\rho(x)}{D(x)\rho(x)}$ for independent many-body diffusion processes (Appendix~\ref{Sec_many_indep_diff}). Hence, the fluctuation relation of Eq.~\eqref{Sym_H} is formally equivalent to the assumption of local detailed balance.

In the following, we investigate the inheritance of the fluctuation relation by the biased and rectified Hamiltonian given the fluctuation relation~\eqref{Sym_H} for the original Hamiltonian $\H$. From the definition of the biased Hamiltonian~\eqref{def_Hbiais_det} and using Eq.~\eqref{Sym_H}, we have
\begin{align}
\H_{\g}(\f,\z) &= \H(\f+\g_1,\z)+\g_2\cdot \z \\
&= \H(\S+\theta (\f+\g_1),\z)+\g_2\cdot \z \\
&= \H(\S+(\theta -1)\g_1+\theta \f+\g_1,\z)+\g_2\cdot \z,
\end{align}
leading to the fluctuation relation for the biased Hamiltonian
\begin{equation} \label{Sym_H_biased}
    \H_{\g}(\f,\z) = \H_{\g}(\S_{\g}+\theta \f,\z),
\end{equation}
where we introduced the \textit{biased affinity} 
\begin{equation}
    \S_{\g} \equiv \S + (\theta-1) \g_1
\end{equation} 
that satisfies $\theta \S_{\g} = - \S_{\g}$. Similarly, from the definition of the rectified Hamiltonian~\eqref{Hr_det} and the fluctuation relation for $\H_{\g}$~\eqref{Sym_H_biased}, the rectified Hamiltonian satisfies
\begin{align}
\H^{\ur}(\f,\z; \g) & = \H_{\g}(\f + \D^\dagger \partial_{\z} W_{\sm}, \z) - \H_{\g}( \D^\dagger \partial_{\z} W_{\sm}, \z) \\
& = \H_{\g}(\S_{\g}+\theta (\f + \D^\dagger \partial_{\z} W_{\sm}), \z) - \H_{\g}( \D^\dagger \partial_{\z} W_{\sm}, \z) \\
& = \H_{\g}(\S_{\g} + (\theta-1) \D^\dagger \partial_z W_{\sm} + \theta \f + \D^\dagger\partial_z W_{\sm} ,z)-\H_{\g}(\D^\dagger\partial_{\z} W_{\sm}, \z)\\
& = \H^{\ur}( \S^{\ur}_{\g} + \theta \f, \z;\g),
\end{align}
leading to the fluctuation relation
\begin{equation}  \label{fluctuation_th_Hamilton}
      \H^{\ur}(\f,\z; \g)  = \H^{\ur}( \S^{\ur}_{\g} + \theta \f, \z;\g),
\end{equation}
where we introduced the \textit{rectified affinity}
\begin{equation}
    \S^{\ur}_{\g} \equiv \S + (\theta-1) (\g_1+  D^\dagger \partial_{\z} W_{\sm})
\end{equation}
that satisfies $\theta \S^{\ur}_{\g} = - \S^{\ur}_{\g}$. Hence, the unbiased, biased and rectified Hamiltonians have a similar fluctuation symmetry with different affinities given respectively by $\S$, $\S_{\gamma}$ and $\S^{\ur}_{\g}$. Note that $(\theta-1) \g_1$ and $(\theta-1) (\g_1+ D^\dagger \partial_{\z} W_{\sm} )$ are (twice) the antisymmetric part of $\g_1$ and $(\g_1+ D^\dagger \partial_{\z} W_{\sm} )$ under $\theta$, and therefore can be separately interpreted as affinities.

\subsection{Dual dynamics}
\label{DualDyn}

The dual dynamics follows from rectifying the time reversed dynamics, i.e. the dynamics with the same Hamiltonian up to a sign change of the momenta $\p$. By definition, this duality transformation is involutive. In the framework of Markov jump processes, this duality corresponds to a similarity transformation based on the stationary probability and applied to the transposed of the rate matrix~\cite{Hatano2001_vol86}. 
In the framework of diffusive processes, the dual dynamics follows from a modification of the Fokker-Plank generator that leads to a reversal of the local probability current~\cite{Chernyak2006_vola}, while the stationary probability density is unchanged. 
In this section, we define the dual dynamics associated with a statistical Hamiltonian and show that it follows from a momentum reversal followed by a canonical transformation. We note that the momentum reversal itself is not a canonical transformation~\cite{Blanchard1988}. We end this section by providing several remarks to emphasize the similarities between a dynamics and its dual.

We define the dual Hamiltonian of $H$ as
\begin{equation}
\hat{H}(\hat \p,\hat \z) = H( \partial_{\hat\z}W_{\um} -\hat \p ,\hat\z), \label{dual-hamiltonien-standard}
\end{equation}
with $W_{\um} = W_{\um}(\z,\g=0)$ the unstable solution of the stationary HJ equation 
\begin{equation}
H(\p = \partial_{\z} W,\z) = 0. \label{HJEqS}
\end{equation}
We denote with an over hat the phase space variables and the Hamiltonian of the dual dynamics. The canonical transformation $(\p,\z)\rightarrow (\hat \p, \hat \z) $ is produced by the generating function (of the second type) 
\begin{equation}
    F_2(\hat \p ,\z) = \hat \p \cdot \z - W_{\um}(\z,0).
\end{equation}
This generates the canonical change of variable 
\begin{eqnarray}
    \p &=&  \partial_{\z} F_2 =  \hat \p - \partial_{\z} W_{\um}, \\ 
    \hat \z &=& \partial_{ \hat \p} F_2 = \z,
\end{eqnarray}
that, when combined with the momentum reversal, leads to the dual Hamiltonian
\begin{equation}
    H(\p,\z) \xrightarrow[\p \rightarrow - \p]{}  H( - \p , \z) = H(\partial_{\hat\z} W_{\um} - \hat \p,\hat \z) = \hat H( \hat \p , \hat \z). \label{reversal+canonical}
\end{equation}

~~

As a first remark, we can relate this procedure to the definition of duality for Markov jump processes: the gauge change plays the role of the similarity transformation while the momentum reversal replaces the transposition of the rate matrix.

~~

Second, the relationship of Eq.~\eqref{reversal+canonical} between a dynamics and its dual shows that $H$ and $\hat{H}$ have the same solutions $0$ and $W_\um$ of the stationary HJ equation
\begin{equation}
    0 = \hat H(\partial_{\hat \z} W ,\z) = H(\partial_{\hat\z} W_{\um} - \partial_{\hat\z} W,\hat \z).
\end{equation}
Given that fixed points are at the intersection of these global solutions $\p=\partial_{\z} W_{\sm}=0$ and $\p = \partial_{\z} W_{\um}$, the max-min formula applied to Eq.~\eqref{reversal+canonical} shows that the two dynamics have the same dominant fixed point (taking a minimum on $\p$ or $-\p$ makes no difference). More generally, since critical manifolds are at the intersection of the two global solutions $\p=\partial_{\z} W_{\sm}=0$ and $\p = \partial_{\z} W_{\um}$ that are the same for the two dynamics and associated with the same value of the Hamiltonians, the Hamiltonian and its dual share the same dominant critical manifold.

However, the two Hamiltonians have opposite stability for their dominant critical manifolds due to the momentum reversal: considering the reduced dual dynamics yields
\begin{eqnarray}
         \dot{\hat \z}_\sm &=& \partial_{\hat \p} \hat H(0,\hat \z) = - \partial_{\p} H(\partial_\z W_\um, \z) = - \dot{\z}_{\um}. \label{OppositeStability}
\end{eqnarray}
In other words, the reduced dual dynamics on its stable manifold is the time-reversal of the reduced original dynamics on its unstable manifold. Eq.~\eqref{OppositeStability} shows as well that the velocities of the two dynamics are opposite in connection with the reversal of currents that must be produced by a duality transformation.

~~

Third, we emphasize that the dual Hamiltonian defines a proper stochastic dynamics, i.e. with null Hamiltonian at $\p = 0$ and with the correct convexity in $\p$ as two dual Hamiltonians have the same convexity in $\p$. The duality does not change the $\z$ coordinate of the Hamiltonian, leaving unchanged the assumption made on this side.

~~

Finally, we remark that detailed-balanced dynamics are self dual. Indeed, assuming detailed balance for the affinities amounts to writing $\bm F = D^\dagger \partial_\z U $, which is to say that $\bm F$ derives from a potential. In this case, the fluctuation relation taken at $\f = 0$ reads
\begin{eqnarray}
 \H(\bm F,\z) = H(0,\z) = 0,
\end{eqnarray}
so that the two global solutions of the HJ equation are simply $ W_{\um} = U$ and $W_\sm = 0 $. 

We then have $\bm F = D^\dagger \partial_\z W_\um$ and, using again the fluctuation relation of Eq.~\eqref{Sym_H}, we find from the definition of the dual Hamiltonian
\begin{equation} \label{dual_detbal}
    \hat H(\hat \p , \hat \z) = H(\partial_{\hat \z} W_\um - \hat \p ,\hat \z) = \H(D^\dagger \partial_{\hat \z} W_\um - D^\dag \hat \p ,\hat \z) = \H( D^\dag \hat \p ,\hat \z) = H(\hat \p , \hat \z).
\end{equation}
The dual dynamics is thus the same as the original one (i.e. it is self-dual). The fact that the duality transformation is the identity transformation for detailed-balanced dynamics (which have null stationary currents) is in complete consistency with the general fact that duality reverses currents.

\section{Application to population processes} \label{application}
Up to now, we have reviewed the modeling of nonlinear stochastic processes using Lagrangian and Hamiltonian dynamics in a rather abstract way. On this basis, we have introduced the rectification of biased processes via a canonical transformation. We now apply this formalism to the case of population processes. This class of processes constitutes a nonlinear generalization of Markov jump processes whose states are described by a set of extensive variables such as energy, number of particles, etc. We will first derive the nonlinear generators of these processes and then move on to a Lagrangian and Hamiltonian description.

\subsection{General population process} \label{poprocesses}

We consider a many-body system modeled by a Markov jump process defined on an infinite lattice. We denote by $X = 1, \dots \Omega$ the states occupied by each particle and $\N$ the state vector where $N_X = 1, \dots, N$ is the number of particles in state $X$. We denoted by $\{\alpha\}$ the set of allowed transitions and assume that for an initial state $\N$ and a transition $\alpha$, the final state $\N'$ is then constrained by $\N' = \N + \d_\alpha$, where $\mathfrak{D}_{X,\alpha}$ is the variation of number of particles in state $X$ after a transition $\alpha$ and $\d_\alpha$ is the column vector with component $(\d_\alpha)_X = \d_{X,\alpha}$.
Let $k_{\N + \d_\alpha,\N}$ the transition rate from $\N$ to $\N+\d_\alpha$. We assume that $k_{\N + \d_\alpha,\N}$ scales with a large parameter $N$ (volume, total number of particle \dots) and that $\N$ is of order $N$: $\N = O(N)$. We can then define a new state variable $\z \equiv \frac{\N}{N}$ and a new intensive rate
\begin{equation} \label{intensive_rate_pop}
k_\alpha(z) \equiv \lim_{N \rightarrow \infty} \frac{k_{\N + \d_\alpha,\N}}{N}.  
\end{equation}
Let us consider the observable $\A_t$ defined by
\begin{equation}
\A_t \equiv \frac{1}{t} \left(
\begin{array}{c}
\sum_{t' \in [0,t]} [\bm \Omega]_{t'}^{t'+\ud t'} \\
\int_0^t \N(t') \ud t'
\end{array} \right),
\end{equation}
where $[\bm \Omega]_{t'}^{t'+\ud t'}$ is the vector function whose component $[\Omega_{\alpha}]_{t'}^{t'+\ud t'}$ is the number of transitions $\alpha$ that have occurred between times $t'$ and $t'+\delta t'$, so that the first component of $\A_t$ counts the number of each transition occurring during the time interval $[0,t]$. The biased transition matrix ruling the evolution of the generating function $G_{\N}(t,\g) = \left\langle \e^{t \g \cdot \A_{t}} \delta_{\N(t),\N} \right\rangle$ reads in the Dirac notation
\begin{equation} \label{kappa_continu}
\bkappa = \sum_{\alpha,\N} k_{\N+\d_\alpha,\N}  \e^{\gamma^1_\alpha}  \ket{\N+\d_\alpha}\bra{\N}  - k_{\N+\d_\alpha,\N}  \ket{\N} \bra{\N}  + \sum_{\N}  \g_2 \cdot \N   \ket{\N} \bra{\N}.  
\end{equation} 
As stated in Section~\ref{Jump}, we compute the generator of the driven process $\K$ by taking the Doob transform of the biased matrix~\eqref{kappa_continu} with respect to its dominant left eigenvector. We obtain
\begin{multline} \label{driven_generator_pop}
\K = \sum_{\alpha,\N} \e^{U_{\N+\d_\alpha}-U_{\N}} k_{\N+\d_\alpha,\N}  \e^{\gamma^1_\alpha}  \ket{\N+\d_\alpha  \N}\bra{\N}  \\
- \sum_{\N} \left(\sum_{\alpha} k_{\N+\d_\alpha,\N} - \g_2 \cdot \N + \un \scgfn \right)  \ket{\N} \bra{\N},
\end{multline}
where $\e^{\U}$ is the left eigenvector of $\bkappa$ associated with its highest eigenvalue $\scgf = \un \scgfn$. Note that when $\g=0$, the biased matrix becomes the original transition rate matrix whose highest eigenvalue $0$ is associated with the left eigenvector whose components are all equal to $1$. Note that the results of Sec.~\ref{SecNjumps} are just a particular case of what is stated up to now. \\

We want now to describe our problem within the Lagrangian and Hamiltonian formalism introduced in Section~\ref{LHdescription}. We focus directly on the biased process since the original one is simply obtained by taking $\g = 0$. {We obtain the biased Lagrangian $\L_{\g}(\bl,\z)$ using the same construction as found in the appendix of Ref.~\cite{Lazarescu2019} in the case of mass-action chemical kinetics, which we apply to the more general rates~\eqref{intensive_rate_pop}.} This procedure consists in computing the biased transition probability
\begin{equation}
    G_{\delta t}(\N_\mathrm{f} \mid \N_\mathrm{i}) \equiv \braket{\N_\mathrm{f} \mid \e^{\delta t \bkappa} \mid \N_\mathrm{i}} 
\end{equation} 
After proving the commutation of the operators appearing in the right-hand side of~\eqref{kappa_continu} in the continuous limit defined by
\begin{equation}
\begin{cases}
N \rightarrow \infty,\\
\delta t \rightarrow 0, \\
N \delta t \rightarrow \infty,
\end{cases}
\end{equation}
we obtain that
\begin{equation}
\braket{\N_\mathrm{f} \mid \e^{\delta t \bkappa} \mid \N_\mathrm{i}} \asymp \e^{-\delta t N \L_{\g}(\bl,\z)} \delta(\dot{\z}-\d \bl),
\end{equation}
where the biased Lagrangian is given by
\begin{equation} \label{Lbiased_derivation}\textbf{}
\L_{\g}(\bl,\z) = \sum_\alpha \left[\lambda_\alpha \ln\left(\frac{\lambda_\alpha}{k_\alpha(\z)} \right) - \lambda_\alpha + k_\alpha(\z) - \gamma^1_\alpha \lambda_\alpha\right] - \g_2 \cdot \z.
\end{equation}
{The variable $\lambda_\alpha$ counts the number of  transitions $\alpha$.}
Taking the Legendre-Fenchel transform of~\eqref{Lbiased_derivation} with respect to $\bl$, we obtain the biased Hamiltonian
\begin{equation} \label{Hbiased_derivation}
\H_{\g}(\f,\z) = \sum_\alpha k_\alpha(\z) \left[\e^{f_\alpha + \gamma^1_\alpha} - 1 \right] + \g_2 \cdot \z,
\end{equation}
where $\f$ is the conjugated variable of $\bl$. We now want to derive the Lagrangian associated with the driven generator $\K$ of Eq~\eqref{driven_generator_pop}. To do so, we assume that for $N \rightarrow \infty$, $\N = O(N)$, there exists for any $\alpha$ a function $W(\z)$ such that
\begin{equation} \label{assump_u}
U_{\N+\d_\alpha}-U_{\N} \simeq \d_\alpha \cdot \partial_{\z}W,
\end{equation}
where the scalar product is performed over the states: $\d_\alpha \cdot \partial_{\z}W \equiv \sum_{X} \d_{X,\alpha} \partial_{z_X}W$. We investigate the nature of the function $W$ by writing the spectral relation between $\U$ and $\bkappa$:
\begin{align}
\e^{\U} \bkappa &= N \scgfn \e^{\U} \\
\sum_\alpha \e^{U_{\N+\d_\alpha}} \kappa_{\N+\d_\alpha,\N} + \e^{U_{\N}} \kappa_{\N,\N} &= N \scgfn \e^{U_{\N}} \\
\sum_\alpha \e^{U_{\N+\d_\alpha}-U_{\N}} \kappa_{\N+\d_\alpha,\N} + \kappa_{\N,\N} &= N \scgfn \\
\sum_\alpha \e^{U_{\N+\d_\alpha}-U_{\N}} k_{\N+\d_\alpha,\N} \e^{\gamma^1_\alpha} - \sum_\alpha k_{\N+\d_\alpha,\N} + \g_2 \cdot \N &= N \scgfn,
\end{align}
where we used Eq.~\eqref{kappa_continu} in the last equation. Taking the continuous limit, and using the assumption~\eqref{assump_u}, we finally obtain
\begin{eqnarray} \label{HJpop}
\sum_\alpha k_\alpha(\z) \left[ \e^{\d_\alpha \cdot \partial_{\z}W + \gamma^1_\alpha} - 1 \right] + \g_2 \cdot \z = \scgfn.
\end{eqnarray}
We recognize the biased Hamiltonian in the left-hand-side of Eq.~\eqref{HJpop}:
\begin{equation}
\H_{\g}(\f = \d^\dagger \partial_{\z}W, \z) = \scgfn,
\end{equation} 
Hence, the function $W$ appears to be Hamilton's characteristic function and corresponds to $W_{\sm}$. Going through the same calculation for the dominant right eigenvector leads to the same HJ equation but implying $W_{\um}$. This is in line with our conjecture on the nonlinear counterpart of the Perron-Frobenius theorem of Sec.~\ref{Sec_conj_PF}.

Following the same procedure used to derive the biased Lagrangian, we obtain that the transition probability $P_{\K,\delta t}(\N_{\mathrm{f}} \mid \N_{\mathrm{i}}) \equiv \braket{\N_{\mathrm{f}} \mid \e^{\delta t \K} \mid \N_{\mathrm{i}}}$ associated with the generator of driven process satisfies a LDP:
\begin{equation}
P_{\K,\delta t}(\N_{\mathrm{f}} \mid \N_{\mathrm{i}}) \apeupn \e^{-\delta t \un \L^{\ur}(\bl,\z;\g)} \delta(\dot{\z}-\d \bl),
\end{equation}
where the detailed rectified Lagrangian is given by
\begin{eqnarray} \label{Lr_derivation}
\L^{\ur}(\bl,\z;\g) = \L_{\g}(\bl,\z) - \bl \cdot \d^\dagger \partial_{\z} W + \scgfn.
\end{eqnarray}
The corresponding rectified Hamiltonian is given by
\begin{equation}
\H^{\ur}(\f,\z;\g) = \sum_\alpha k_\alpha(\z) \left[\e^{f_\alpha + \gamma^1_\alpha + \d_\alpha \cdot \partial_{\z} W} - 1 \right] + \g_2 \cdot \z - \scgfn = \H_{\g}(\f+\d^\dagger \partial_{\z} W,\z) - \H_{\g}(\d^\dagger \partial_{\z} W,\z).
\end{equation}
As expected, the rectified Lagrangian and Hamiltonian obtained for our population process are consistent with the definitions of Eqs.~(\ref{Hr_det}--\ref{Lr_det}). The Lagrangian--Hamiltonian description is thus just another face of the same coin and the rectification in one formalism or another is equivalent. 

We considered in this section general population processes with unspecified nonlinear rates. In the following, we look at specific models. In the first example, we come back to linear many-body Markov jump processes, this time from the point of view of the Lagrangian--Hamiltonian description. The second and third examples deal with nonlinear systems.

\subsection{$\un$ independent Markov jump processes} \label{Sec_many_indep_jump}

In this section, we apply the results of Section \ref{poprocesses} to the case of $N$ independent Markov jump processes introduced in Section~\ref{SecNjumps}. In this case, our observables $\z$ and $\bl$ are respectively the empirical density $\bmu$ \eqref{emp_density_Npart} and the empirical transition current $\om$ \eqref{emp_current_Npart}. The transitions $\alpha {=(m'm)}$ {occur between two microstates $m \rightarrow m'$}, the operator $\d$ becomes the incidence matrix $\bm D$ with component:
\begin{equation}
D_{n,\,(m'm)} =
    \begin{cases}
        1 &         \text{if } n=m',\\
            -1 & \text{if } n=m,\\
            0 & \text{otherwise},
    \end{cases}
\end{equation}
while the intensive rates $\kk(\bmu)$ \eqref{intensive_rate_pop} are linear in $\bmu$ and can be explicitly written using \eqref{rate_Npart}:
\begin{equation}
k_{nm}(\bmu) = \tilde{k}_{nm} \mu_m.
\end{equation}
From Eqs.~(\ref{Lbiased_derivation}--\ref{Hbiased_derivation}), the detailed biased Lagrangian reads
\begin{equation} \label{L_biaise_jumps}
\L_{\g}(\om, \bmu) = \sum_{n, m \neq n} \left[ \omega_{nm} \ln \left( \frac{\omega_{nm}}{k_{nm} \mu_m} \right) - \omega_{nm} + k_{nm} \mu_m  \right] - \g_1 \cdot \om - \g_2 \cdot \bmu.
\end{equation}
with $\g_1 \cdot \om \equiv \sum_{n, m \neq n} \gamma^1_{nm} \omega_{nm}$, while the detailed biased Hamiltonian reads
\begin{equation} \label{H_biaise_jumps}
\H_{\g}(\f,\bmu) = \sum_{n, m \neq n} k_{nm} \mu_m \left[\e^{f_{nm} + \gamma^1_{nm}} - 1 \right] + \g_2 \cdot \bmu. 
\end{equation}
with $\f$ the conjugate variable of $\om$. Note that the unbiased Lagrangian and Hamiltonian are recovered when taking $\g=0$. Using the fact the function $\U$ appearing in the left eigenvector $\e^{\U}$ of $\bkappa$ is related to the function $\u$ appearing the left eigenvector $\e^{\u}$ of the single-particle biased matrix $\tilde{\bkappa}$ through $U_{\N} = \N \cdot \u$ as seen in Sec.~\eqref{SecNjumps}, one finds that Hamilton's characteristic function $W_{\sm}$ and $\u$ are related by
\begin{equation}
\partial_{\bmu}W_{\sm} = \u.
\end{equation}
The rectified Hamiltonian reads then
\begin{equation}
\H^{\ur}(\f,\bmu;\g) = \H_{\g}(\f + \bm D^\dagger \u,\bmu) - \H_{\g}(\bm D^\dagger \u,\bmu),
\end{equation}
leading to
\begin{equation}
\H^{\ur}(\f,\bmu; \g) = \sum_{n, m \neq n} \tilde{K}_{nm} \mu_m \left[\e^{f_{nm}} - 1 \right],
\end{equation}
which corresponds to the Hamiltonian of a non-biased process with generator $\K$ with component $K_{nm} = \tilde{K}_{nm} \mu_m$. This illustrates the fact that the rectification of biased Hamiltonians is equivalent to the rectification of biased generators in the linear operator formalism using the Doob transform.

\subsection{Interacting processes: the Brownian Donkey} \label{Sec_BK}
In this section, we apply the results of Section~\ref{poprocesses} to the Brownian Donkey model involving $N$ interacting unicyclic machines~\cite{Cleuren2001, vroylandt2020efficiency}. Each machine consists of a two-level system jumping between a lower state $\downarrow$ of energy $0$ and a higher state $\uparrow$ of energy $\ebd > 0$ via two heat reservoirs labeled by $\nu = 1,2$ and of inverse temperature $\beta^\nu$. Two machines interact via an interaction energy $\frac{V}{N}$ only when they are in different states. {We denote by $N_{\uparrow}$ the number of machines in the high energy state. When the Brownian Donkey is in state $N_{\uparrow}$, it has a total energy $U_{N_{\uparrow}} = N_{\uparrow} \ebd + N_{\uparrow}(\un-N_{\uparrow})\frac{V}{\un}.$ We assume that during an infinitesimal time $\delta t$, two machines cannot jump at the same time and only one transition occurs so that jumping from an initial state $N_{\uparrow}$ to a final state $N_{\uparrow}'$ imposes that $N_{\uparrow}' = N_{\uparrow} \pm 1.$ The transition rate $k^\nu_{\epsilon+N_{\uparrow},N_{\uparrow}}$ from state $N_{\uparrow}$ to state $N_{\uparrow}+\epsilon$, with $\epsilon = \pm 1$, via the reservoir $\nu$ is given by 
\begin{equation}
k^\nu_{N_{\uparrow}+\epsilon,N_{\uparrow}} = \un \left(\frac{1+\epsilon}{2} - \epsilon \frac{N_{\uparrow}}{\un} \right) \exp{\left[-\frac{\beta^\nu}{2}\left(E_a + U_{N_{\uparrow}+\epsilon} - U_{N_{\uparrow}} + \epsilon (-1)^\nu F \right)\right]},    
\end{equation}
with $E_a$ an activation energy and $F$ a non-conservative force.} For this model, the observable $\z$ becomes then the density of machines in the high energy state $N_{\uparrow}/N$ and $\bl$ with component $\lambda_{\epsilon}^\nu$ becomes the current density of machines passing from the high (low) energy level to the low (high) energy level when $\epsilon = -1$ ($\epsilon = +1$) via channel $\nu$. Both are related by $\dot{z} = \D \bl \equiv \sum_{\epsilon,\nu} \epsilon \lambda_\epsilon^\nu$ where $\D$ is here the line vector operator defined by $\D_\epsilon^\nu \equiv \epsilon$, $\forall \nu$. The transition rates $\kk(z)$ in the continuous limit reads
\begin{equation}
k_\epsilon^\nu(z) = \left(\frac{1+\epsilon}{2} - \epsilon z\right) \e^{-\frac{\beta^\nu}{2}\left(E_a + \epsilon \ebd + \epsilon V (1 - 2 z) + \epsilon (-1)^\nu F \right)}.
\end{equation}
From Eqs.~(\ref{Lbiased_derivation}--\ref{Hbiased_derivation}), the detailed biased Lagrangian and Hamiltonian read:
\begin{align} \label{LbiaiseBrownian}
\L_{\g}(\bl,z) &= \sum_{\epsilon, \nu} \left[ \lambda_\epsilon^\nu \ln \left( \frac{\lambda_\epsilon^\nu}{k_\epsilon^\nu(z)} \right) - \lambda_\epsilon^\nu + k_\epsilon^\nu(z) \right] - \g_1 \cdot \bl - \gamma_2 z, \\
\H_{\g}(\f,z) &= \sum_{\epsilon, \nu}  k_\epsilon^\nu(z) \left[ \e^{f_\epsilon^\nu + \gamma_{1,\epsilon}^\nu} - 1 \right] + \gamma_2 z, \label{HbiaiseBrownian}
\end{align}
with $ \g_1 \cdot \bl \equiv \sum_{\epsilon, \nu} \gamma_{1,\epsilon}^\nu \lambda_\epsilon^\nu$. For this model, we can compute explicitly the standard Lagrangian using Eq.~\eqref{L_stand}:
\begin{equation} \label{LbiasedBD_stand}
L_{\g}(\dot{z},z) = - \sqrt{\dot{z}^2 +\varphi(z,\g_1)} + \sum_{\epsilon,\nu} k_\epsilon^\nu(z) - \dot{z} \ln \left[ \frac{- \dot{z} + \sqrt{\dot{z}^2 +\varphi(z,\g_1)}}{2 \sum_\nu k_-^\nu(z) \e^{{\gamma_{1,-}}^\nu}}    \right] - \gamma_2 z,
\end{equation}
with $\varphi(z,\g_1) \equiv 4 \prod_{\epsilon}\sum_\nu k_\epsilon^\nu(z) \e^{ \gamma_{1,\epsilon}^\nu}$, recovering the result of Ref.~\cite{vroylandt2019}. Taking the Legendre-Fenchel transform of~\eqref{LbiasedBD_stand}, we obtain the standard biased Hamiltonian
\begin{equation} \label{HbiasedBD_stand}
H_{\g}(p,z) = \sum_{\epsilon, \nu}  k_\epsilon^\nu(z) \left[ \e^{\epsilon p + \gamma_{1,\epsilon}^\nu} - 1 \right] + \gamma_2 z.
\end{equation}
Note that the detailed Hamiltonian~\eqref{HbiaiseBrownian} and the standard Hamiltonian~\eqref{HbiasedBD_stand} are indeed related by:
\begin{equation}
H_{\g}(p,z) = \H_{\g}(\f=\D^\dagger p,z).
\end{equation}
\begin{figure}
\includegraphics[scale=0.45,trim=0.5cm 0cm 1cm 0cm,clip = true]{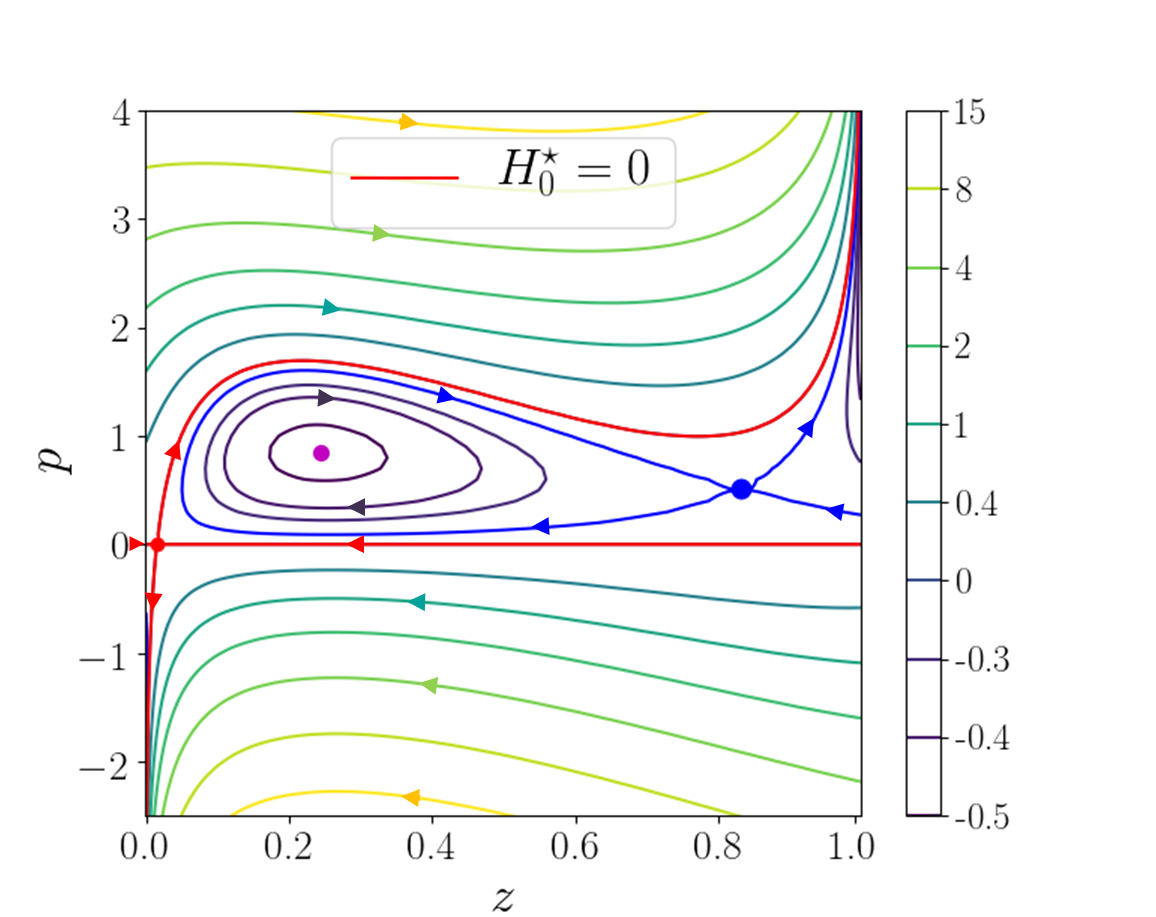}
\includegraphics[scale=0.45,trim=1.3cm 0cm 2cm 0cm,clip = true]{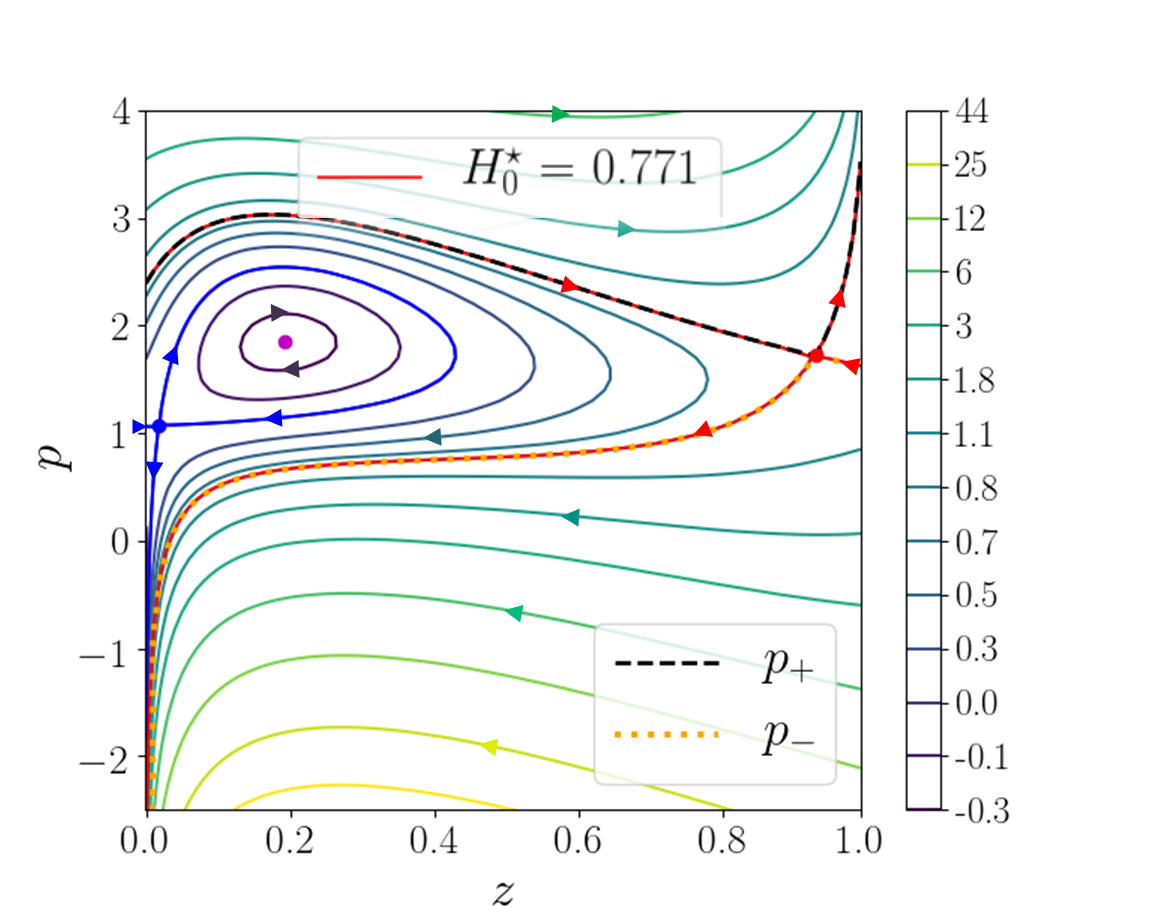} 
\caption{(Left) Trajectories of the original Hamiltonian ($\g = 0$).
(Right) Trajectories of the biased Hamiltonian ($\gamma_{1,\epsilon}^\nu = \epsilon$ with $\epsilon = \pm 1$ and $\nu = 1,2$, and $\gamma_2 = 1$). The dashed black line corresponds to the solution $p_+$ and the gold dotted line corresponds to the solution $p_-$. \\
In both figures, there are three fixed points represented by the three colored points. The red point corresponds to the dominant fixed point of coordinates $(z_0^\star, p_0^\star)$, and the red trajectory is associated with the max-min value of the Hamiltonian. \\
Both figures are obtained for $\ebd = 0.8$, $V = 2$, $E_a = 1$, $F = 1$, $\beta_1 = 1$, $\beta_2 = 2$. \label{fig_traj_BK_biased_original} }
\end{figure}
For this model, the solutions of the implicit equation
\begin{equation} \label{W_BD}
H_{\g}(p(z,\g),z) = \scgfn
\end{equation}
can be explicitly computed provided that $\left(\scgfn - \gamma_2 z + \sum_{\epsilon,\nu} k_\epsilon^\nu(z)\right)^2 \geq \varphi(z,\g_1)$. In this case, Eq.~\eqref{W_BD} admits two solutions $p_\pm \equiv p_\pm(z,\g)$ with
\begin{equation}
p_\pm(z,\g) \equiv \ln \left[ \frac{\scgfn - \gamma_2 z + \sum_{\epsilon,\nu} k_\epsilon^\nu(z) \pm \sqrt{\left(\scgfn - \gamma_2 z + \sum_{\epsilon,\nu} k_\epsilon^\nu(z)\right)^2 - \varphi(z,\g_1)}}{2 \sum_\nu k_+^\nu(z) \e^{ \gamma_{1,+}^\nu}} \right]. \label{sol_HJ_BD}
\end{equation}
Note that the solution $p_-$ exists only if $\scgfn-\gamma_2 z+~\sum_{\epsilon,\nu} k_\epsilon^\nu(z)\geq~0$, defining a domain of validity for $\gamma_2$. The SCGF $\scgfn$ coincides with the value $H_0^\star = \max_{z} \min_{p} \Hg(p,z)$ of the biased Hamiltonian at the dominant fixed point $(p_0^\star,z_0^\star)$, represented by a red point in Fig.~\ref{fig_traj_BK_biased_original}. {We may remark in passing that this system exhibits a first-order dynamical phase transition at some value of $\g$ between the two values illustrated above, which is made clear by the fact that the dominant fixed point switches from left to right, as described (in a slightly different context) in Ref.~\cite{Lazarescu2019}. As we have mentioned in \ref{Sec_conj_PF}, the transition point itself does not fit in our conjecture, and in particular there are no globally stable or unstable solutions of HJ passing through the fixed points in that case.}

\subsubsection*{Stability of the solutions}

As illustrated in Fig.~\ref{fig_traj_BK_biased_original}, the solutions $p_\pm$ of Eq.~\eqref{sol_HJ_BD} have the following stability properties: \\
$\bullet$ For $z \in [0, z_0^\star]$:
\begin{equation}  \label{stab1BD}
\begin{cases}
\textrm{The branch } p_+ \textrm{ is stable},  \\
\textrm{The branch } p_- \textrm{ is unstable}  \\
\end{cases}
\end{equation}
$\bullet$ For $z \in [z_0^\star,1]$:
\begin{equation}  \label{stab2BD}
\begin{cases}
\textrm{The branch } p_+ \textrm{ is unstable},  \\
\textrm{The branch } p_- \textrm{ is stable}.  \\
\end{cases}
\end{equation}
It follows from Eqs.~(\ref{stab1BD}--\ref{stab2BD}) that HJ equation at $E = \scgfn$ admits one globally stable solution $p_{\sm} \equiv \partial_{z} W_{\sm}$ and one globally unstable solution $p_{\um} \equiv \partial_{z} W_{\um}$ with
\begin{align}
\partial_{\z} W_{\sm} &\equiv 
\begin{cases}
p_+ ~~~ \text{if } z \in [0,z_0^\star] \\
p_- ~~~ \text{if } z \in [z_0^\star,1]
\end{cases}, \\
\partial_{\z} W_{\um} &\equiv
\begin{cases}
p_- ~~~ \text{if } z \in [0,z_0^\star] \\
p_+ ~~~ \text{if } z \in [z_0^\star,1]
\end{cases}.
\end{align}
The global stability (resp. instability) of the dynamics at the manifold $p=p_{\sm}$ (resp. $p = p_{\um}$) is illustrated in Fig.~\ref{fig_traj_BK_biased_original}. Indeed, the two orbits of the globally stable (resp. unstable) manifold converge to (resp. exit from) the red fixed point. In the non-biased case, the globally stable solution is $p_{\sm}(z,\g=0)$ as required.

\subsubsection*{Rectification}

The rectified Lagrangian is given by
\begin{equation} \label{stand_L_BD}
\L^{\ur}(\bl,z;\g) = \L_{\g}(\bl,z) - \bl \cdot \D^\dagger p_{\sm} + \scgfn,
\end{equation}
with $\left( \D^\dag p_{\sm} \right)_\epsilon^\nu \equiv \epsilon p_{\sm}$, and the rectified Hamiltonians by
\begin{align}
\H^{\ur}(\f,z;\g) &= \H_{\g}(\f+ \D^\dagger p_{\sm},z) - \scgfn, \\
H^{\ur}(p,z;\g) &= H_{\g}(p + p_{\sm},z) - \scgfn.
\end{align}
%------------------------------------------------------
\begin{figure}
\begin{center}
\includegraphics[scale=0.45,trim=0cm 0cm 0cm 0cm,clip = true]{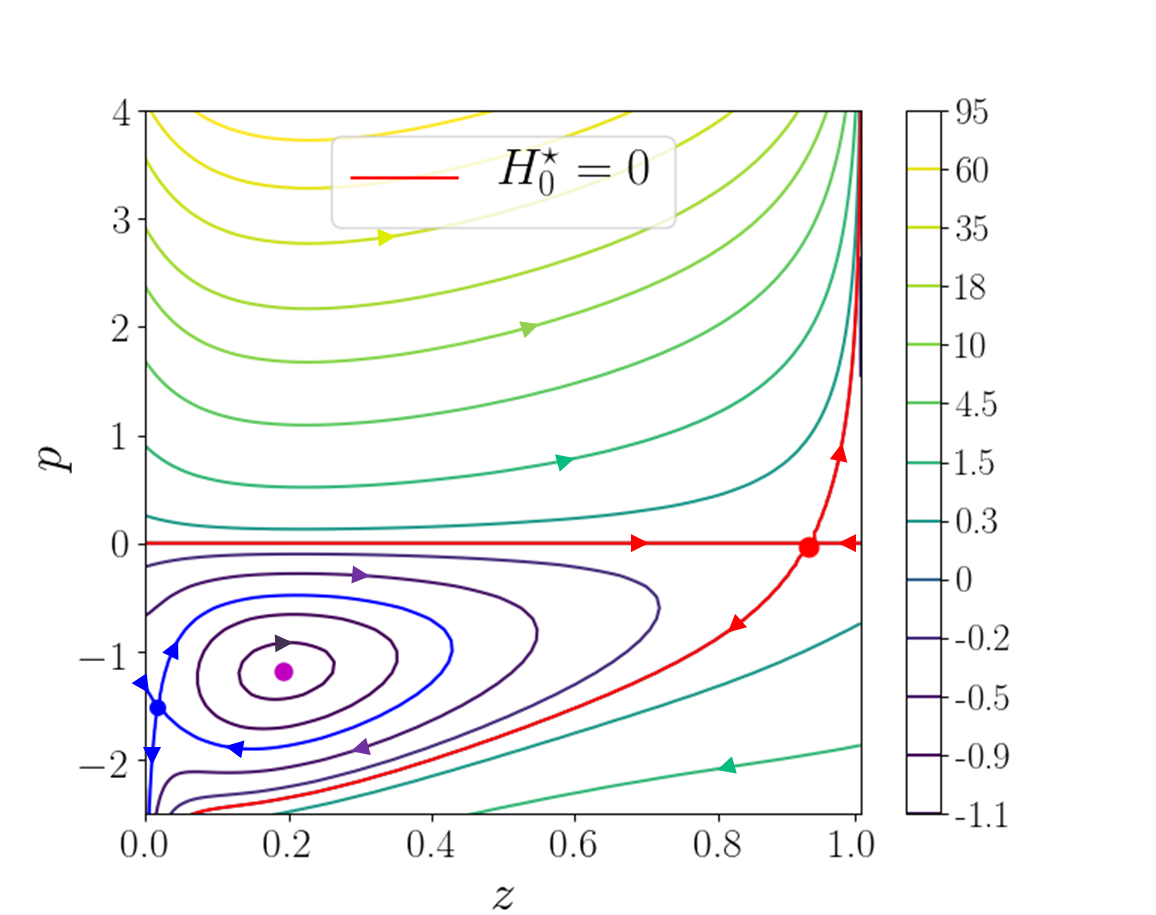}
\caption{Trajectories of the rectified Hamiltonian. There are three fixed points represented by the three colored dot. The red trajectory is associated with the max-min value of the rectified Hamiltonian. As expected, this value is zero. \\
%The figures are obtained for $\ebd = 0.8$, $N=1000$, $V = 2$, $E_a = 1$, $F = 1$, $\beta_1 = 1$, $\beta_2 = 2$, $\gamma_{1,\epsilon}^1 = \epsilon$, $\gamma_{1,\epsilon}^2 = \epsilon$ and $\gamma_2 = 1$. 
We used the parameters of Fig.~\ref{fig_traj_BK_biased_original} (right).\label{fig_traj_BK_rectified_original} }
\end{center}
\end{figure}
%-------------------------------------------------
For this model, we can compute explicitly the standard rectified Lagrangian by using relation~\eqref{L_stand} on Eq.~\eqref{stand_L_BD}:
\begin{equation} \label{LrBD_stand}
L^{\ur}(\dot{z},z;\g) = - \sqrt{\dot{z}^2 +\varphi(z,\g_1+\D^\dag p_{\sm})} + \sum_{\epsilon,\nu} k_\epsilon^\nu - \dot{z} \ln \left[ \frac{- \dot{z} + \sqrt{\dot{z}^2 +\varphi(z,\g_1+\D^\dag p_{\sm})}}{2 \sum_\nu k_-^\nu \e^{{\gamma^\nu_{1,-}}-p_{\sm}}}    \right] - \gamma_2 z + \scgfn,
\end{equation}
which can be written as
\begin{equation}
    L^{\ur}(\dot{z},z;\g) = L_{\g^{\ur}}(\dot{z},z) + \scgfn,
\end{equation}
with $\g^{\ur} \equiv (\g_1+\D^\dag p_{\sm},\gamma_2)$.

\subsubsection*{Fluctuation symmetry}
\begin{figure} 
\includegraphics[scale=0.22]{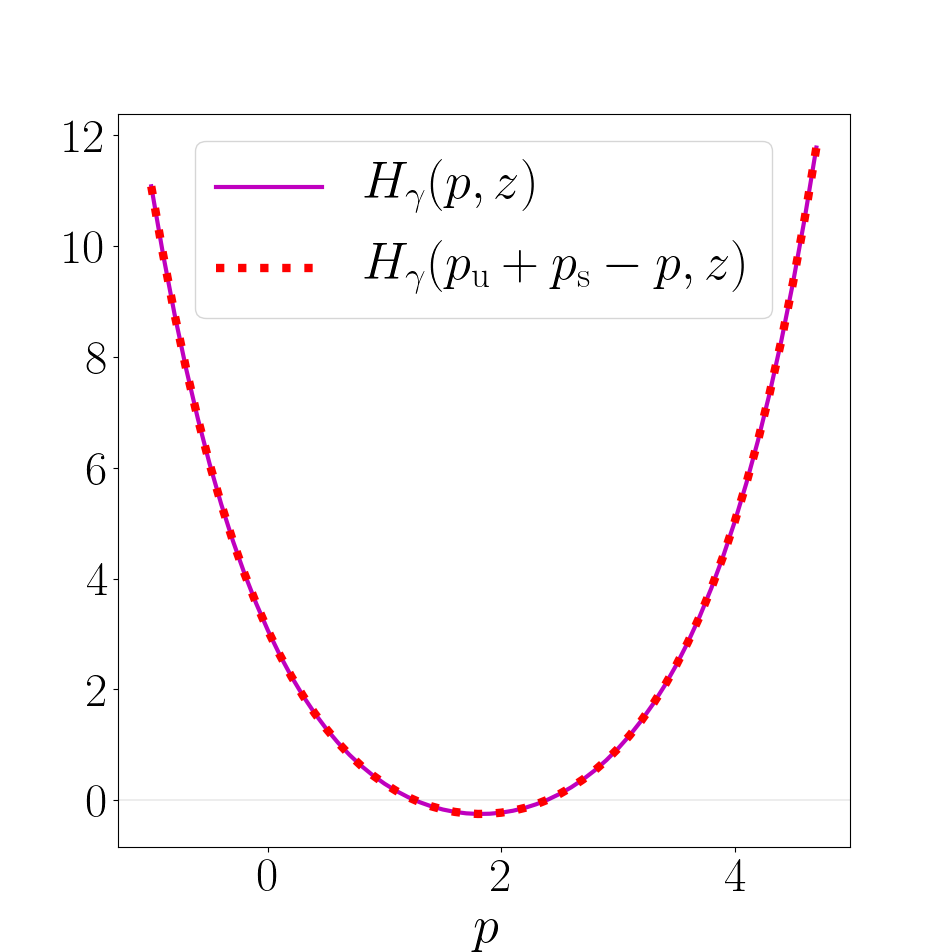}
\includegraphics[scale=0.22]{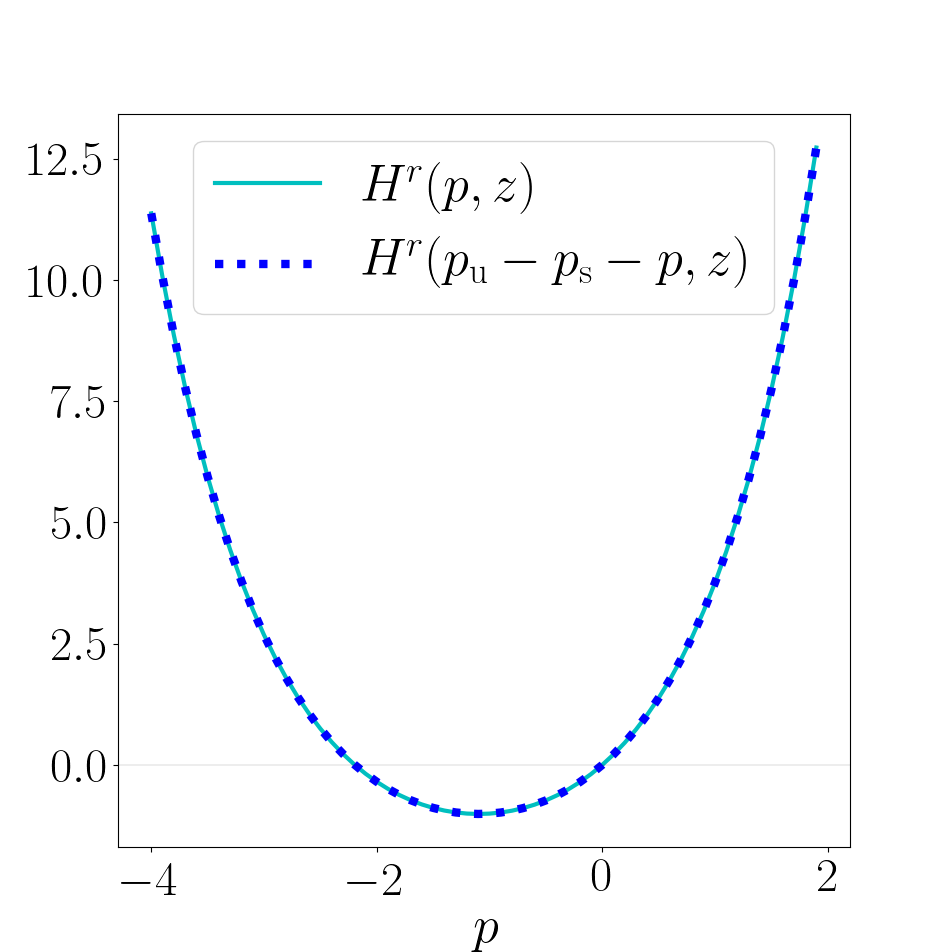}
\includegraphics[scale=0.22]{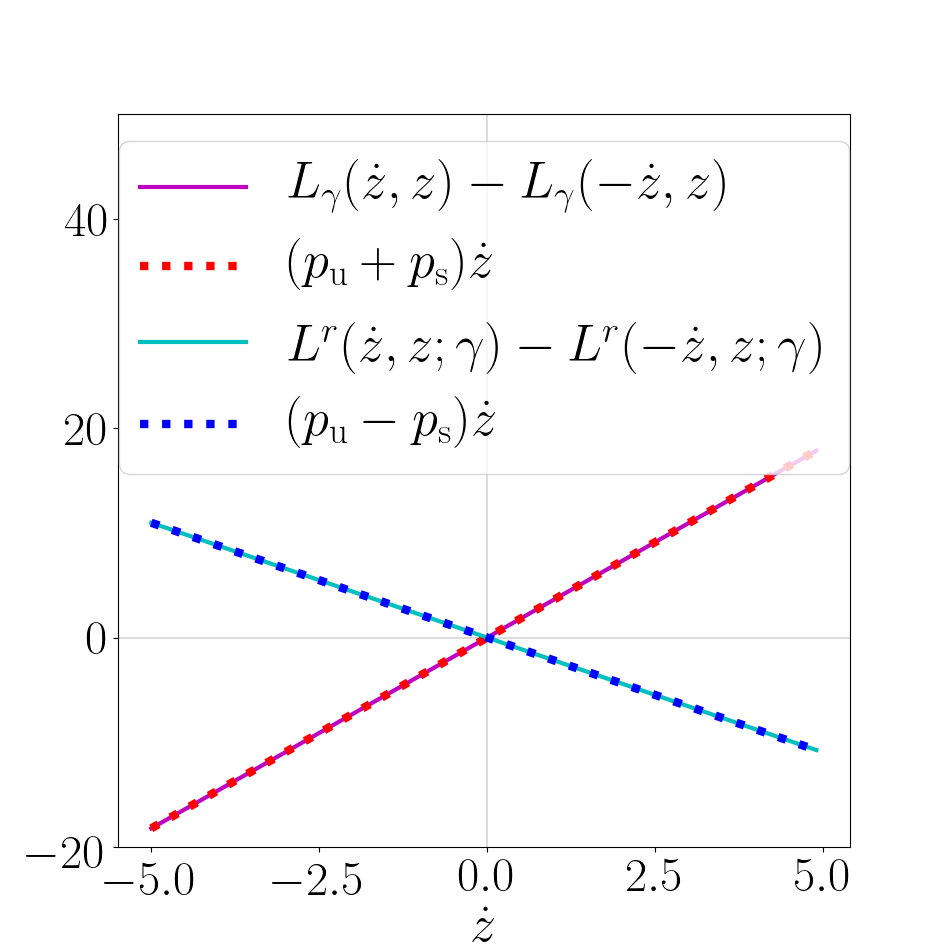}
\caption{(Left) Fluctuation symmetry for the biased Hamiltonian~\eqref{Sym_H_BD}. (Middle) Fluctuation symmetry for the rectified Hamiltonian~\eqref{Sym_Hr_BD}. (Right) Fluctuation symmetries for the biased and rectified Lagrangians~(\ref{Sym_L_BK}--\ref{Sym_Lr_BK}). \\
The figures are obtained for $z = 0.3, \ebd = 0.8$, $V = 2$, $E_a = 1$, $F = 1$, $\beta_1 = 1$, $\beta_2 = 2$, $\gamma_{1,\epsilon}^\nu = \epsilon$ ($\epsilon = \pm 1$ and $\nu = 1,2$) and $\gamma_2 = 1$.} \label{fig_sym_H_BK}
\end{figure}

The biased Hamiltonian has a fluctuation symmetry:
\begin{equation}
\H_{\g}(\f,z) = \H_{\g}(\S_{\g} + \theta \f,z),
\end{equation}
with
\begin{align}
(\S_{\g})_\epsilon^\nu & \equiv  \epsilon \ln \frac{k_-^\nu \e^{\gamma_{1,-}^\nu}}{k_+^\nu \e^{\gamma_{1,+}^\nu}},  \\
\theta f_\epsilon^\nu & \equiv f_{-\epsilon}^\nu.
\end{align}
As seen in Sec.~\ref{Sec_fluct_th}, this symmetry is inherited by the rectified Hamiltonian through
\begin{equation}
\H^{\ur}(\f,z;\g) = \H^{\ur}(\S^{\ur}_{\g} + \theta \f,z;\g),
\end{equation}
with
\begin{align}
\S^{\ur}_{\g} \equiv \S_{\g} + (\theta-1) \D^\dag p_{\sm}. 
\end{align}
The fluctuation symmetry reads for the standard biased and rectified Hamiltonians:
\begin{align} \label{Sym_H_BD}
H_{\g}(p,z) &= H_{\g}(p_{\um} + p_{\sm} - p,z), \\
H^{\ur}(p,z;\g) &= H^{\ur}(p_{\um} - p_{\sm} - p, z; \g),
\end{align} \label{Sym_Hr_BD}
where we used $\theta \D^\dag = - \D^\dag$. We check numerically this symmetries in Fig.~\ref{fig_sym_H_BK} (left and middle).
Let's finally remark that for Lagrangians these symmetries yield
\begin{align} \label{Sym_L_BK}
L_{\g}(\dot{z},z) - L_{\g}(-\dot{z},z) &= (p_{\um} + p_{\sm}) \dot{z}, \\
L^{\ur}(\dot{z},z;\g) - L^{\ur}(-\dot{z},z;\g) &= (p_{\um} - p_{\sm}) \dot{z},  \label{Sym_Lr_BK}
\end{align}
as shown numerically in Fig.~\ref{fig_sym_H_BK} (right).

\subsubsection*{Duality}

Taking $\g=0$, the unbiased Hamiltonians reads
\begin{align}
\H(\f,z) &= \sum_{\epsilon, \nu}  k_\epsilon^\nu(z) \left[ \e^{f_\epsilon^\nu} - 1 \right] \\
H(p,z) &= \sum_{\epsilon, \nu}  k_\epsilon^\nu(z) \left[ \e^{\epsilon p} - 1 \right].
\end{align}
The HJ equation $H(\partial_{z}W,z) = 0$ admits two solutions
\begin{align}
\partial_{z} W_{\sm}(z) &= 0,  \\
\partial_{z} W_{\um}(z) &= \ln \frac{\sum_\nu k_-^\nu(z)}{\sum_\nu k_+^\nu(z)}.
\end{align}
The dual Hamiltonian~\eqref{dual-hamiltonien-standard} reads then
\begin{equation}
\hat{H}(\hat p,\hat z) = H( \partial_{\hat z}W_{\um} -\hat p ,\hat z) = H(\hat p,\hat z),
\end{equation} 
as expected from Eq.~\eqref{dual_detbal} and from the fact that the standard Hamiltonian has the fluctuation symmetry
\begin{equation}
H(\partial_{z} W_{\um} - p, z) = H(p,z).
\end{equation}

\subsection{Chemical reaction network} \label{Sec_chimie}
\begin{figure}
\includegraphics[scale=0.45,trim=0.cm 0cm 1cm 1cm,clip = true]{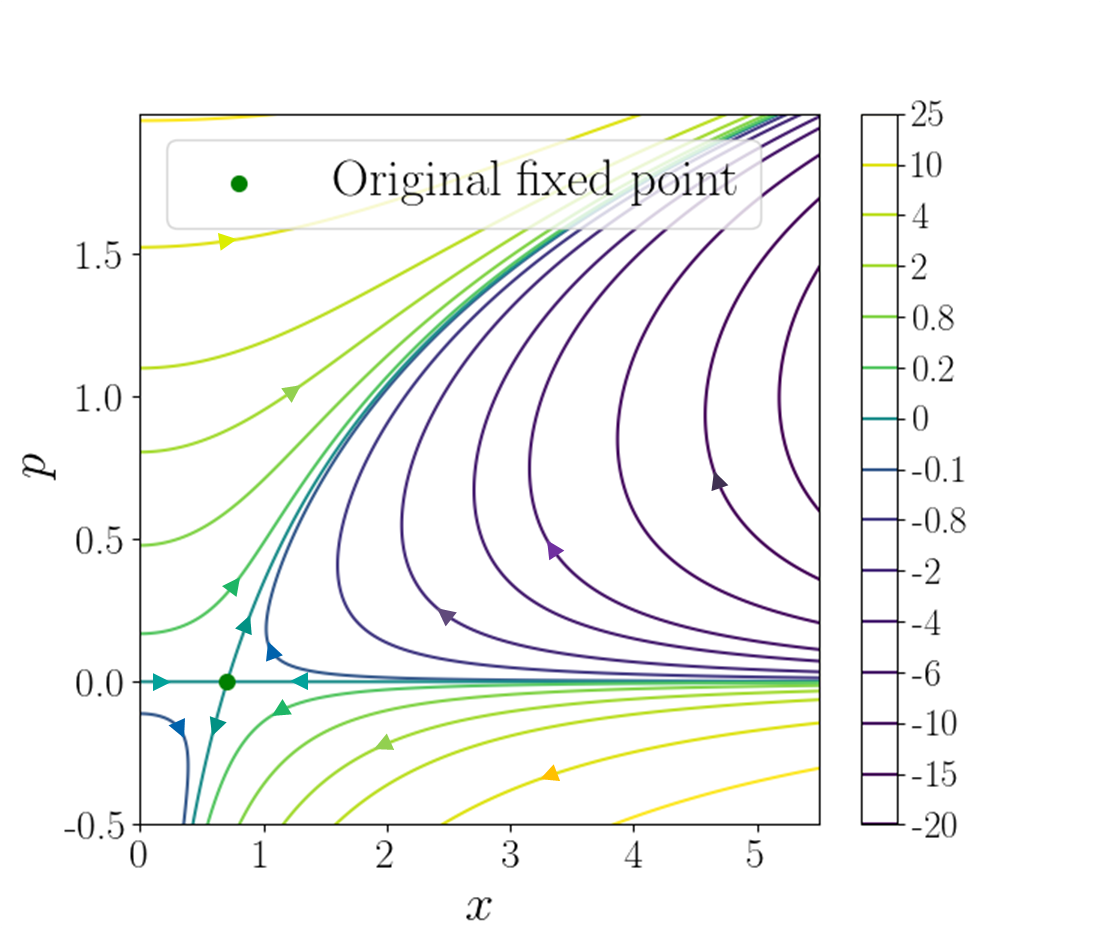}
\includegraphics[scale=0.45,trim=0.2cm 0cm 2cm 1cm,clip = true]{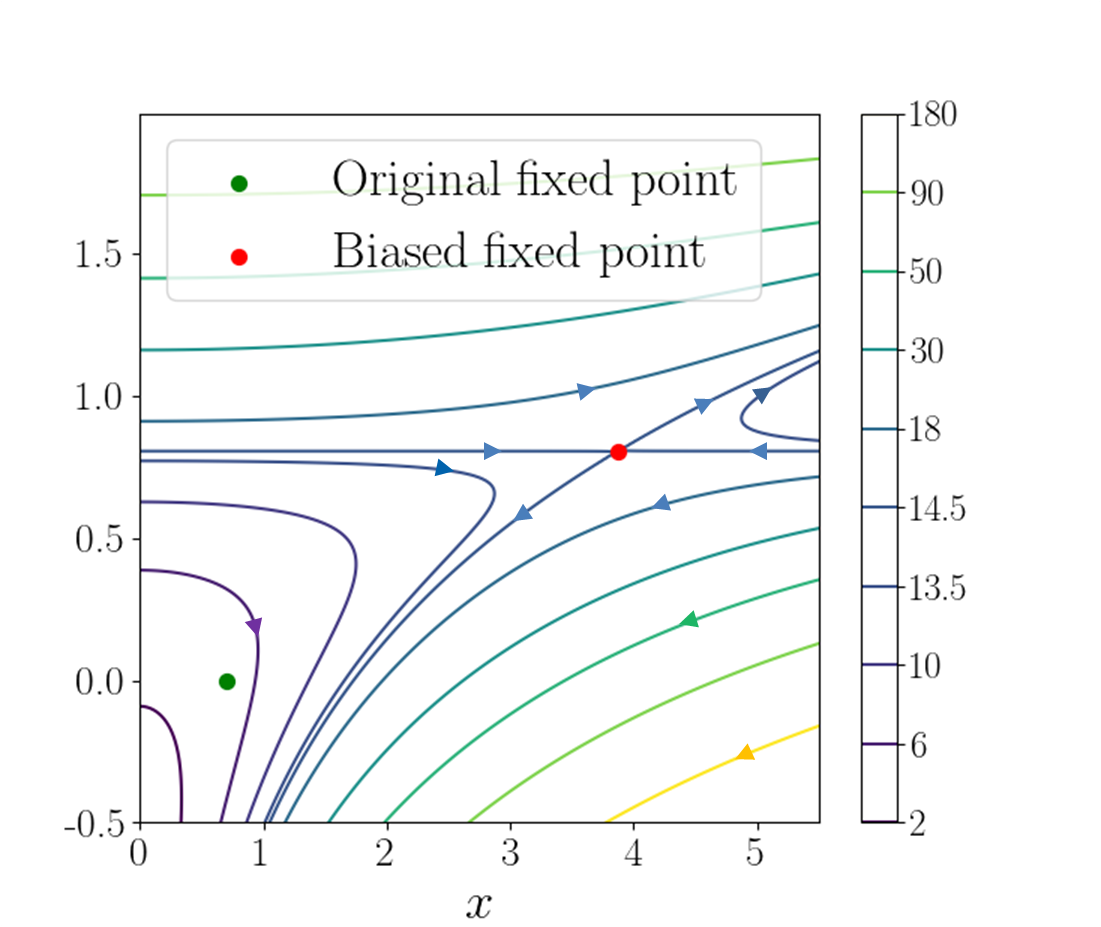} 
\caption{(Left) Trajectories of the original Hamiltonian ($\g = 0$). The coordinates of the fixed point (green point) are $\left(x_{\textsc{nb}} = \sqrt{(\pa a + \mb b)/\delta}, p_{\textsc{nb}} = 0\right)$. 
(Right) Trajectories of the biased Hamiltonian. The coordinates of the fixed point (red point) are $\left(x_0^\star = \sqrt{\alpha \beta/\delta^2},p_0^\star = (1/2) \ln (\beta/\delta)\right)$. \\
The left figure is obtained for $\pa a + \mb b = 0.5$, $\delta = 1$, for which $x_{\textsc{nb}} = 0.707$. The right figure is obtained for $\alpha = 3$, $\beta = 5$, $\delta = 1$, $\pa a + \mb b = 0.5$ for which $x_0^\star = 3.873$ and $p_0^\star =  0.805$.  \label{fig_traj_chimie_biased_original}}
\end{figure}

We now illustrate the results of Section \ref{poprocesses} on a chemical system modeled by the following chemical reactions: 
\begin{equation}
A \underset{\mathfrak{K}_{-1}}{\overset{\mathfrak{K}_{+1}}{\rightleftharpoons}} 2X \underset{\mathfrak{K}_{-2}}{\overset{\mathfrak{K}_{+2}}{\rightleftharpoons}} B,
\end{equation}
where $\mathfrak{K}_{\er}$ is the kinetic constant of reaction $\er$ ($\epsilon = \pm$, $\r = 1,2$) and the species $A$ and $B$ are chemostatted and have constant concentrations $a$ and $b$. We denote by $x$ the concentration of the species $X$. The observables $\z$ and $\bl$ represent here respectively the concentration $x$ and the chemical current such that $N \delta t \lambda_{\er}$ is the number of times reaction $\er$ occurs during an infinitesimal time $\delta t$ for a volume $N$. They are related by $\dot{x} = \D \bl$, where $\D$ is the line vector whose component $\D_{\er} \equiv -2\epsilon(-1)^{\r}$ is the variation of the number of species $X$ when reaction $\er$ occurs. We choose the transition rates $k_{\er}(\z)$ according to the mass-action law, stating that the rates are directly proportional to the product of the concentrations of the reactants:
\begin{eqnarray}
k_{+1} &=& \pa a, \\
k_{-1} &=& \ma x^2, \\
k_{+2} &=& \pb x^2, \\
k_{-2} &=& \mb b.
\end{eqnarray}
For simplicity, the observable $\A$ is chosen to be the chemical current $\frac{1}{t} \int_0^t \bl(t') \ud t'$ of conjugate variable $\g \equiv (\{ \gamma_{\er} \})$. The standard biased Hamiltonian reads then
\begin{equation}
\Hg(p,x) = \sum_{\er} k_{\er}(x) \left[ \e^{-2 (-1)^{\r} \epsilon p + \gamma_{\er}} - 1 \right].
\end{equation}
For clarity we introduce:
\begin{eqnarray} 
\alpha &\equiv& \pa a \e^{\gamma_{+1}} + \mb b \e^{ \gamma_{-2}}, \\
\beta &\equiv& \ma \e^{\gamma_{-1}} + \pb \e^{\gamma_{+2}}, \\
\delta &\equiv& \ma + \pb,
\end{eqnarray}
so that the biased Hamiltonian \eqref{hamilton_biaisé} simplifies to
\begin{equation} \label{Hbiased_chimie}
\Hg(p,x) = \alpha \e^{2p} + \beta x^2 \e^{-2p} - \delta x^2 - (\pa a + \mb b).
\end{equation}
Hamilton's Equations for the biased Hamiltonian are given by
\begin{equation} \label{hamilton_biaisé}
\begin{cases}
\dot{x} = \frac{\partial \Hg}{\partial p} = 2\alpha \e^{2p} -2 \beta x^2 \e^{-2p}, \\ 
- \dot{p} = \frac{\partial \Hg}{\partial x} = 2\beta x \e^{-2p}  - 2\delta x.
\end{cases}
\end{equation}
This system admits one critical manifold consisting of a fixed point of coordinates 
\begin{equation} \label{fp_Hbiaisé}
\begin{cases}
x_0^\star = \sqrt{\frac{ \alpha \beta}{\delta^2}}, \\
p_0^\star = \frac{1}{2} \ln \frac{\beta}{\delta}, 
\end{cases}
\end{equation}
In Fig~\ref{fig_traj_chimie_biased_original}, we plot the phase portrait both in the non-biased and biased cases. The dominant trajectory in the long-time limit corresponds to $\Hg(p,x) = \scgfn$ with
\begin{equation}
\scgfn = \Hg(p_0^\star,x_0^\star) = \frac{\alpha \beta}{\delta} - (\pa a + \mb b).
\end{equation}
It is easy to check that $\scgfn(\g=0)=0$, as required in the non-biased case. Biasing modifies the position of the fixed point from a concentration $x_{\textsc{nb}}$ to a new concentration $x_0^\star$ and shifts the variable $p$ of the fixed point from $0$ to $p_0^\star \neq 0$, which is consistent with the fact that the SCGF is non-zero when $\g \neq 0$ and that the biased Hamiltonian does not vanish at $p=0$.
%-----------------------------------------------------------------------------------------------
%-----------------------------------------------------------------------------------------------
The implicit equation
\begin{equation} \label{HJ_chimie}
H_{\g}(p(x,\g),z) = \scgfn
\end{equation}
admits two solutions $p_{\sm,\um} \equiv p_{\sm,\um}(x,\g)$ with: 
\begin{equation}
\begin{cases}
p_{\sm}(x,\g) \equiv p_0^\star = \frac{1}{2} \ln \frac{\beta}{\delta} , \\
p_{\um}(x,\g) \equiv \frac{1}{2} \ln \frac{x^2}{{x_0^\star}^2 \e^{-2p^\star}} = \frac{1}{2} \ln \frac{\delta x^2}{\alpha}.
\end{cases}
\end{equation}
The solution $p_{\sm}$ corresponds to the globally stable solution and contains two orbits that converge toward the fixed point, and the $p_{\um}$ corresponds to the globally unstable solution and contains two orbits that leave the fixed point (see Fig.~\ref{fig_traj_chimie_biased_original}). Notice that in the non-biased case, we have indeed that $p_{\sm}(\g=0) = 0$. Finally, we can compute the rectified Hamiltonian from Eq.~\eqref{Hr_stand}:
\begin{equation}
H^{\ur}(p,x;\g) = \delta {x_0^\star}^2 \left( \e^{2p} - 1 \right) + \delta x^2 \left(\e^{-2p} - 1 \right),
\end{equation}
which can be rewritten as
\begin{equation}
H^{\ur}(p,x;\g) = \sum_{\er} K_{\er}(\z) \left[ \e^{-2(-1)^{\r}\epsilon p} - 1 \right],
\end{equation}
where $K_{\er}(x) \equiv k_{\er}(x) \e^{\gamma_{\er} - 2(-1)^{\r} p_{\sm}}$ is the rectified intensive rate obtained from the Doob transform of the biased transition rate in the linear operator formalism. As expected, the rectified Hamiltonian respects the structure of the original unbiased Hamiltonian with new rates. 

\begin{figure} 
\begin{center}
\includegraphics[scale=0.45]{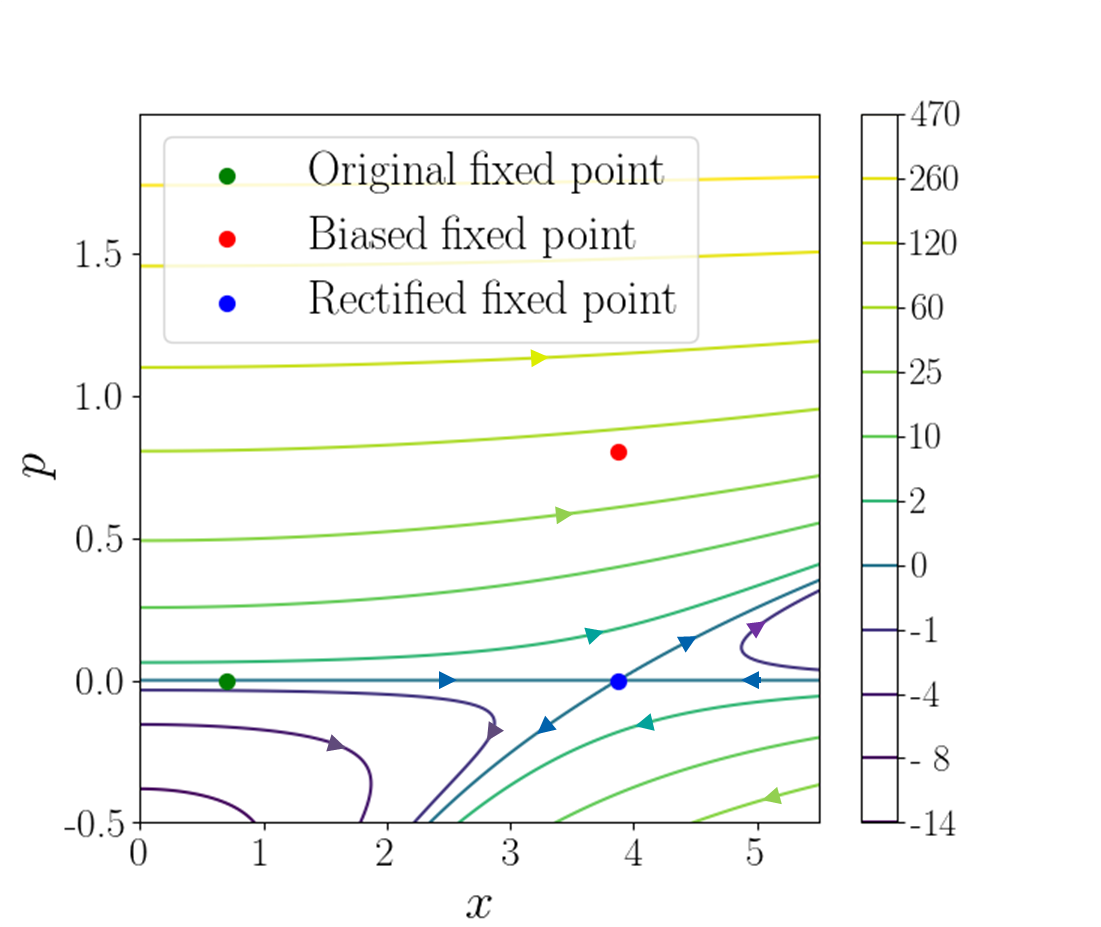}
\caption{Trajectories of the rectified Hamiltonian. The coordinates of the fixed point (blue point) are  $\left(x_0^\star = \sqrt{\alpha \beta/\delta^2},p=0\right)$.
%The figures are obtained for $\alpha = 3$, $\beta = 5$, $\delta = 1$, $A = 0.5$, $\pa a + \mb b = 0.5$.
We used the parameters of Fig.~\ref{fig_traj_chimie_biased_original} (right).
\label{fig_traj_chimie_rectified}}
\end{center}
\end{figure}

Figs.~\ref{fig_traj_chimie_biased_original} and \ref{fig_traj_chimie_rectified} offer a visualization of the effect of biasing and rectification on the fixed point. Starting from the original Hamiltonian with fixed point $(x_{\textsc{nb}},p_{\textsc{nb}} = 0)$, we bias the dynamics via the parameter $\g$ to impose a new typical value of the observable $\A$ in the long-time limit. This translates into a new dominant trajectory, hence a new fixed point $(x_0^\star,p_0^\star)$. Yet, the biased Hamiltonian does not vanish at $p=0$, and the globally stable $p_{\sm} = p_0^\star$ is not zero, as required for a norm-conserving Markov process. The rectification allows to build a proper statistical Hamiltonian by shifting the globally stable manifold $p_{\sm}$ to $0$, leading to a new fixed point at $p=0$ while keeping the concentration $x_0^\star$, compatible with the imposed value of the observable $\A$.
%-----------------------------------------------------------------------------------------
%-----------------------------------------------------------------------------------------
\subsubsection*{Fluctuation symmetry}

\begin{figure} 
\includegraphics[scale=0.22]{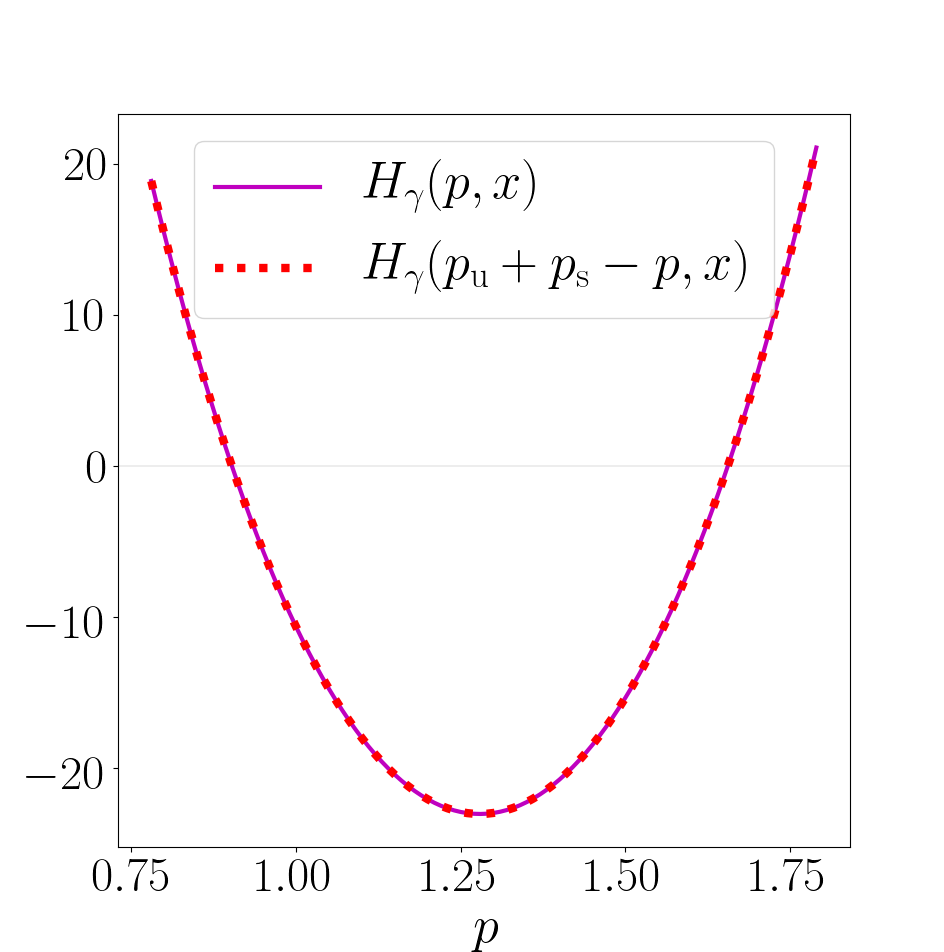}
\includegraphics[scale=0.22]{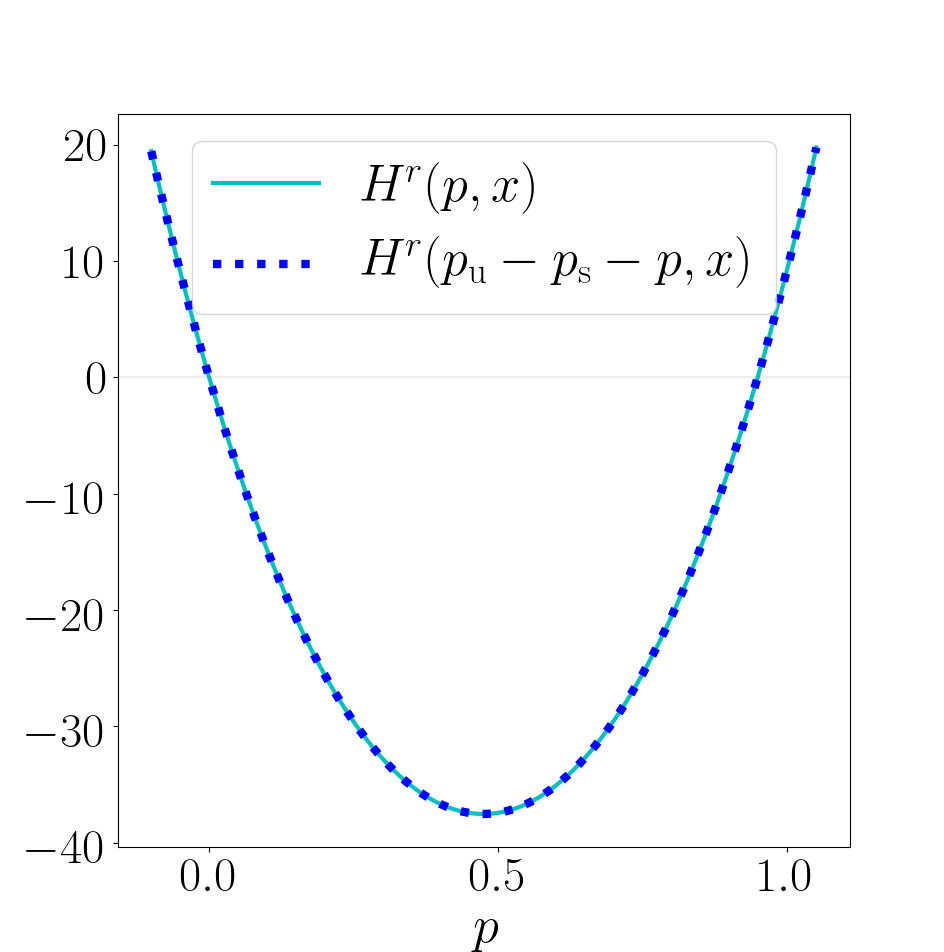}
\includegraphics[scale=0.22]{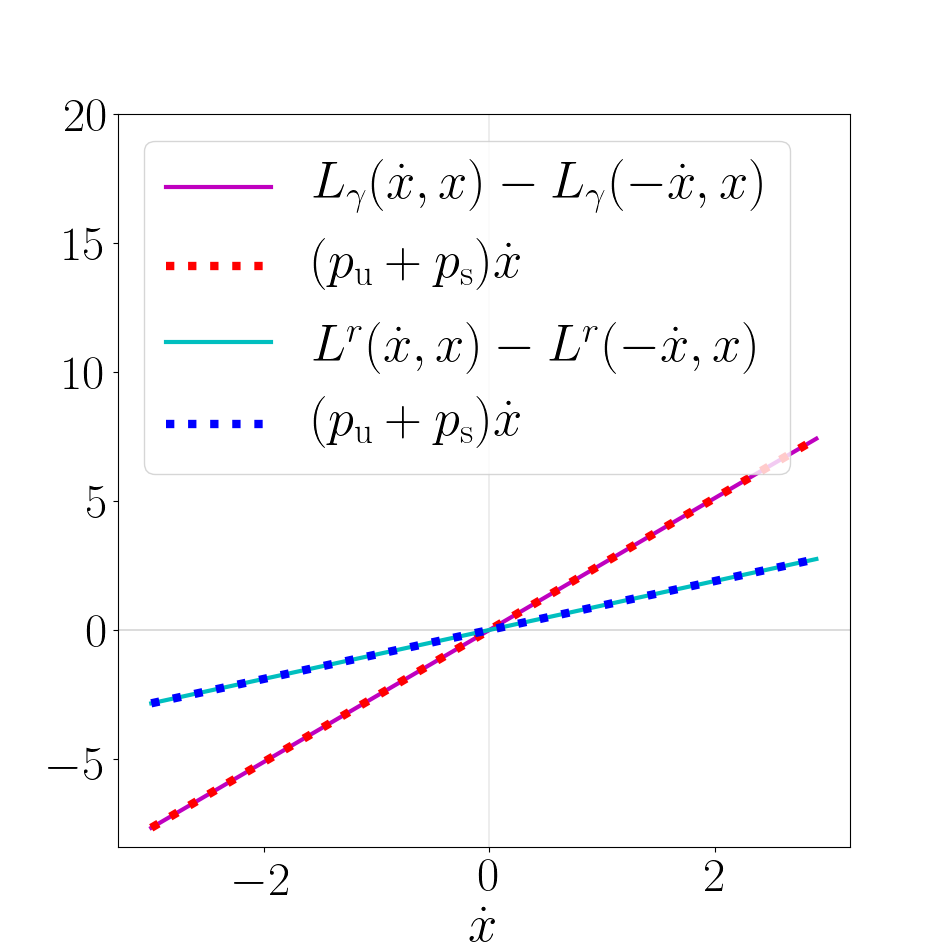}
\caption{(Left) Fluctuation symmetry for the biased Hamiltonian~\eqref{sym_chimie_biased}. (Middle) Fluctuation symmetry for the rectified Hamiltonian~\eqref{sym_chimie_rectified}. (Right)  Fluctuation symmetries for the biased and rectified Lagrangians~(\ref{sym_L_biased_rectified}--\ref{sym_L_rectifedd_rectified}). \\
The figures are obtained for $\alpha = 3$, $\beta = 5$, $\delta = 1$, $\pa a + \mb b = 0.5$ and $x = 10$. \label{fig_sym_H_chimie}}
\end{figure}

The biased Hamiltonian has a fluctuation symmetry:
\begin{equation} \label{sym_chimie_biased}
\Hg(p,x) = \Hg(p_{\um} + p_{\sm} - p,x),
\end{equation}
which implies a fluctuation symmetry for the rectified Hamiltonian 
\begin{equation} \label{sym_chimie_rectified}
H^{\ur}(p,x;\g) = H^{\ur}(p_{\um} - p_{\sm} - p ,x; \g).
\end{equation}
We check numerically this symmetries in Fig.~\ref{fig_sym_H_chimie} (left and middle). For this system, the standard biased Lagrangian can be computed explicitly and we obtain
\begin{equation}
L_{\g}(\dot{x},x) = \frac{1}{2} \dot{x} \ln \left( \frac{\dot{x} + \sqrt{\dot{x}^2 + 16\alpha \beta x^2}}{4 \alpha} \right) - \frac{1}{2} \sqrt{\dot{x}^2 + 16\alpha \beta x^2} + \delta x^2 + \pa a + \mb b,
\end{equation}
while the standard rectified Lagrangian reads
\begin{equation}
L^{\ur}(\dot{x},x) = \frac{1}{2} \dot{x} \ln \left( \frac{\dot{x} + \sqrt{\dot{x}^2 + 16 (\delta x {x_0^\star})^2}}{4 \delta {x_0^\star}^2} \right) - \frac{1}{2} \sqrt{\dot{x}^2 + 16 (\delta x {x_0^\star})^2} + \delta (x^2 + {x_0^\star}^2),
\end{equation}
Both Lagrangians satisfy the following fluctuation symmetry:
\begin{align} \label{sym_L_biased_rectified}
L_{\g}(\dot{x},x) - L_{\g}(-\dot{x},x) &= (p_{\um} + p_{\sm}) \dot{x}, \\
L^{\ur}(\dot{x},x;\g) - L^{\ur}(-\dot{x},x;\g) &= (p_{\um} - p_{\sm}) \dot{x}, \label{sym_L_rectifedd_rectified}
\end{align}
as shown numerically in Fig~\ref{fig_sym_H_chimie} (right).
%-----------------------------------------------------------------------------------------
%-----------------------------------------------------------------------------------------

\section{Conclusion}

%REVIEW & RESULTS
In this work, we have extended to nonlinear Markov processes described by Lagrangian or Hamiltonian the well-known techniques used to study Markov processes under conditioning in the linear operator formalism. The spectral problem that must be solved to determine the effective driven process becomes a Hamilton-Jacobi equation in the Lagrangian--Hamiltonian formalism. Accordingly, an equivalent of the Perron-Frobenius (resp. Krein-Rutman) theorem that tightly constrains the eigenvector (resp. eigenfunction) should exist in the latter formalism. For a class of \textit{statistical} Hamiltonians which describe random processes, we have conjectured that, for a critical value of the Hamiltonian $E^\star$, there exists two global solutions of the stationary HJ equation with opposite stability for the corresponding reduced dynamics. Above $E^\star$, no orbits contain or reach a critical manifold, while below $E^\star$ the solutions are not global. Considering its degree of generality, this conjecture is challenging to prove in general, but our confidence stems from the fact that its contents seem to be necessary in order to have well-defined nonlinear processes emerging from a large-size limit.

Thanks to this conjecture, we have introduced the rectification of biased Lagrangians and Hamiltonians using a canonical transformation. We have shown that the generating function of this canonical transformation involves the stable Hamilton's characteristic function at the critical value $E^\star$, similarly to how the eigenvector for the highest eigenvalue appears in the Doob transform within the linear operator formalism. In practice, the rectification transforms the momenta and the critical value $E^\star$ to produce a proper Hamiltonian from the biased Hamiltonian. 

Given the significance of fluctuation relations for physical currents in nonequilibrium physics, we have also examined this symmetry in our framework. We have shown that biasing (or biasing and then rectifying) preserves the fluctuation relation although for modified affinities. Finally, we have defined a dual Hamiltonian as a rectification of the Hamiltonian with reversed momenta (associated with the time reversed dynamics). In this case, it is the unstable solution of the HJ equation associated with the original Hamiltonian that appears in the generating function of the rectifying canonical transformation, and its existence is also guaranteed as a consequence of our conjecture.

~~

{The following table gives a summary of the correspondence between some of the objects, conjectures and results presented in this paper, and their simpler equivalents for linear jump processes relying on the Perron-Frobenius theorem.}

{\begin{center}
\begin{tabular}{ |c|c| } 
 \hline
 Linear operator formalism & Hamiltonian formalism \\ 
 \hline\hline
 Biased Markov matrix $\bkappa_{\g}$ & Biased statistical Hamiltonian $\Hg$ \\
 \hline
 Nondegenerate largest eigenvalue $\mathcal{E}_{\g}$ & Nondegenerate min-max value $H_0^\star$ \\ 
 \hline
 Unique left/right eigenvectors $\mathbf{v}_{l,r}$ & Unique stable/unstable solutions to HJ $W_{\sm,\um}$ \\ 
 \hline
 $\mathbf{v}_{l,r}$ are positive & $W_{\sm,\um}$ are global\\
 \hline
 SCGF $\scgf=\mathcal{E}_{\g}$&SCGF $\scgfn=H_0^\star$\\
 \hline
 Rectified matrix $\mathbf{v}_l \bkappa_{\g}\mathbf{v}_l^{-1}-\mathcal{E}_{\g}$&Rectified Hamiltonian $H_{\g}(\bm{p}+\partial_{\bm{z}} W_{\sm},\bm{z}) - H_0^\star$\\
 \hline
 Rectified stationary state $\mathbf{v}_l \mathbf{v}_r$&Rectified quasipotential $W_{\um}-W_{\sm}$ (only fixed point)\\
 \hline
\end{tabular}
\end{center}}

~~

This work is a first step in the study of conditioned nonlinear Markov processes, and many interesting questions remain open, beyond that of proving our conjecture.

First and foremost, we have only looked at very simple examples, and in particular we were only able to describe the precise dynamics of one dimensional models whose critical manifolds can only be fixed points. The next logical step is to analyse systems which can sustain limit cycles, such as the Brusselator or other chemical networks \cite{Nguyen2018}. Such models have been considered in their non-biased version in the past \cite{Assaf2010}, but to our knowledge the question of their behaviour under conditioning, and even whether their limit cycles are maintained under biasing, are still open. It goes without saying that the same questions can be posed for strange attractors, though in this case they might be near impossible to tackle analytically, and a numerical approach could be better suited \cite{Gaspard2004}.

Another natural extension of our results would be to consider biased Hamiltonians that depend on time explicitly (either through the unbiased Hamiltonian and/or through the observable used for the bias). Preliminary results have been obtained for time periodic Hamiltonians \cite{These-Lydia}. In this case, the stationary Hamilton-Jacobi equation is replaced by its time-dependent version. The periodicity of the problem shall impose that the solution of the time dependent HJ equation equals itself after one period up to a constant that is related to the SCGF~\cite{Chabane2020}. The latter condition should determine uniquely the SCGF and the Hamilton's principal function that in turn can be used to define a canonical transformation realizing the rectification. Applications of these ideas remain be tested at first in the quasistatic approximation for instance.

On the numerical side of the field, a fairly recent and quite significant advancement was made with the invention of so-called cloning algorithms \cite{Giardina2011_vol145} or algorithms for adaptative sampling of large deviations \cite{Ferre2018vol172}, which allow to efficiently simulate the dynamics of biased linear Markov processes and extract both the SCGF and the dominant eigenvectors even for very large systems. A good question and potential application of our approach is whether this can be extended to nonlinear processes in a way which would be more efficient than running them on a large but finite version of the system. Disposing of such an algorithm could also be useful outside of the field of nonlinear Markov processes by providing techniques to solve the HJ equation for statistical Hamiltonians.

\subsection*{Acknowledgement}

A.L. was supported by the Belgian Excellence of Science (EOS) initiative through the project 30889451 PRIMA Partners in Research on Integrable Systems and Applications.
% Appendices-----------------------------------------

\appendix
\section*{Appendix}
\addcontentsline{toc}{section}{Appendices}

This appendix deals with the biasing and rectification of Markov diffusion processes. We first review the case of a single process in the linear operator framework, and then we study the case of $\un$ independent processes within the Lagrangian--Hamiltonian formalism.

\section{Single diffusion process}  \label{Sec_single_diff}

We consider a diffusion process described by the following Langevin equation
\begin{equation} \label{diffusive_process}
\dot{x}_t = b(x_t) + \sigma(x_t) \circ \xi_t,
\end{equation}
where $x_t$ is a random variable and $\xi_t$ a Gaussian white noise of mean $\braket{\xi_t} = 0$ and variance $\braket{\xi_t \xi_{t'}}\,=\,\delta(t-t')$. The stochastic integrals are defined according to the mid-point Stratonovich convention, referred to by a circle $\circ$. The drift $b$ and the diffusion $\sigma$ are functions of $x_t$ and do not depend explicitly on time. The probability density $\varrho(x,t)$ satisfies the Fokker–Planck equation 
\begin{equation} \label{FK1}
\frac{\partial \varrho(x,t)}{\partial t} = - \nabla J(x,t),
\end{equation}
with $\nabla$ the derivative with respect to $x$ and 
\begin{equation} \label{jdiffusion}
J^{\varrho}(x,t) \equiv \hat{b}(x) \varrho(x,t) - \frac{1}{2} \sigma(x)^2 \nabla \varrho(x,t),
\end{equation}
where we introduced the modified drift $\hat{b}(x) \equiv b(x) - \frac{1}{2} \sigma(x) \nabla \sigma(x)$. The Fokker-Planck equation appears to be a continuity equation that conserves the normalization of the probability density, i.e. $\int \ud x \varrho(x,t) =1 , \forall t$. From now on, unspecified integrations are implicitly on $x$, e.g.  $\int \varrho =1$.
One can rewrite the Fokker-Planck equation as
\begin{equation}
\frac{\partial \varrho}{\partial t} = \fp \varrho,
\end{equation}
where $\fp$ is the Fokker-Planck operator in the Stratonovich convention, defined by its action on a function $\varphi$:
\begin{equation} \label{FK_operator}
\fp \varphi(x) \equiv - \nabla \left[ \hat{b}(x) \varphi(x) \right] + \frac{1}{2} \nabla \left[ \sigma(x)^2 \nabla \varphi(x) \right]
\end{equation}
The adjoint Fokker-Planck operator $\fp^\dagger$ is given by
\begin{equation}
\fp^\dagger \varphi(x) = \hat{b}(x) \nabla \varphi(x)  + \frac{1}{2} \nabla \left[ \sigma(x)^2 \nabla \varphi(x) \right],
\end{equation}
both operators being related by
\begin{equation}
\int   \left(\fp \varphi\right) \psi = \int   \left(\fp^\dagger \psi\right) \varphi,
\end{equation}
for any functions $\varphi$, $\psi$. 
We denote a path by $[x_t]$, with $x_t$ the state of the system at time $t$. The path probability of $[x_t]$ between the initial time $0$ and the final time $t$ within the Stratonovich convention reads~\cite{Lau2007}
\begin{equation} \label{pathproba_diffusion}
\uP_{b,\sigma,\varrho(0)}[x_t] = \varrho(x_0,0) \exp \left\{ - \int_0^t \ud t' \left[ \left(\dot{x}_{t'} - \hat{b}(x_{t'}) \right)^2 + \frac{1}{2} \nabla b(x_{t'}) \right]  \right\},
\end{equation}
Many dynamical and thermodynamic observables such as heat, matter currents, work, entropy production, etc. can be written as linear combinations of the empirical occupancy $\tilde{\rho}_t$ and the empirical current $\tilde{j}_t$.
The function $\tilde{\rho}_t(x)$ counts the rate of occupancy of the position $x$ along the trajectory $[x_t]$:
\begin{equation} \label{emp_occupancy_1diff}
    \tilde{\rho}_t(x) = \frac{1}{t} \int_{0}^{t} \ud\t \delta(x_\t-x),
\end{equation}  
while the function $\tilde{j}_t(x)$ informs on the time-averaged local velocity at $x$~\cite{Chetrite2014}:
\begin{equation} \label{emp_curr_1diff}
    \tilde{j}_t(x) = \frac{1}{t} \int_0^t \ud \t \delta(x_\t-x) \circ \dot x_\t, 
\end{equation}
where the circle $\circ$ refers to the Stratonovich  convention. We would like to condition our original Markov process by filtering the ensemble of paths to select those leading to a chosen value of $\tilde \A_t(x) = ( \tilde{j}_t(x), \tilde{\rho}_t(x))$, for each state $x$. This defines the conditioned process for which we aim to find an equivalent Markov process in the long-time limit, namely the driven process~\cite{ChetriteHDR}.
This process is described by the microcanonical path probability~\cite{Chetrite2014}
\begin{equation} \label{pathprobamicrodiff}
\uPmicrodiff = \uPkconddiff.
\end{equation}
To explicit the generator of the driven process, let us introduce the generating function
\begin{equation} \label{gen_func_diffusion}
G_{\g}(x,t) \equiv \left\langle \e^{t \g \cdot \tilde\A_{t} } \delta(x_t - x) \right\rangle_{\varrho(0)},
\end{equation}
where $\braket{\cdots}_{\varrho(0)}$ is the path average based on \eqref{pathproba_diffusion}, $\g(x) \equiv \left(  \gamma_1(x) , \gamma_2(x) \right)$ is the conjugate variable of $\tilde\A_t(x)$ and where the dot stands for the scalar product $\g \cdot \tilde\A_{t} = \int \ud x \left[ \gamma_1(x) \tilde{j}_t(x) + \gamma_2(x) \tilde{\rho}_t(x) \right]$. From now on, we drop in the notation the $x$-dependency of the functions for clarity. The generating function \eqref{gen_func_diffusion} evolves according to
\begin{equation}
\dot{G}_{\g} = \Lambda_{\g} G_{\g},
\end{equation}
where the biased Fokker-Planck operator $\Lambda_{\g}$ is given by~\cite{Chetrite2014}:
\begin{equation} \label{biased_FP_op_single}
\Lambda_{\g} \varphi \equiv (- \nabla + \gamma_1) ( \hat{b} \varphi ) + \frac{1}{2} (- \nabla + \gamma_1) \left[ \sigma^2 (-\nabla + \gamma_1) \varphi \right] + \gamma_2 \varphi,
\end{equation}
One can compute its adjoint operator
\begin{equation}  \label{adjoint_biased_FP_op_single}
\Lambda_{\g}^\dagger \varphi = \hat{b} (\nabla + \gamma_1) \varphi  + \frac{1}{2} (\nabla + \gamma_1) \left[ \sigma^2 (\nabla + \gamma_1) \varphi \right] + \gamma_2 \varphi.
\end{equation}
The biased Fokker-Planck operator $\Lambda_{\g}$ generates a Markov process that is not norm-conserving since $\int \Lambda_{\g} \varrho \neq 0$. As with Markov jump processes, we can build the operator of the driven process $\fpr$ by taking the Doob transform of the biased operator $\Lambda_{\g}$ associated with its dominant left eigenfunction:
\begin{equation} \label{doob_diffusion}
\fpr \varphi \equiv  l \Lambda_{\g} (l^{-1} \varphi) - l (\Lambda_{\g}^\dagger l^{-1}) \varphi, 
\end{equation}
with $l \equiv l(x)$ being the left eigenfunction of $\Lambda_{\g}$ for the highest eigenvalue $\scgfn$
\begin{equation} \label{eigenv_diffusion}
\Lambda_{\g}^\dagger l = \scgfn l.
\end{equation}
Since the Krein-Rutman theorem ensures the positivity of $l$, we introduce a new function $u \equiv u(x)$ such that $l \equiv \e^u$. It follows from \eqref{eigenv_diffusion}:
\begin{equation} \label{scgf_diffusion}
\scgfn = \int \e^{-u} (\Lambda_{\g}^\dagger \e^u) \rho.
\end{equation}
Computing explicitly Eq.~\eqref{doob_diffusion} using Eqs.~(\ref{biased_FP_op_single}--\ref{adjoint_biased_FP_op_single}) and $l = \e^u$, we finally find that the driven Fokker-Planck operator is given by
\begin{equation} \label{FK_driven_generator}
\fpr \varphi =  - \nabla \left[ \hat{B}_\gamma \varphi  - \frac{1}{2} \sigma^2 \nabla \varphi \right],
\end{equation}
where we introduced the rectified drift
\begin{equation} \label{driven_drif}
\hat{B}_{\g} \equiv  \hat{b} + \sigma^2 (\nabla u + \gamma_1).
\end{equation}
The driven process is thus a diffusive process obeying the same stochastic equation \eqref{diffusive_process} as the original process but with a new drift \eqref{driven_drif}. Note that the dependence of $u$ on $\g$ is made implicit.

\section{$\un$ independent diffusion processes}  \label{Sec_many_indep_diff}

We consider $\un$ independent and identical systems, each one modeled by a time-homogeneous Markov diffusion process of time-independent drift $b$ and diffusion coefficient $\sigma$. We denote by $\nu \in \{1, 2, \dots \un\}$ the $\nu^\text{th}$ system and by $x^\nu_t$ the stochastic process of the system $\nu$ which evolves according to the Langevin equation
\begin{equation} \label{Ndiffusive_process}
\dot{x}^\nu_t = b(x_t^\nu) + \sigma(x_t^\nu) \xi^\nu_t,
\end{equation}
We are interested in the empirical occupation density:
\begin{equation}
\rho(x,t) = \frac{1}{N} \sum_{\nu=1}^N \delta(x^\nu_t-x) ,
\end{equation}
and the empirical current:
\begin{equation}
j(x,t) = \frac{1}{N} \sum_{\nu=1}^N \delta(x^\nu_t-x) \circ \dot{x}^\nu_t,
\end{equation}
playing respectively the role of the variables $\z$ and $\bl$ in our general framework of section \ref{LHdescription}. The empirical occupation density gives the density of systems being at a state in $[x,x+\ud x[$ at time $t$, and the empirical current measures the density of systems performing a displacement between $x$ and $x+\ud x$ within the time interval $[t,t+\ud t[$. Both variables are related by
\begin{equation}  \label{contineq_Ndiff}
\dot{\rho}(x,t) = - \nabla j(x,t),
\end{equation}   
with $- \nabla$ playing the role of $\D$. Notice that these observables are related to the empirical occupancy  $\tilde{\rho}_t^\nu$~\eqref{emp_occupancy_1diff} and the empirical transition current $\tilde{j}_t^\nu$~\eqref{emp_curr_1diff} for a single process through 
\begin{align}   \label{link_single_global_diff}
\begin{split}
\frac{1}{t} \int_0^t \ud \t \,  \rho(\t) &= \frac{1}{\un} \sum_{\nu = 1}^N \tilde{\rho}_t^\nu, \\
\frac{1}{t} \int_0^t \ud \t \, j(\t) &= \frac{1}{\un} \sum_{\nu = 1}^N \tilde{j}_t^\nu,
\end{split}
\end{align}
where the superscript $\nu$ indicates that the empirical occupancy or current are those for the trajectory of the $\nu^{\text{th}}$ system.

\subsection{Stochastic equation for the empirical occupation density}

We aim to give a coarse-grained description of the global system by deriving the stochastic equation for $\rho$. To do so, we compute the quantity: $\Delta x^\nu~\equiv~x^\nu_{t+\Delta t}~-~x^\nu_t$. Using the Langevin equation \eqref{diffusive_process} for $x^\nu_t$, we get:
\begin{equation} \label{step21}
\Delta x^\nu = \int_t^{t+\Delta t} \ud \t \left[ b(x^\nu_\t) + \sigma(x_\t^\nu) \circ \xi^\nu_\t \right].
\end{equation}
For an infinitesimal $\Delta t$, we have in the Stratonovich convention~\cite{Lau2007}:
\begin{align} \label{step22}
\int_t^{t+\Delta t} \ud \t \, b(x^\nu_\t)  & \simeq b (x^\nu_t + \frac{\Delta x^\nu}{2} ) \Delta t, \\
\int_t^{t+\Delta t} \ud \t \, \left[ \sigma(x_\t^\nu) \xi^\nu_\t \right] & \simeq \sigma(x^\nu_t + \frac{\Delta x^\nu}{2} ) \int_t^{t+\Delta t} \ud \t \, \xi^\nu_\t.
\end{align}
Expanding up to order $\Delta t$ and using the identity $(\int_t^{t+\Delta t} \xi_t \ud t)^2 = \Delta t$ when $\Delta t \rightarrow 0$~\cite{Gardiner1985, sekimoto2010stochastic}, it follows
\begin{align}
\Delta x^\nu & \simeq b(x^\nu_t) \Delta t + \sigma(x^\nu_t) \int_t^{t+\Delta t} \xi^\nu_{\t} \ud \t + \frac{1}{2} \label{dx} \sigma(x^\nu_t) \nabla \sigma(x^\nu_t) \Delta t, \\ 
(\Delta x^\nu)^2 & \simeq \sigma(x^\nu_t)^2 \Delta t. \label{dx2}
\end{align}
We now compute $\rho(x,t+\Delta t) = \frac{1}{N} \sum_{\nu=1}^N \delta(x^\nu_{t+\Delta t}-x)$. Let $\varphi$ be a test function, then
\begin{align}
\int \ud x \varphi(x) \rho(x,t+\Delta t) &= \frac{1}{N} \sum_{\nu = 1}^N \varphi(x_{t+\Delta t}^\nu) \\
&= \frac{1}{N} \sum_{\nu = 1}^N \varphi(x_{t}^\nu + \Delta x^\nu) \\
&\simeq  \frac{1}{N} \sum_{\nu = 1}^N \varphi(x_t^\nu) + \frac{1}{N} \sum_{\nu = 1}^N \Delta x^\nu \varphi'(x_t^\nu) + \frac{1}{N} \sum_{\nu = 1}^N \frac{1}{2}  (\Delta {x^\nu})^2 \varphi''(x_t^\nu), \label{aux}
\end{align}
where we used Taylor's formula around $x^\nu_t$ up to second order in $\Delta x^\nu$ in the last equation. Using Eqs. (\ref{dx} -- \ref{dx2}) and the fact that $\frac{1}{N} \sum_{\nu=1}^N \varphi(x_t^\nu) = \int \ud x \varphi(x) \rho(x,t)$, Eq.~\eqref{aux} gives
\begin{equation}
\int \ud x \varphi(x) \dot{\rho}(x,t) = \int \ud x \varphi(x) \left\{ - \nabla \left[\hat{b}(x) \rho(x,t) - \frac{1}{2} \sigma(x)^2 \nabla \rho(x,t) + \sigma(x) \sqrt{\frac{\rho(x,t)}{N}} \eta(x,t)    \right]  \right\}, \label{langevinNderiv}
\end{equation}
with $\dot{\rho}(x,t) = \lim_{\Delta t \rightarrow 0} \frac{\rho(x,t+\Delta t) - \rho(x,t)}{\Delta t}$ and where we introduced 
\begin{equation}
\eta(x,t) \equiv  \frac{1}{\sqrt{\un \, \rho(x,t)} } \sum_{\nu=1}^\un \delta(x-x_t^\nu) \bar{\xi}^\nu_t,
\end{equation}
with $\bar{\xi}^\nu_t \equiv \lim_{\Delta t \rightarrow 0}  \frac{1}{\Delta t}  \int_t^{t+\Delta t} \ud \t \xi_\t^\nu$. The stochastic process $\eta$ is a Gaussian white noise in time and space~\cite{Dean1996, Bouchet2016} with mean and variance
\begin{eqnarray}
\braket{\eta(x,t)} &=& 0, \\
\braket{\eta(x,t) \eta(x',t')} &=& \delta(x-x')  \delta(t-t').
\end{eqnarray}
Since \eqref{langevinNderiv} is valid for any function $\varphi$, we obtain the stochastic equation for the density $\rho$:
\begin{equation}  \label{deaneq}
\dot{\rho}(x,t) = - \nabla \left[\hat{b}(x) \rho(x,t) - \frac{1}{2} \sigma(x)^2 \nabla \rho(x,t) + \sigma(x) \sqrt{\frac{\rho(x,t)}{N}} \eta(x,t)    \right].
\end{equation}
Eq.~\eqref{deaneq} is known as the \textit{Dean equation}.

\subsection{Derivation of the Lagrangians and Hamiltonians}

In order to obtain the Lagrangian $\L(j,\rho)$, we compute the conditional probability $P_{\delta t}(j \mid \rho)$ using Eqs.~(\ref{contineq_Ndiff}, \ref{deaneq}):
\begin{equation}
P_{\delta t}(j \mid \rho) = \prod_x \left\langle\delta \left[j - \hat{b} \rho + \frac{1}{2} \sigma^2 \nabla \rho - \sigma \sqrt{\frac{\rho}{\un}} \eta \right]\right\rangle_{\eta}.
\end{equation} 
The continuous product $\prod_x$ runs over the states $x_\ell \equiv \ell \delta x$ with $\ell$ an integer and $\delta x$ an infinitesimal space step, and $\braket{\cdots}_\eta $ is the average over the noise $\eta$:
\begin{equation}
\left\langle \mathcal{O}(\eta) \right \rangle_\eta \equiv \frac{1}{\mathsf{N}} \int \mathcal{O}(\eta) \e^{-\frac{1}{2} \delta t \delta x \eta^2} \ud \eta,
\end{equation}
with $\mathsf{N}$ the normalization factor and $O(\eta)$ an arbitrary function of $\eta$. We have dropped the $(x,t)$-dependency in all functions for clarity. It follows
\begin{align}
P_{\delta t}(j \mid \rho) &=  \frac{1}{\mathsf{N}} \int \ud \eta \prod_{x} \e^{-\frac{1}{2} \delta t \delta x \eta^2} \delta \left[j - \hat{b} \rho + \frac{1}{2} \sigma^2 \nabla \rho - \sigma \sqrt{\frac{\rho}{\un}} \eta \right] \nonumber \\
&= \frac{1}{\mathsf{N}} \int \ud \eta \prod_{x} e^{-\frac{1}{2} \delta t \delta x \eta^2} \frac{\sqrt{\un}}{\sigma \sqrt{\rho}} \delta \left[ \eta - \frac{ j - \hat{b} \rho +  \frac{1}{2} \sigma^2 \nabla \rho}{\frac{1}{\sqrt{\un}} \sigma \sqrt{\rho}} \right] \nonumber \\
&= \frac{1}{\mathsf{N}} \exp{ \left[ - \delta t \int  \frac{\un}{2\sigma^2 \rho} \left(j - \hat{b} \rho +  \frac{1}{2} \sigma^2 \nabla \rho \right)^2 + \frac{1}{2} \int \ln\left(\frac{\un}{\rho \sigma^2}\right)\right]},
\end{align} 
where we used in the second equality the relation $\delta(\varphi(y)) = \lvert \varphi'(y_0) \rvert^{-1} \delta(y-y_0)$, for any smooth test function $\varphi$ and any root $y_0$ of $\varphi(y)=0$. In the limit of large $\un$, the last term in the exponential is asymptotically dominated by $\un$ and we obtain
\begin{equation}
P_{\delta t}(j \mid \rho) \apeupn \e^{- \un \delta t \L(j,\rho)},
\end{equation}
where the Lagrangian is given by
\begin{equation} \label{LDF_diffusion}
\L(j,\rho) = \int \frac{1}{2\sigma^2 \rho} \left(j - J^\rho \right)^2,
\end{equation}
with $J^\rho = \hat b \rho - \frac{1}{2} \sigma^2 \nabla \rho$. Computing the Legendre transform of $\L$ with respect to $j$ yields the detailed Hamiltonian
\begin{equation} \label{Hamilonian_diffusion}
\H(f,\rho) = \int f \left[ \frac{1}{2} \sigma^2 f \rho + J^\rho \right].
\end{equation}
We are interested in the observable
\begin{equation} \label{observable_Ndiff}
\A_{t}(x) \equiv  \frac{\un}{t} \left(
\begin{array}{c}
\int_0^t \ud \t j(x,\t)  \\
\int_0^t \ud \t \rho(x,\t) 
\end{array} \right).
\end{equation}
Using the results of Section~\ref{Sec_biasedHL}, the dynamical fluctuations of $\A$ are encoded in the biased Lagrangian and Hamiltonian
\begin{align} \label{L_biaise_diffusion}
\L_{\g}(j,\rho) &= \L(j,\rho) - \gamma_1 \cdot j - \gamma_2 \cdot \rho, \\
\H_{\g}(f,\rho) &= \int (f+\gamma_1) \left[ \frac{1}{2} \sigma^2 (f+\gamma_1) \rho + J^\rho \right] + \gamma_2 \cdot \rho = \H(f+\gamma_1,\rho) + \gamma_2 \rho, \label{H_biaise_diffusion}
\end{align}
where $\gamma_1(x)$ (resp. $\gamma_2(x)$) is conjugated to the first (resp. second) component of $\A(x)$.

\subsection{SCGF and HJ equation}

In order to derive the rectified Hamiltonian, we first need to translate the spectral properties of the biased generator from the linear operator formalism to the Hamiltonian formalism. Because of the independence of the $\un$ processes, it suffices to look at a single process. Indeed, we can relate the SCGF $\scgf$ of the global system to the SCGF $\scgfn$ of the single process by:
\begin{align}
\scgf &= \lim_{t \rightarrow \infty} \frac{1}{t} \ln \left\langle \e^{t \g \cdot \A_t} \right\rangle_{x_0} \\
& = \lim_{t \rightarrow \infty} \frac{1}{t} \ln \left\langle \e^{\un \int_0^t \ud \t \gamma_1 \cdot j(\t) + \un \int_0^t \ud \t \gamma_2 \cdot \rho(\t)} \right\rangle_{x_0} \label{globaltosingle} \\
&= \lim_{t \rightarrow \infty} \frac{1}{t} \ln \left\langle \e^{t \sum_{\nu = 1}^N \left( \gamma_1 \cdot \tilde{j}_t^\nu + \gamma_2 \cdot \tilde{\rho}_t^\nu \right)} \right\rangle_{x_0} \label{globaltosingle0} \\
&= \un \lim_{t \rightarrow \infty} \frac{1}{t} \ln \left\langle \e^{t 
\left( \gamma_1 \cdot \tilde{j}_t^\nu + \gamma_2 \cdot \tilde{\rho}_t^\nu \right)} \right\rangle_{x_0} \label{globaltosingle1} \\
&= \un \scgfn,
\end{align}
where we used Eq.~\eqref{observable_Ndiff} in Eq.~\eqref{globaltosingle}, Eq.~\eqref{link_single_global_diff} in Eq.~\eqref{globaltosingle0} and the fact that the $\un$ processes are independent and identically distributed in Eq.~\eqref{globaltosingle1}. Computing explicitly the right-hand-side of Eq.~\eqref{scgf_diffusion}, we find
\begin{equation}
\scgfn = \H_{\g}(f=\nabla u, \rho),
\end{equation}
with $\nabla = (-\nabla)^\dag$. As expected, the function $u$ appearing in the left eigenfunction of the single-process biased Fokker-Planck operator is the solution of the HJ equation and we write $u = \partial_{\rho} W_{\sm}$. 

\subsection{Rectified Hamiltonian}

The rectified Hamiltonian follows from Eq.~\eqref{Hr_det}:
\begin{equation} \label{Hdriven_diffusion}
\H^r(f,\rho;\g) = \H_{\g}(f+\nabla u, \rho) - \H_{\g}(\nabla u, \rho),
\end{equation}
leading after explicit computation to
\begin{equation} \label{H_doob_diffusion}
\H^{r}(f,\rho;\g) = \int  f \left[ \frac{1}{2} \sigma^2 f \rho + J^{\ur,\rho}_{\g} \right],
\end{equation}
with $J^{\ur,\rho}_{\g} \equiv \hat{B}_{\g} - \frac{1}{2} \sigma^2 \nabla \rho$, where the rectified drift $\hat{B}_{\g}$ is defined in Eq.~\eqref{driven_drif}. Unsurprisingly, the rectified Hamiltonian corresponds to an unbiased Hamiltonian associated with the drift $\hat{B}_{\g}$ of the driven process obtained from a Doob transform as seen in Eq.~\eqref{driven_drif}. This illustrates the fact that the rectification of biased Hamiltonians is equivalent to the rectification of biased generators in the linear operator formalism using the Doob transform. 

%Biblio ---------------------------------------------
\bibliographystyle{ieeetr}
%\bibliography{/home/lydia/Bureau/Biblio/Biblio}
\bibliography{BibNonlinear}
\addcontentsline{toc}{section}{Bibliography}

\end{document}